\documentclass{article}

\usepackage{arxiv}

\usepackage[utf8]{inputenc} 
\usepackage[T1]{fontenc}    
\usepackage{hyperref}       
\usepackage{url}            
\usepackage{booktabs}       
\usepackage{amsfonts}       
\usepackage{nicefrac}       
\usepackage{microtype}      
\usepackage{lipsum}
\usepackage{graphicx}
\graphicspath{ {./images/} }

\usepackage{siunitx}
\usepackage{bm}
\usepackage{amsmath, amssymb} 
\usepackage{mathrsfs}
\usepackage{graphicx}        
\usepackage{tikz}            
\usepackage{pgfplots}        
\pgfplotsset{compat=1.18}    
\usepackage{caption}         
\usepackage{subcaption}      
\usepgfplotslibrary{groupplots}

\usepackage{placeins}

\usetikzlibrary{intersections}
\usetikzlibrary{positioning}
\usepgfplotslibrary{fillbetween}

\definecolor{rwth1}{RGB}{0,84,159}      
\definecolor{rwth2}{RGB}{142,186,229}   
\definecolor{rwth3}{RGB}{0,97,101}      
\definecolor{rwth4}{RGB}{0,152,161}     
\definecolor{rwth5}{RGB}{87,171,39}     
\definecolor{rwth6}{RGB}{189,205,0}     
\definecolor{rwth7}{RGB}{255,237,0}     
\definecolor{rwth8}{RGB}{246,168,0}     
\definecolor{rwth9}{RGB}{227,0,102}     
\definecolor{rwth10}{RGB}{204,7,30}     
\definecolor{rwth11}{RGB}{161,16,53}    
\definecolor{rwth12}{RGB}{97,33,88}     
\definecolor{rwth13}{RGB}{122,111,172}  

\definecolor{rwthb1}{HTML}{e8f1fa}      
\definecolor{rwthb2}{HTML}{c7ddf2}      
\definecolor{rwthb3}{HTML}{8ebae5}      
\definecolor{rwthb4}{HTML}{407fb7}      
\definecolor{rwthb5}{HTML}{00549f}      

\definecolor{rwtho1}{HTML}{fff7ea}      
\definecolor{rwtho2}{HTML}{feeac9}      
\definecolor{rwtho3}{HTML}{fdd48f}      
\definecolor{rwtho4}{HTML}{fabe50}      
\definecolor{rwtho5}{HTML}{f6a800}      

\definecolor{lightblue}{RGB}{173,216,230}

\definecolor{losscyan}{RGB}{0,180,180}
\definecolor{lossmagenta}{RGB}{220,20,120}
\definecolor{lossamber}{RGB}{255,190,40}

\definecolor{rOne}{RGB}{0,180,180}   
\definecolor{rTwo}{RGB}{220,20,120}  

\definecolor{rcol1}{HTML}{8C9FB1}   
\definecolor{rcol2}{HTML}{005287}    
\definecolor{rcol3}{HTML}{C50F3C}     
\definecolor{rcol4}{HTML}{B9D9EC}   

\title{A Complement to Neural Networks for Anisotropic Inelasticity at Finite Strains}

\author{
Hagen Holthusen \\
  Institute of Applied Mechanics\\
  University of Erlangen-Nuremberg\\
  Erlangen, 91058, Germany \\
  \texttt{hagen.holthusen@fau.de} \\
   \And
Ellen Kuhl \\
  Department of Mechanical Engineering\\
  Stanford University\\
  Stanford, CA 94305, United States \\
  \texttt{ekuhl@stanford.edu} \\
}

\begin{document}
\maketitle
\begin{abstract}
  We propose a complement to constitutive modeling that augments neural networks with material principles to capture anisotropy and inelasticity at finite strains. 
The key element is a dual potential that governs dissipation, consistently incorporates anisotropy, and—unlike conventional convex formulations—satisfies the dissipation inequality without requiring convexity.  

Our neural network architecture employs invariant-based input representations in terms of mixed elastic, inelastic and structural tensors.
It adapts Input Convex Neural Networks, and introduces Input Monotonic Neural Networks to broaden the admissible potential class. 
To bypass exponential-map time integration in the finite strain regime and stabilize the training of inelastic materials, we employ recurrent Liquid Neural Networks. 

The approach is evaluated at both material point and structural scales. 
We benchmark against recurrent models without physical constraints and validate predictions of deformation and reaction forces for unseen boundary value problems. 
In all cases, the method delivers accurate and stable performance beyond the training regime. 
The neural network and finite element implementations are available as open-source and are accessible to the public via \href{https://doi.org/10.5281/zenodo.17199965}{Zenodo.org}.
\end{abstract}


\section{Introduction}
Artificial intelligence (AI) has increasingly permeated diverse areas of computational mechanics. 
Its main advantage lies in its intrinsic ability to handle complex and high-dimensional data while remaining continuously adaptable. 
Applications range from inverse design \cite{lu2021physics,JADOON2025106161}, topology optimization \cite{GHOULI2025106210}, multiscale modeling \cite{STOLLBERG2025117808,DANOUN2024117192,Kalina2023}, and multiphysics problems \cite{Abueidda2024,fuhg2024polyconvexneuralnetworkmodels}, to model order reduction \cite{kehls2025autoencoderbasednonintrusivemodelorder,hesthaven2018non}, the identification of material behavior \cite{Flaschel2021,Linka2021,Klein2022}, and real-time digital twins \cite{CHIACHIO2022104333}. 
In all these contexts, the development of improved models requires more data to enhance predictive accuracy and deepen our understanding of the underlying mechanics. 
At the same time, however, larger datasets generally increase the complexity of the models themselves. 
Overcoming this apparent contradiction highlights the particular promise of AI in mechanics.

Beyond its methodological benefits, the current momentum in AI research—both in terms of algorithmic development and in its application across disciplines—represents an additional advantage \cite{duede2024oilwaterdiffusion,Hajkowicz_2023}. 
One may question whether AI always constitutes the most efficient solution, or whether in certain cases its application may be disproportionate to the complexity of the problem. 
Nevertheless, disregarding the accumulated expertise and its adaptation to computational mechanics would be a missed opportunity.

Compared to fields such as the social sciences, AI in mechanics benefits from the existence of uncertainty-free physical laws. 
Recent years have therefore seen considerable efforts to incorporate these laws into AI models. 
Although early applications of neural networks to material modeling date back more than 30 years \cite{Ghaboussi1991}, a major boost was provided by the introduction of physics-informed neural networks (PINNs) \cite{RAISSI2019686}, where governing equations are weakly enforced through additional loss function terms. 
An alternative paradigm designs neural network architectures that satisfy physical principles \textit{a priori}, thereby guaranteeing that predictions adhere to the underlying laws. 
Prominent approaches include the unsupervised EUCLID framework \cite{Flaschel2021}, Thermodynamics-based Artificial Neural Networks (TANNs) \cite{MASI2021104277,MASI2022115190}, Constitutive Artificial Neural Networks (CANNs) \cite{Linka2021,Linka2023}, and Physics-Augmented Neural Networks (PANNs) \cite{Klein2022,Linden2023}. 
While all of these methods enforce physical consistency, they differ in the balance between interpretability and expressivity: CANNs favor sparse networks that allow tracing each computational path, whereas PANNs rely on denser architectures with potentially greater representational power. 
Although originally developed for elastic materials, recent research has focused on extending these frameworks to (anisotropic) inelastic behavior—representing a natural progression towards handling ever-increasing model complexity in order to explain increasingly rich datasets.

\subsection{State-of-the-art in neural networks for inelastic and anisotropic material modeling}

\paragraph{Inelastic materials.}
The EUCLID framework has been extended in recent years to cover a broad spectrum of inelastic phenomena. Early developments demonstrated its applicability to the discovery of plasticity \cite{Flaschel2022}, with particular emphasis on non-associative, pressure-sensitive plasticity \cite{Xu2025}, as well as to the modeling of linear viscoelastic material behavior at small strains \cite{MARINO2023104643}. In a more general context, \cite{FLASCHEL2023115867} embedded EUCLID in the theoretical setting of Generalized Standard Materials (GSM) \cite{Halphen1975,HACKL1997667}, a widely recognized framework for capturing inelasticity. GSM is based on the assumption of an inelastic potential complementing the Helmholtz free energy, thereby ensuring thermodynamic consistency. 

The idea of postulating a potential that governs the evolution of inelastic deformations has inspired a variety of neural network architectures aiming to replicate this principle; see, for example, \cite{FLASCHEL2025106103,HUANG2022104856,Holthusen2024,BENADY2024116967,Rosenkranz2024,Dettmer2024}. While GSM typically relies on potentials expressed in terms of inelastic strain rates, one may alternatively consider potentials formulated in terms of stresses. This stress-based perspective is often employed when dealing with yield functions, such as the von Mises criterion, and naturally connects to convex analysis. Neural networks exploiting such formulations have recently been applied to a wide range of inelastic processes, including finite viscoelasticity \cite{ASAD2023116463,Holthusen2025}, elasto-plasticity \cite{Jadoon2025,Boes2025}, fracture mechanics \cite{DAMMA2025117937}, and growth and remodeling phenomena \cite{Holthusen2025growth}.
Notably, stress-based potentials and inelastic strain rate-based potentials share an intrinsic connection: if both are convex, they can be transformed into one another by means of a Legendre transformation, a relation that also holds at finite strains \cite{Holthusen2025}.

In the spirit of \cite{Jadoon2025}, elasto-plasticity at small strains was recently extended to intrinsically incorporate anisotropy. Two strategies were proposed in \cite{jadoon2025thermodynamicallyconsistenthybridpermutationinvariant}: the first augments the classical Hill-48 yield criterion with an Input Convex Neural Network (ICNN) \cite{Amos2017}, thereby preserving convexity and guaranteeing stability, while the second relies entirely on a neural network in the principal stress space, where anisotropy is embedded through additional permutation-invariant networks.

In the training of neural networks for inelastic material modeling, the availability of high-quality data is a central challenge. While most physics-informed networks are formulated at the material point level, the primary experimental data typically available consists of macroscopic force–displacement curves. Consequently, finite element model updating strategies are frequently employed to generate synthetic data for training, also in the inelastic regime \cite{KUMAR2025118159}. 

Beyond GSM-inspired formulations, several alternative paradigms have emerged for tackling inelastic material behavior with neural networks. A first direction focuses on dynamical system representations: \cite{Jones2022} introduced a neural ordinary differential equation framework that models inelastic stress response via internal state variables, effectively replacing classical evolution equations by a data-driven continuous-time formulation. 
Building on this idea, \cite{jones2025attentionbasedneuralordinarydifferential} extended the neural ODE framework by incorporating attention mechanisms, thereby enhancing its capability to capture complex history-dependent inelastic processes with improved generalization across diverse loading paths.
Similarly, recurrent architectures have been explored, as in \cite{BORKOWSKI2022106678}, where a physics-informed recurrent neural network was developed for multiaxial plasticity, ensuring thermodynamic consistency by enforcing monotonic plastic work. Building upon this time-series perspective, long short-term memory (LSTM) networks were proposed in \cite{LI2025105325} to capture strain path dependence in crystal plasticity, which offers the possibility of orders of magnitude speed-ups compared to traditional simulations while preserving accuracy in path-dependent effects such as the Bauschinger phenomenon. 

A second direction emphasizes operator learning. The work of \cite{GUO2025118358} proposed the History-Aware Neural Operator (HANO), combining Fourier neural operators with hierarchical self-attention to achieve discretization-invariant modeling of path-dependent inelasticity, including anisotropic damage. This approach alleviates the limitations of recurrent models by avoiding hidden state inconsistencies and by generalizing across varying load histories. In parallel, graph-based models have been employed to capture spatially distributed responses: \cite{Maurizi2022} demonstrated that graph neural networks can accurately predict stress, strain, and displacement fields in materials and structures, highlighting their potential for bridging microstructure-informed modeling with structural-scale predictions. 

\paragraph{Anisotropic materials.}
The modeling of anisotropy with neural networks has gained increasing attention due to the ubiquity of direction-dependent material behavior in both engineering and biological applications \cite{patel2025generalautomatedmethodbuilding}. A central idea is to encode preferred material directions through structural tensors, thereby ensuring frame invariance and compatibility with material symmetry groups. Several approaches have been developed along these lines. For instance, \cite{TAC2022115248} introduced polyconvex neural ordinary differential equations for tissue mechanics, ensuring thermodynamic admissibility and convexity while capturing the nonlinear anisotropic response of skin. Complementary to this, probabilistic machine learning strategies have been proposed by \cite{FUHG2022114915}, who employed Gaussian process regression and physics-informed sampling techniques to build anisotropic hyperelastic models that respect material symmetries and thermodynamic consistency. More recently, \cite{KALINA2025117725} presented a framework of Physics-Augmented Neural Networks (PANNs) that leverage generalized structural tensors of higher order. This formulation allows for the simultaneous calibration of network parameters and structural tensors, thereby enabling the detection and accurate representation of complex anisotropy arising from heterogeneous microstructures.

In addition to tensor-based formulations, automated model discovery has been explored for highly anisotropic engineering materials. A notable example is the work of \cite{MCCULLOCH2024461}, who combined biaxial testing with constitutive neural networks to uncover the unique anisotropic mechanical signature of warp-knitted fabrics. Their study emphasized the sensitivity of constitutive laws to microstructural directions and demonstrated that data-driven discovery can outperform classical orthotropic models.

Anisotropy is particularly crucial in biomechanics, where soft tissues inherently exhibit direction-dependent behavior. 
In \cite{Tac2022}, operator-learning strategies were applied to model anisotropic biological tissues, while physics-informed neural networks have been employed to capture the anisotropic hyperelastic response of the human myocardium \cite{Osman2025,Martonov2025} as well as the constitutive behavior of pulmonary arteries \cite{Vervenne2025}. These studies underline the importance of integrating physics-based constraints with data-driven methods to achieve both predictive accuracy and generalizability in biologically inspired anisotropic material models.

Overall, these works demonstrate that neural networks provide a versatile toolkit for modeling anisotropy, ranging from physics-augmented formulations with structural tensors, to automated discovery of textile mechanics, and advanced PINN-based approaches for cardiovascular tissues. Together, they illustrate the broad applicability of AI-enhanced constitutive modeling in addressing anisotropy across scales and domains.

\medskip
This literature review is by no means exhaustive, as the integration of machine learning into computational material modeling is currently being studied with great intensity. A particularly comprehensive overview is provided in the recent reviews by \cite{fuhg2024reviewdatadrivenconstitutivelaws,watson2025machinelearningphysicsknowledge}, which systematically discuss data-driven constitutive laws for solids.

\subsection{Research gap and aim of the study}

\paragraph{Research gap.}
In recent years, substantial progress has been made in the development of physics-embedded neural networks for modeling inelastic and anisotropic materials. 
Nevertheless, most approaches remain confined to the small-strain regime and typically impose convexity of the underlying inelastic potential. 
The systematic incorporation of anisotropy, both in the Helmholtz free energy and in the corresponding inelastic potential, has received less consistent attention. 
Furthermore, physics-embedded neural network architectures for inelasticity continue to encounter issues of stability and robustness during training. 
As a result, a comprehensive framework for anisotropic inelasticity in the finite-strain regime that is robust throughout training is still missing.

\paragraph{Aim of this study.}
The present contribution aims to provide a complement to the existing theoretical foundations by establishing a consistent strategy for incorporating anisotropy into the architecture of neural networks for inelastic materials at finite strains. 
This complement is formulated independently of a specific network architecture, thereby broadening its applicability. 
To further account for the increasing complexity of materials--such as architectured microstructures--we propose an extension to non-convex yet physically admissible formulations of the inelastic potential. For stabilizing the training procedure, we complement recent advances in time-integration schemes for inelastic materials by extending them to the finite-strain regime. Together, these complementary developments are expected to substantially advance the use of neural networks in computational material modeling.

\subsection{Outline}
This contribution is structured as follows: 
In Section~\ref{sec:illustrative_example}, we introduce the underlying thermodynamic requirement in a simplified setting. 
We demonstrate how this condition can be satisfied in one dimension and explain why convexity of the dual potential is a sufficient, but not a necessary, constraint.  
Section~\ref{sec:NetworkTypes} reviews several neural network architectures. 
We briefly recap Input Convex Neural Networks and introduce Input Monotonic Neural Networks, whose combination generalizes the findings from the illustrative example to a broader class of network architectures. 
In addition, we present Liquid Neural Networks as a specialized class of recurrent architectures.  
The continuum mechanical foundations of finite strain anisotropic inelasticity are developed in Section~\ref{sec:Fundamentals}. 
Sections~\ref{sec:Helmholtz} and \ref{sec:DualPotential} then specify how to represent the Helmholtz free energy and the dual potential in terms of invariants. 
Building on these foundations, Section~\ref{sec:ConstiNN} integrates the theoretical principles with the neural architectures introduced earlier. 
Here, we also describe the prediction of inelastic variable updates using Liquid Neural Networks.  
The effectiveness of the proposed framework is demonstrated in Section~\ref{sec:results} through benchmark problems at both the material point and structural level. 
We investigate isotropic and anisotropic responses and compare the results against recurrent neural networks that lack physics-based constraints.  
Section~\ref{sec:discussion} provides a critical assessment of the approach, outlining its limitations and highlighting open research questions related to anisotropic inelasticity modeled with neural networks. 
Finally, Section~\ref{sec:conclusion} summarizes the key findings and conclusions of this work.

\section{Preliminary Problem Description}
\label{sec:Preliminary}
\begin{figure}[h]
    \centering
    \includegraphics{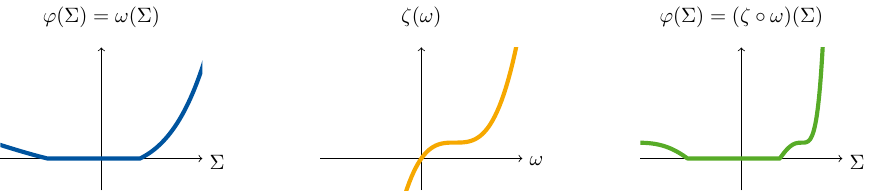}
    \caption{
        Schematic illustration of possible constructions for a function $\varphi(\Sigma)$ satisfying Inequality~\eqref{eq:dissipation_1D}.  
        The function $\omega$ is \textit{convex}, \textit{non-negative}, and \textit{zero-valued at the origin}, whereas $\zeta$ is \textit{monotonically increasing} and \textit{zero-valued} with respect to $\omega$.  
        Their composition $(\zeta \circ \omega)$ ensures that the sign of the subgradient $\partial_\Sigma \varphi$ coincides with the sign of $\Sigma$ in the negative and positive regimes, respectively.
    }
\label{fig:potential_1D}
\end{figure}
In this section, we introduce the fundamental mathematical problem underlying finite-strain inelasticity. 
This preliminary discussion will facilitate a deeper understanding of the challenges involved and motivate the design of the neural network architectures employed throughout this manuscript.
\subsection{Notation}
Let $a$, $\bm{a}$, and $\bm{A}$ denote tensors of order zero (scalars), one (vectors), and two (second-order tensors), respectively. 
To distinguish tensor notation from standard linear algebra notation, we write $\mathbf{a}$ and $\mathbf{A}$ for vectors and matrices in the usual sense.
The set of unit vector is denoted by $\mathbb{S}^{n-1} := \{\, \bm{v} \in \mathbb{R}^n : \|\bm{v}\| = 1 \,\}$.
Further, the set of real $n \times n$ matrices is denoted by $\mathbb{M}^{n\times n}$, and the group of invertible matrices is given by $\mathrm{GL}(n,\mathbb{R}) = \{\, \bm{A} \in \mathbb{M}^{n\times n} : \det \bm{A} \neq 0 \,\}$, with the subgroup of orientation-preserving transformations $\mathrm{GL}^+(n,\mathbb{R}) = \{\, \bm{A} \in \mathrm{GL}(n,\mathbb{R}) : \det \bm{A} > 0 \,\}$. 
The set of symmetric matrices is denoted by $\mathrm{Sym}^{n\times n} = \{\, \bm{A} \in \mathbb{M}^{n\times n} : \bm{A}^T = \bm{A} \,\}$, with the subset of symmetric positive definite matrices $\mathrm{Sym}_+^{n\times n} = \{\, \bm{A} \in \mathrm{Sym}^{n\times n} : \bm{x}^T \bm{A}\bm{x} > 0 \ \forall \bm{x}\neq \bm{0} \,\}$.
Any two tensors $\bm{A}, \bm{B} \in \mathbb{M}^{n\times n}$ related by $\bm{A}=\bm{P}\,\bm{B}\bm{P}^{-1}$ for $\bm{P} \in \mathrm{GL}(n,\mathbb{R})$ are called \textit{similar}, i.e., they share the same eigenvalues.
The cofactor of a tensor $\bm{A} \in \mathrm{GL}(n,\mathbb{R})$ is defined as $\mathrm{cof}\,\bm{A} = \det(\bm{A})\, \bm{A}^{-T}$, and $\mathrm{cof}\,\bm{A} = \det(\bm{A})\, \bm{A}^{-1}$ if $\bm{A} \in \mathrm{Sym}_+^{n\times n}$.
The set of orthogonal tensors is given by $\mathrm{O}(n) = \{\, \bm{Q} \in \mathbb{M}^{n\times n} : \bm{Q}^T \bm{Q} = \bm{I}, \ \det \bm{Q} = \pm 1 \,\}$, while the subgroup of proper orthogonal tensors (rotations) is denoted by $\mathrm{SO}(n) = \{\, \bm{Q} \in \mathrm{O}(n) : \det \bm{Q} = 1 \,\}$. 
The second-order identity tensor is expressed as $\bm{I} = \delta_{ij}\, \bm{e}_i \otimes \bm{e}_j$ with $\{\bm{e}_i\}$ denoting the Cartesian basis. 
For consistency, the symbol $\mathbf{1}$ will be used whenever the identity is required in a more general context.
In addition, the deviatoric operator $\mathrm{dev}$ is defined as $\mathrm{dev}\bm{A} = \bm{A} - \tfrac{1}{3}\mathrm{tr}(\bm{A})\, \bm{I}$. 
The inner product of two objects is denoted by $\langle \cdot, \cdot \rangle$, and a superposed dot $\dot{(\bullet)}$ indicates the total time derivative.
In tensor notation, $:$ denotes a double contraction, while $\cdot$ refers to a single contraction.
Finally, $\partial_A f$ denotes the derivative of $f$ with respect to an argument $A$ (understood as the gradient in the smooth case, or as the subgradient in the nonsmooth case).
Moreover, the operator $\mathcal{N}$ indicates that a function is represented by a neural network.

Throughout this manuscript, we do not introduce new symbols for functions whose arguments have been modified; instead, we use the same symbol whenever the context makes the meaning unambiguous.
\subsection{One-dimensional illustrative example}
\label{sec:illustrative_example}
We begin by defining two scalar-valued functions
\[
\psi: \mathbb{R}^2 \rightarrow \mathbb{R}, \quad (\varepsilon, \varepsilon_i) \mapsto \psi(\varepsilon, \varepsilon_i), \quad 
\varphi: \mathbb{R} \rightarrow \mathbb{R}, \quad \Sigma \mapsto \varphi(\Sigma),
\]
where $\varepsilon$ denotes the total strain and $\varepsilon_i$ is an internal (inelastic) variable that accounts for history dependency. 
We further introduce the following constitutive relations
\begin{equation}
    \sigma = \frac{\partial \psi}{\partial \varepsilon}, \qquad
    \Sigma = -\frac{\partial \psi}{\partial \varepsilon_i},
\label{eq:state_laws_1D}
\end{equation}
where $\Sigma$ is referred to as the \textit{thermodynamically consistent driving force} conjugate to $\dot{\varepsilon}_i$ and $\sigma$ denotes the mechanical stress state. 
The essential thermodynamic requirement is
\begin{equation}
    \Sigma\,\dot{\varepsilon}_i \geq 0.
\end{equation}
This inequality expresses the non-negativity of the internal dissipation and must be satisfied at all times. 

To close the system, we postulate that the evolution of the internal variable $\varepsilon_i$ can be derived from the scalar potential $\varphi$, i.e.,
\begin{equation}
    \Sigma\,\dot{\varepsilon}_i = \Sigma\,\partial_\Sigma \varphi \geq 0.
\label{eq:dissipation_1D}
\end{equation}
where $\partial_\Sigma \varphi$ is generally nonsmooth, i.e., the subgradient of $\varphi$.
The central question is then how to construct $\varphi$ such that the dissipation inequality~\eqref{eq:dissipation_1D} is always satisfied.

\paragraph{Convex functions.}  
A standard choice in material modeling is to assume that $\varphi$ is \textit{convex}, \textit{non-negative}, and \textit{zero-valued at the origin} \cite{Germain1983}. 
Such a function ensures that the dissipation inequality is satisfied automatically, due to the properties of subgradients \cite{Rockafellar1970}, namely
\begin{equation}
    \omega(0) \geq \omega(\Sigma) - \langle \partial_\Sigma \omega, \Sigma \rangle,
\label{eq:subderivative_1D}
\end{equation}
where $\omega$ denotes a convex function with $\omega(0)=0$ and $\omega(\Sigma) \ge 0$.  
Inequality~\eqref{eq:dissipation_1D} is then satisfied \textit{a priori}.  
A schematic illustration of such a potential is given in Figure~\ref{fig:potential_1D}.

\paragraph{Monotonic functions.}  
While convexity is \textit{sufficient}\footnote{In combination with zero-valued at the origin and non-negativity.}, it is not \textit{necessary} to guarantee positive dissipation.  
Indeed, Equation~\eqref{eq:dissipation_1D} only requires that the signs of $\Sigma$ and $\partial_\Sigma \varphi$ coincide. 
This can be achieved if $\varphi$ is \textit{monotonically decreasing} for $\Sigma<0$ and \textit{monotonically increasing} for $\Sigma>0$\footnote{Note that convex, zero-valued, and non-negative functions are a special case of this class.}.  
To formalize this, let $\zeta:\mathbb{R} \rightarrow \mathbb{R}$ be a monotonically increasing, zero-valued function.  
We can then define
\[
\varphi(\Sigma) = (\zeta \circ \omega)(\Sigma).
\]
By the chain rule, the dissipation reads
\begin{equation}
    \Sigma\,\partial_\Sigma \varphi = \Sigma\, \left(\frac{\partial \zeta}{\partial \omega}\, \partial_\Sigma \omega \right)
    = \underbrace{\frac{\partial \zeta}{\partial \omega}}_{\ge 0} \underbrace{(\partial_\Sigma \omega \, \Sigma)}_{\ge 0} \ge 0,
\label{eq:phi_1D}
\end{equation}
which is guaranteed by the monotonicity of $\zeta$ with respect to $\omega$ and the convexity property~\eqref{eq:subderivative_1D}.  
This generalized construction allows greater flexibility in designing constitutive potentials while maintaining thermodynamic consistency.
Figure~\ref{fig:potential_1D} illustrates the different possibilities: the blue (left) graph represents the convex (more constraining) option, while the green graph is monotonically increasing in the positive regime and monotonically decreasing within the negative regime (more flexible).
\section{Neural Network Architectures}
\label{sec:NetworkTypes}
\begin{figure}[ht]
    \centering
    \includegraphics{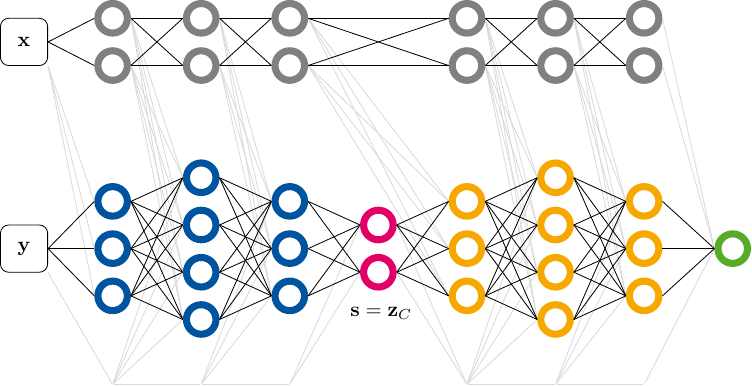}
    \caption{Schematic illustration of the composition network $\mathcal{N}_\circ$.
The blue part represents the Input Convex Neural Network $\mathcal{N}_c$, which is convex in $\mathbf{y}$ but not in $\mathbf{x}$.
Each layer takes as input the output of the previous layer and of the parallel network in gray carrying $\mathbf{x}$ as well as $\mathbf{y}$ itself.
Its final output $\mathbf{z}_C$ highlighted in red serves as the input $\mathbf{s}$ to the Input Monotonic Neural Network $\mathcal{N}_m$ shown in orange, whose architecture mirrors that of $\mathcal{N}_c$.
To guarantee a non-negative $\mathbf{z}_C$, the activation between $\mathbf{z}_{C-1}$ and $\mathbf{z}_C$ is chosen as ReLU.
    }
    \label{fig:composition_network}
\end{figure}
In the previous Section~\ref{sec:Preliminary}, we demonstrated how to construct a function $\varphi(\Sigma)$ that satisfies the reduced dissipation inequality \textit{a priori}. 
In the setting of finite strain inelasticity, these considerations can be extended in a straightforward manner to three-dimensional problems. 
Since our ultimate goal is to approximate both the Helmholtz free energy $\psi$ and the dual potential $\varphi$, this section introduces the neural network architectures employed for this purpose. 
Moreover, inelastic materials additionally require the solution of evolution equations for history-dependent internal variables. 
To this end, we further make use of a specialized recurrent architecture known as \textit{Liquid Neural Networks}.

\paragraph{Input Convex Neural Networks.}
We begin by introducing (partially) Input Convex Neural Networks (ICNNs) $(\mathbf{x},\mathbf{y}) \mapsto \mathcal{N}_c(\mathbf{x},\mathbf{y})$, originally proposed in \cite{Amos2017}. 
These networks are designed such that the output is convex with respect to $\mathbf{y}$ while remaining unconstrained in $\mathbf{x}$.  
The operator $\mathcal{N}_c$ denotes a neural network with hidden layers $l=0,\ldots,C-1$ and an output layer $C$.  
The key idea is to employ two parallel computational branches: one for the convex variables and one for the non-convex variables.  
The non-convex branch may feed into the convex branch in a way that preserves convexity with respect to $\mathbf{y}$.  
The architecture proposed in \cite{Amos2017} reads
\begin{equation}
\begin{aligned}
    \mathbf{u}_{l+1} &= k_l^c \left( \mathbf{W}_l'\,\mathbf{u}_l + \mathbf{b}_l' \right), \quad \mathbf{u}_0 = \mathbf{x}, \\
    \mathbf{z}_{l+1} &= g_l^c \Big( \mathbf{W}_l^z \big( \mathbf{z}_l \odot [ \mathbf{W}_l^{zu}\, \mathbf{u}_l + \mathbf{b}_l^{zu} ]_+ \big) 
        + \mathbf{W}_l^y \big( \mathbf{y} \odot [ \mathbf{W}_l^{yu}\, \mathbf{u}_l + \mathbf{b}_l^{yu} ] \big) + \mathbf{W}_l^u\, \mathbf{u}_l + \mathbf{b}_l^z \Big), \\
    \mathbf{z}_{1} &= g_0^c \Big( \mathbf{W}_0^y \big( \mathbf{y} \odot [ \mathbf{W}_0^{yu}\, \mathbf{x} + \mathbf{b}_0^{yu} ] \big) + \mathbf{W}_0^u\, \mathbf{x} + \mathbf{b}_0^z \Big),
\end{aligned}
\label{eq:ICNN}
\end{equation}
where $\odot$ denotes the Hadamard product and $[\bullet]_+$ the non-negative part, typically implemented using the ReLU activation function.  
To ensure that each component of the final output layer $\mathbf{z}_C$ is convex in $\mathbf{y}$, the weights $\mathbf{W}^z$ must satisfy $\mathbf{W}^z \in \mathbb{R}_{\geq 0}$, while the activation functions $g_l^c$ are required to be \textit{convex} and \textit{non-decreasing}.  
The remaining weights $\mathbf{W}$ and biases $\mathbf{b}$ are unconstrained, and the activations $k_l^c$ are not restricted.  

In the context of finite strain inelasticity, we employ ICNNs to approximate both the Helmholtz free energy and the dual potential.  
Since these functions are zero-valued at the origin, an adjustment to the architecture is necessary.  
Moreover, if the input $\mathbf{y}$ is already passed through a convex function $f$, i.e.\ $(\mathbf{x}, f(\mathbf{y})) \mapsto \mathcal{N}_c(\mathbf{x},\mathbf{y})$, then also $\mathbf{W}^y$ must satisfy $\mathbf{W}^y \in \mathbb{R}_{\geq 0}$.  
To incorporate these aspects, we slightly modify the architecture~\eqref{eq:ICNN}, yielding
\begin{equation}
\begin{aligned}
    \mathbf{u}_{l+1} &= \tilde{k}_l^c \left( \mathbf{W}_l'\,\mathbf{u}_l + \mathbf{b}_l' \right), \quad \mathbf{u}_0 = \mathbf{x}, \\
    \mathbf{z}_{l+1} &= \tilde{g}_l^c \Big( \mathbf{W}_l^z \big( \mathbf{z}_l \odot [ \mathbf{W}_l^{zu}\, \mathbf{u}_l + \mathbf{b}_l^{zu} ]_+ \big) 
        + \mathbf{W}_l^y \big( \mathbf{y} \odot [ \mathbf{W}_l^{yu}\, \mathbf{u}_l + \mathbf{b}_l^{yu} ]_+ \big) + \mathbf{b}_l^c \Big), \\
    \mathbf{z}_{1} &= \tilde{g}_0^c \Big( \mathbf{W}_0^y \big( \mathbf{y} \odot [ \mathbf{W}_0^{yu}\, \mathbf{u}_0 + \mathbf{b}_0^{yu} ]_+ \big) + \mathbf{b}_0^c \Big),
\end{aligned}
\label{eq:ICNN_adapted}
\end{equation}
with $\mathbf{b}_l^c := \mathbf{W}_l^u\, \mathbf{u}_l + \mathbf{b}_l^z$, and where $\mathbf{W}^z, \mathbf{W}^y \in \mathbb{R}_{\geq 0}$.  
The modified activations
\begin{equation}
    \tilde{k}_l^c := k_l^c((\bullet)_l) - k_l^c(\mathbf{b}_l'), 
    \qquad
    \tilde{g}_l^c := g_l^c((\bullet)_l) - g_l^c(\mathbf{b}_l^c),
\label{eq:shift_activation}
\end{equation}
ensure that $\mathbf{u}_L = \mathbf{0}$ if $\mathbf{x}=\mathbf{0}$ and $\mathbf{z}_C = \mathbf{0}$ if $\mathbf{y}=\mathbf{0}$.  
This guarantees zero-valuedness of the scalar functions to be approximated.
Note that we made the architectural choice that $\mathbf{z}_C = \mathbf{0}$ if $\mathbf{y}=\mathbf{0}$ holds independently of the value of $\mathbf{x}$.
Lastly, it is worth noting that although the constraint $\mathbf{W}^y \in \mathbb{R}_{\geq 0}$ appears to limit the expressive power of ICNNs, architectures of the form~\eqref{eq:ICNN} can still be represented by augmenting the input as $(\mathbf{x}, \mathbf{y}, -\mathbf{y}) \mapsto \mathcal{N}_c(\mathbf{x},\mathbf{y})$.

\paragraph{Input Monotonic Neural Networks.}
Next, we propose a new architecture, termed \textit{Input Monotonic Neural Networks} (IMNNs), in analogy to ICNNs.  
These networks are designed to produce outputs that are monotonically increasing with respect to a specific subset of inputs $\mathbf{s}$, while remaining unconstrained in other inputs $\mathbf{v}$.  
Formally, IMNNs implement mappings $(\mathbf{v},\mathbf{s}) \mapsto \mathcal{N}_m(\mathbf{v},\mathbf{s})$.  
The architecture again employs two parallel branches:
\begin{equation}
\begin{aligned}
    \mathbf{r}_{l+1} &= \tilde{k}_l^m \left( \mathbf{W}_l^*\, \mathbf{r}_l + \mathbf{b}_l^* \right), \quad \mathbf{r}_0 = \mathbf{v}, \\
    \mathbf{p}_{l+1} &= \tilde{g}_l^m \Big( \mathbf{W}_l^p \big( \mathbf{p}_l \odot [ \mathbf{W}_l^{pr}\,\mathbf{r}_l + \mathbf{b}_l^{pr} ]_+ \big) 
        + \mathbf{W}_l^s \big( \mathbf{s} \odot [ \mathbf{W}_l^{sr}\,\mathbf{r}_l + \mathbf{b}_l^{sr} ]_+ \big) + \mathbf{b}_l^m \Big), \\
    \mathbf{p}_{1} &= \tilde{g}_0^m \Big( \mathbf{W}_0^w \big( \mathbf{s} \odot [ \mathbf{W}_0^{sr}\,\mathbf{r}_0 + \mathbf{b}_0^{sr} ]_+ \big) + \mathbf{b}_0^m \Big),
\end{aligned}
\end{equation}
where $\mathbf{b}_l^m := \mathbf{W}_l^r\, \mathbf{r}_l + \mathbf{b}_l^p$.  
If $\tilde{g}_l^m$ are monotonically increasing activation functions and $\mathbf{W}^p, \mathbf{W}^s \in \mathbb{R}_{\geq 0}$, then each output component of $\mathbf{p}_M$ is guaranteed to be monotonically increasing in each component of $\mathbf{s}$; see Appendix~\ref{app:IMNN}.
As before, shifted activations analogous to Equation~\eqref{eq:shift_activation} are used to enforce zero-valuedness.  
A monotonically decreasing output can be obtained by simply replacing $\mathbf{s}$ with $-\mathbf{s}$.

Noteworthy, a three-layer architecture with min-max pooling, which enforces monotonicity, was first proposed in \cite{Sill1997MonotonicN} and later extended in \cite{Daniels2010} to partially input specific monotonic networks.
More recently, \cite{you2017deeplatticenetworkspartial} combined three types of layers—calibrators, linear embeddings, and lattices—within a deep lattice framework to enforce monotonicity.

\paragraph{Composition of ICNNs and IMNNs.}
Section~\ref{sec:Preliminary} and Figure~\ref{fig:potential_1D} demonstrated that the dual potential $\varphi$ can be expressed as a composition of convex and monotonic functions.  
Accordingly, we construct a neural network architecture $\mathcal{N}_\circ$ by composing ICNNs and IMNNs: 
\begin{equation}
\mathcal{N}_\circ(\mathbf{x},\mathbf{y}) = \mathcal{N}_m \big( \mathbf{u}_{C-1}, \mathbf{z}_C \big), 
\quad \text{with } (\mathbf{u}_{C-1}, \mathbf{z}_C) = \mathcal{N}_c(\mathbf{x},\mathbf{y}).
\label{eq:composition_network}
\end{equation}
Here, $\mathbf{u}_{C-1}$ is chosen instead of $\mathbf{u}_C$, since $\mathbf{u}_C$ is typically not computed in ICNNs as it does not influence the output $\mathbf{z}_C$.
Figure~\ref{fig:composition_network} schematically illustrates the composition.

\paragraph{Liquid Neural Networks.}
Unlike purely elastic materials, inelasticity requires not only an additional dual potential but also the solution of evolution equations for the internal variables.  
Explicit time discretization schemes are computationally efficient but often unstable, while implicit schemes are stable but computationally costly and thus impractical for neural network training, which typically requires many epochs.  
An alternative, explored in \cite{Rosenkranz2024} for feed-forward networks and long-short-term memory (LSTM) networks, is to approximate the update of the internal variables using auxiliary networks \cite{Asad2023}.  

In this work, we employ \textit{Liquid Neural Networks} (LiNNs) \cite{Hasani2021}, a class of recurrent networks that update their hidden state $\mathbf{h}$ using dynamics inspired by differential equations of continuous systems.  
This formulation naturally reflects the temporal evolution typical for inelastic internal variables.  
In continuous time, the update reads
\begin{equation}
    \frac{\mathrm{d}\mathbf{h}(t)}{\mathrm{d}t} 
    = - \bm{\alpha}(\mathbf{h}(t), \mathbf{q}(t)) \odot \mathbf{h}(t) 
    + \mathbf{f}(\mathbf{h}(t), \mathbf{q}(t)),
\label{eq:continous_LiNN}
\end{equation}
with $\alpha_i \geq 0$ and $\mathbf{q}$ denoting an external input.  
Since $\bm{\alpha}$ and $\mathbf{f}$ may vary with time, these networks are called \textit{liquid}.  

For numerical implementation, Equation~\eqref{eq:continous_LiNN} can be discretized within $t\in[t_n, t_{n+1}]$, for example by a (semi-)explicit Euler scheme
\begin{equation}
    \mathbf{h}_{n+1} = \big( \mathbf{1} - \Delta t\, \bm{\alpha}(\mathbf{h}_n, \mathbf{q}_{n+1}) \big) \odot \mathbf{h}_n 
    + \Delta t\, \mathbf{f}(\mathbf{h}_n, \mathbf{q}_{n+1}),
    \label{eq:LiNN_EM}
\end{equation}
or by a (semi-)explicit exponential time integration scheme \cite{Cox2002}
\begin{equation}
    \mathbf{h}_{n+1} = \exp(-\Delta t\,\bm{\alpha}(\mathbf{h}_n, \mathbf{q}_{n+1})) \odot \mathbf{h}_n 
    + \Delta t\, \xi(-\Delta t\,\bm{\alpha}(\mathbf{h}_n, \mathbf{q}_{n+1})) \odot \mathbf{f}(\mathbf{h}_n, \mathbf{q}_{n+1}),
    \quad \xi(x) = \frac{\exp(x)-1}{x}
\end{equation}
where $\Delta t := t_{n+1}-t_n$.

In the neural network formulation, both $\bm{\alpha}$ and $\mathbf{f}$ are learned by independent feed-forward networks, i.e.\ $(\mathbf{h}_n, \mathbf{q}_{n+1}) \mapsto \mathcal{N}_\alpha(\mathbf{h}_n, \mathbf{q}_{n+1})$ and $(\mathbf{h}_n, \mathbf{q}_{n+1}) \mapsto \mathcal{N}_f(\mathbf{h}_n, \mathbf{q}_{n+1})$.  
There are no constraints on weights, biases, or activation functions, except that we again apply shifted activations (analogous to Equation~\eqref{eq:shift_activation}) to ensure that the state remains unchanged if both the previous state and external input are zero.
Further, the last activation function of $\mathcal{N}_\alpha$ should ensure the non-negativity of $\bm{\alpha}$.
Although both discretization schemes are viable, we observed that the (semi-)explicit Euler method yields superior stability for the problems considered in this work.  
Accordingly, we restrict our investigation to this scheme.
\section{Fundamentals of Constitutive Relations}
\label{sec:Fundamentals} 
\begin{figure}[ht]
    \centering
    \includegraphics{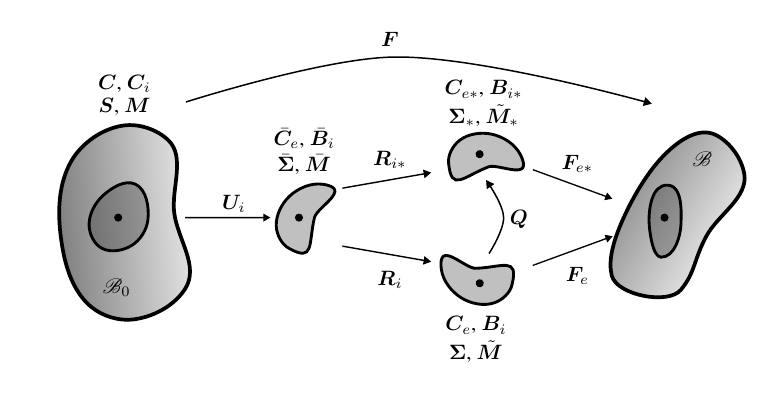}
    \caption{
    Motion from the reference configuration $\mathscr{B}_0$ to the current configuration $\mathscr{B}$.
    An infinitesimal material element is described by the deformation gradient $\bm{F}$,
    which admits a multiplicative decomposition into elastic $\bm{F}_e$ and inelastic $\bm{F}_i$ parts.
    Due to the non-uniqueness, one may equivalently write $\bm{F}=\bm{F}_{e*}\,\bm{F}_{i*}$ with $\bm{F}_{e*}=\bm{F}_e\,\bm{Q}^T$ and $\bm{F}_{i*}=\bm{Q}\,\bm{F}_i$ with $\bm{Q}\in\mathrm{SO}(3)$.
    From the polar decompositions of $\bm{F}_i$ and $\bm{F}_{i*}$, it follows that $\bm{R}_i^*=\bm{Q}\bm{R}_i$.
    Noteworthy, the shown quantities referred to one intermediate configuration share their eigenvalues with the corresponding quantities in any other intermediate configuration.}
    \label{fig:map}
\end{figure}
After introducing the neural network architectures in the previous section, which will later serve as functional representations of $\varphi$ and $\psi$, we now turn to the foundations of constitutive modeling for inelastic materials at finite strains.  
Our objective is to formulate a framework for anisotropic materials that captures the essential kinematics and satisfies the dissipation inequality.

Noteworthy, the term \textit{anisotropy} is not unambiguous in inelasticity. 
Unlike in elasticity, where anisotropy does not necessarily result in coaxiality of the stress–strain response, inelasticity exhibits additional sources of anisotropic behavior.
For example, kinematic hardening or the coupling between elastic and inelastic strains can both induce effective anisotropy.
In this context, one may further distinguish between \textit{initial} anisotropy (present in the undeformed material, e.g., due to texture or fibers) and \textit{induced} anisotropy (evolving during inelastic deformation).
Here, we use anisotropy exclusively to denote the presence of preferred directions—either in the energy (e.g., due to embedded fibers) or in the dual potential (e.g., Hill’s plasticity).
The latter case is closely related to what is commonly referred to as \textit{distortional hardening}.

In this contribution, we limit ourselves to the class of initial anisotropy that can be characterized by several line or outward normal vectors $\bm{n}^i$ perpendicular to a plane.
For brevity, we derive the equations for one family of vectors $\bm{n}$ (transversal isotropy), however, the extension to more such classes of anisotropy is straightforward.

To this end, we will first introduce the kinematics of the multiplicative decomposition and the material principles that the Helmholtz free energy must satisfy. We will then derive the evolution equation for isotropic materials based on a dual potential and extend this concept to anisotropic materials.
The scalar functions, namely the energy and the dual potential, will be discovered by the neural networks introduced in the previous section as the manuscript progresses.

\paragraph{Kinematics.}  
Let $\bm{\chi}: \mathscr{B}_0 \subset \mathbb{R}^3 \rightarrow \mathscr{B}$ denote the motion that maps the reference configuration $\mathscr{B}_0$ onto the current configuration $\mathscr{B}$ in Euclidean space. 
The deformation is characterized by the deformation gradient 
\[
\bm{F} = \mathrm{Grad}\,\bm{\chi}, \quad J := \det \bm{F}. 
\]
To ensure an admissible, orientation-preserving motion one requires $J > 0$ in $\mathscr{B}_0$, such that $\bm{F} \in \mathrm{GL}^+(3,\mathbb{R})$. 
In the finite strain regime, inelastic effects are commonly described by postulating a multiplicative decomposition\footnote{As shown in \cite{Itskov2004}, the multiplicative decomposition is superior to the additive decomposition in case of finite elasto-plasticity regardless of the elastic material law.} of the deformation gradient \cite{Eckart1948,Kroener1959,Sidoroff1974,Rodriguez1994,Christ2009}, irrespective of the specific inelastic mechanism, 
\begin{equation}
  \bm{F} = \bm{F}_e \bm{F}_i = \bm{F}_{e*} \bm{F}_{i*}.
\end{equation}
Here, $\bm{F}_e$ and $\bm{F}_i$ denote the elastic and inelastic parts, respectively. Constitutive admissibility requires $J_e := \det \bm{F}_e > 0$ and $J_i := \det \bm{F}_i > 0$. The decomposition is not unique, since any rotation $\bm{Q} \in \mathrm{SO}(3)$ yields an equivalent representation with $\bm{F}_{e*} = \bm{F}_e \bm{Q}^T$ and $\bm{F}_{i*} = \bm{Q} \bm{F}_i$ (see, e.g., \cite{Casey2016}). Since $\bm{F}_e, \bm{F}_i \in \mathrm{GL}^+(3,\mathbb{R})$, both admit polar decompositions, $\bm{F}_e = \bm{R}_e \bm{U}_e$ and $\bm{F}_i = \bm{R}_i \bm{U}_i$, with $\bm{R}_e, \bm{R}_i \in \mathrm{SO}(3)$ and $\bm{U}_e, \bm{U}_i \in \mathrm{Sym}_+^{3\times 3}$. It follows that the elastic right stretch tensor in the rotated configuration, $\bm{U}_{e*} = \bm{Q} \bm{U}_e \bm{Q}^T$, is orthogonally similar to $\bm{U}_e$, while the inelastic stretch remains unchanged, $\bm{U}_{i*} = \bm{U}_i$.
Figure~\ref{fig:map} shows the various mappings involved.

\paragraph{Material principles.}
For anisotropic materials, we introduce a symmetry class \(\mathscr{S}\subset\mathrm{O}(3)\) consisting of all orthogonal tensors \(\bm{Q}\) relative to which the material response remains invariant. 
In this work, following the structural-tensor approach (cf.\ \cite{Boehler1979,Spencer1984}), we focus on a single preferred direction \(\bm{n}\in\mathbb{S}^2\) in the reference configuration and set the structural tensor \(\bm{M}=\bm{n}\otimes\bm{n}\).
In analogy to \cite{Schroeder2003}, we introduce the group $\mathscr{G} = \left\{ \bm{Q}\in\mathrm{SO}(3) : \bm{Q}\bm{n}=\bm{n} \right\}$,
i.e., the set of proper rotations leaving \(\bm{n}\) invariant.
As mentioned at the beginning, we set \(\mathscr{S}=\mathscr{G}\), and thus, the material is invariant about \(\bm{n}\).
For isotropic materials, in contrast, the symmetry class is \(\mathscr{S}=\mathrm{SO}(3)\) (often \(\mathrm{O}(3)\) is admitted as well).

We derive the constitutive equations in the sense of hyperelasticity and postulate the existence of a Helmholtz free energy
\[
  \psi:\ \mathrm{GL}^+(3,\mathbb{R})\times \mathrm{GL}^+(3,\mathbb{R})\times \mathcal{M}\ \to\ \mathbb{R}, 
  \qquad (\bm{F},\bm{F}_i,\bm{M})\ \mapsto\ \psi(\bm{F},\bm{F}_i,\bm{M}),
\]
where \(\mathcal{M}=\{\bm{M}=\bm{n}\otimes\bm{n} : \bm{n}\in\mathbb{S}^2\}\).
For inelastic materials with multiplicative split \(\bm{F}=\bm{F}_e\bm{F}_i\), the free energy is required to satisfy the following principles:
\begin{itemize}
  \item Objectivity (frame indifference)
  \[
    \psi(\bm{F},\bm{F}_i,\bm{M}) \;=\; \psi(\bm{Q}\bm{F},\bm{F}_i,\bm{M})
    \qquad \forall\,\bm{Q}\in\mathrm{SO}(3).
  \]
  \item Material symmetry
  \[
    \psi(\bm{F},\bm{F}_i,\bm{M}) \;=\; \psi(\bm{F}\bm{Q}^T,\ \bm{F}_i\bm{Q}^T,\ \bm{Q}\bm{M}\bm{Q}^T)
    \qquad \forall\,\bm{Q}\in\mathscr{G}.
  \]
  \item Indifference to the choice of intermediate configuration
  \[
    \psi(\bm{F},\bm{F}_i,\bm{M}) \;=\; \psi(\bm{F},\ \bm{Q}\bm{F}_i,\ \bm{M})
    \qquad \forall\,\bm{Q}\in\mathrm{SO}(3).
  \]
\end{itemize}
Consequently, $\psi$ can be expressed as a scalar-valued isotropic function in terms of the right Cauchy--Green tensors \(\bm{C}=\bm{F}^T\bm{F}\) and \(\bm{C}_i=\bm{F}_i^T\bm{F}_i\) and the structural tensor \(\bm{M}\)
\begin{equation}
  \psi(\bm{F},\bm{F}_i,\bm{M})
  \;=\;
  \psi(\bm{C},\bm{C}_i,\bm{M})
  \;=\;
  \psi\!\left(\bm{Q}\bm{C}\bm{Q}^T,\ \bm{Q}\bm{C}_i\bm{Q}^T,\ \bm{Q}\bm{M}\bm{Q}^T\right)
  \quad \forall\,\bm{Q}\in\mathrm{SO}(3).
\label{eq:Helmholtz_C_Ci_M}
\end{equation}
Further, we impose volumetric growth (coercivity) conditions
\begin{gather}
\psi \to \infty \quad \text{as}\quad J \searrow 0 \ \text{or}\ J \to \infty, \label{eq:growth_J}\\
\psi \to \infty \quad \text{as}\quad J_e \searrow 0 \ \text{or}\ J_e \to \infty. \label{eq:growth_Je}
\end{gather}
In addition, non-essential conditions are employed: the stress is usually normalized such that it vanishes for \(\bm{F}\bm{F}_i^{-1}=\bm{I}\), and we set the reference level \(\psi(\bm{I},\bm{I},\bm{M})=0\).

To ensure existence of minimizers, we design \(\psi\) to be polyconvex in \(\bm{F}\) \cite{Ball1976,Ball1977}, i.e., there exists a function
$W=W\big(\bm{F},\operatorname{cof}\bm{F},J;\,\bm{F}_i,\bm{M}\big)$ that is convex in \((\bm{F},\operatorname{cof}\bm{F},J)\) (with \(\bm{F}_i\) and \(\bm{M}\) acting as parameters) and satisfies
\[
  \psi(\bm{F},\bm{F}_i,\bm{M})=W(\bm{F},\operatorname{cof}\bm{F},J;\,\bm{F}_i,\bm{M}).
\]
For background on polyconvex hyperelastic formulations in terms of the deformation gradient for anisotropic solids we refer the reader to \cite{Schroeder2003}; see also \cite{Marsden1994} for classical symmetry principles and \cite{Svendsen2001} for a discussion of isomorphism concepts in inelastic anisotropy.

\paragraph{Thermodynamic consistency for isotropic materials.}
Considering Equation~\eqref{eq:Helmholtz_C_Ci_M}, we evaluate the Clausius--Planck inequality
$-\dot{\psi} + \tfrac{1}{2}\,\bm{S}:\dot{\bm{C}} \;\geq\; 0,$
where $\bm{S}$ denotes the second Piola--Kirchhoff stress tensor.  
Following the arguments of \cite{Coleman1961,Coleman1963,Coleman1967}, this yields the state law
$\bm{S} = 2\, \partial_{\bm{C}}\psi$
as well as the reduced dissipation inequality
\begin{equation}
    \mathscr{D} := \bm{\Sigma} : \bm{D}_i \;\geq\; 0, 
    \qquad 
    \bm{\Sigma} := -2\,\bm{F}_i\,\frac{\partial\psi}{\partial\bm{C}_i}\,\bm{F}_i^T,
\label{eq:reduced_dissipation}
\end{equation}
where $\bm{D}_i$ denotes the symmetric part of $\dot{\bm{F}}_i\bm{F}_i^{-1}$ and is related to $\bm{C}_i$ by $\dot{\bm{C}}_i=2\,\bm{F}_i^T\bm{D}_i\bm{F}_i$.
As discussed in \cite{Dettmer2004}, $\bm{\Sigma}$ shares the same eigenvalues as the Kirchhoff stress $\bm{F}\bm{S}\bm{F}^T$ if $\psi$ depends solely on the elastic part of the deformation\footnote{Alternatively, one may derive the reduced dissipation inequality with respect to $\dot{\bm{C}}_i$ and its conjugated force (see \cite{Reese2021}). 
However, this approach loses the physical interpretation of $\bm{\Sigma}$, which is considered a disadvantage here}.
Note that neither $\bm{D}_i$ nor $\bm{\Sigma}$ is invariant under the choice of the intermediate configuration; indeed,
$\bm{D}_{i*}=\bm{Q}\bm{D}_i\bm{Q}^T,\ 
\bm{\Sigma}_{*}=\bm{Q}\bm{\Sigma}\bm{Q}^T$
for any $\bm{Q}\in\mathrm{SO}(3)$.
However, the resulting dissipation is invariant with respect to the intermediate configuration, i.e.,
$\bm{\Sigma}_{*}:\bm{D}_{i*}=\bm{\Sigma}:\bm{D}_{i}.$

We postulate the existence of a dual potential \cite{Rice1971}
\[
\varphi: \mathrm{Sym}^{3\times 3} \;\longrightarrow\; \mathbb{R}, 
\qquad \bm{\Sigma} \;\mapsto\; \varphi(\bm{\Sigma}).
\]
Similar to the Helmholtz free energy, the dual potential must satisfy the principles of objectivity, material symmetry, and invariance with respect to the choice of intermediate configuration.  
The first two properties are automatically satisfied, while the latter implies that $\varphi$ can be expressed as a scalar-valued isotropic function.  
We then postulate that the evolution equation follows from the dual potential
\begin{equation}
    \bm{D}_i \;\in\; \partial_{\bm{\Sigma}} \varphi,
\label{eq:Evolution}  
\end{equation}
which in general is nonsmooth \cite{Germain1998}.

As shown, for instance, in \cite{Holthusen2025}, it suffices to formulate a potential $\omega$ that is convex, non-negative, and vanishes at the origin in terms of the invariants $\mathcal{S}_\varphi(\bm{\Sigma}):=(S_1,S_2,S_3)$ with $S_2 := \sqrt{S_2'}$ and $S_3 := \sqrt[3]{S_3'}$, where\footnote{See Appendix~\ref{app:numerics} for a note on the computation of roots.}
\begin{equation}
S_1 := \mathrm{tr}\,\bm{\Sigma}, 
\qquad 
S_2' := \tfrac{1}{2}\,\mathrm{tr}\big((\mathrm{dev}\,\bm{\Sigma})^2\big), 
\qquad 
S_3' := \tfrac{1}{3}\,\mathrm{tr}\big((\mathrm{dev}\,\bm{\Sigma})^3\big),
\label{eq:stress_invars_isotropic}
\end{equation}
and by setting $\varphi=\omega$; see Appendix~\ref{app:consistency} for details\footnote{%
If $\varphi$ is convex in $\bm{\Sigma}$, the dual potential can be interpreted as the Legendre transform of the dissipation potential known from Generalized Standard Materials \cite{Halphen1975}; see, e.g., \cite{Holthusen2025}.
}.

As outlined in Section~\ref{sec:Preliminary} and Figure~\ref{fig:potential_1D}, this concept can be further generalized by introducing a monotonically increasing function $\zeta$ and defining
\[
\varphi(\mathcal{S}_\varphi) = (\zeta \circ \omega)(\mathcal{S}_\varphi).
\]
Since thermodynamic consistency already holds for $\varphi=\omega$, it is preserved under this composition provided that $\zeta$ is monotonically increasing; cf. Section~\ref{sec:Preliminary} or see Appendix~\ref{app:consistency}.

\paragraph{Extension to anisotropic materials.}
We now extend the methodology developed for inelastic isotropic materials to the anisotropic case.  
To this end, we postulate a dual potential
\[
\varphi: \mathrm{Sym}^{3\times 3} \times \mathbb{M}^{3 \times 3} \;\longrightarrow\; \mathbb{R}, 
\qquad (\bm{\Sigma},\tilde{\bm{M}}) \;\mapsto\; \varphi(\bm{\Sigma},\tilde{\bm{M}}),
\]
where $\tilde{\bm{M}} = \bm{F}_i \star \bm{M}$ denotes an appropriate push-forward of $\bm{M}$ onto the intermediate configuration such that
$\bm{Q}\bm{F}_i \star \bm{M} = \bm{Q}\,\tilde{\bm{M}}\,\bm{Q}^T \quad \forall\, \bm{Q}\in\mathrm{SO}(3)$,
see \cite{Dafalias1987}.  
The push-forward must be chosen such that $\varphi$ remains invariant under the choice of the intermediate configuration.

Let us introduce a set of extended invariants
\[
\mathcal{T}_\varphi(\bm{\Sigma}, \tilde{\bm{M}}) := \big(\mathcal{S}_\varphi(\bm{\Sigma}),\,\mathcal{A}_\varphi(\bm{\Sigma},\tilde{\bm{M}})\big),
\]
where $\mathcal{A}_\varphi(\bm{\Sigma},\tilde{\bm{M}}) := \big(A_1(\bm{\Sigma},\tilde{\bm{M}}),\dots,A_m(\bm{\Sigma},\tilde{\bm{M}})\big)$ accounts for anisotropy.  
Similarly to $\mathcal{S}_\varphi$, we assume that each anisotropic invariant $A_k$ satisfies
\begin{equation}
    \frac{\partial A_k}{\partial \bm{\Sigma}} : \bm{\Sigma} \;=\; A_k, 
    \qquad k=1,\dots,m.
\end{equation}

Thermodynamic consistency can then be established by the same arguments as in the isotropic case for
\begin{equation}
\varphi(\mathcal{T}_\varphi) = (\zeta \circ \omega)(\mathcal{T}_\varphi).
\label{eq:dual_potential_T}
\end{equation}
A suitable choice of invariants $\mathcal{A}_\varphi$ will be introduced in Section~\ref{sec:DualPotential}; see Appendix~\ref{app:consistency} for the corresponding proof of consistency.
\subsection{Helmholtz free energy}
\label{sec:Helmholtz}
\begin{figure}
    \centering
\begin{tikzpicture}
  \begin{groupplot}[
    group style={
      group size=2 by 1,
      horizontal sep=2cm,
    },
    width=5cm,
    height=4cm,
    axis lines=middle,
    axis line style={->},
    xmin=-1, xmax=3,
    ymin=-1, ymax=3,
    samples=200,
    domain=-3:3,
    xtick={1},
    ytick=\empty,
    enlargelimits=false,
    every axis x label/.style={
      at={(current axis.right of origin)}, anchor=west, yshift=-2pt
    },
    every axis y label/.style={
      at={(current axis.above origin)}, anchor=south, xshift=2pt
    },
  ]

    \nextgroupplot[
      title={$\beta_1$, $\beta_2$}, 
      domain=-1:3,
      ymin=-2, ymax=4,
      xlabel={ $A_i$, $A$}
    ]
    \addplot[rwth4, line width=2pt, samples=200, domain=0.01:3] {x - 1 - ln(x)};

    \nextgroupplot[
      title={$\beta_3$},
      domain=-1:3,
      ymin=-2, ymax=4,
      xlabel={$\mathrm{tr}\bm{A}$}
    ]
    \addplot[orange, line width=2pt, samples=200] {ln(cosh(max(x-1,0)))};

  \end{groupplot}
\end{tikzpicture}
    \caption{Illustration of the normalization functions for the isotropic $\mathcal{S}_\psi$ and anisotropic $\mathcal{A}_\psi$ sets for the Helmholtz free energy $\psi$.}
    \label{fig:normalization_func}
\end{figure}
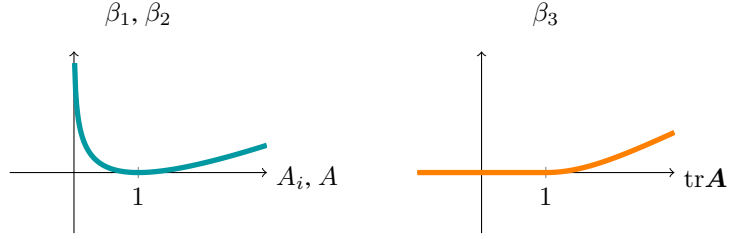
In the previous section, we have introduced several principles the Helmholtz free energy must respect. 
As a consequence, we express it in terms of invariants, which we now define. 
Inelastic materials are usually described relative to the intermediate configuration through elastic and inelastic invariants. 
To include anisotropy, the structural tensor is pushed to the intermediate configuration by an admissible mapping $\bm{F}_i \star \bm{M}$, as discussed previously.
The structural vector $\tilde{\bm{n}}$ in the intermediate configuration is obtained via a covariant or contravariant mapping \cite{Dafalias1987}
\begin{equation}
    \tilde{\bm{n}} \in 
    \left\{\frac{\bm{F}_i\,\bm{n}}{\|\bm{F}_i\,\bm{n}\|}, \quad
    \frac{\mathrm{cof}\bm{F}_i\,\bm{n}}{\|\mathrm{cof}\bm{F}_i\,\bm{n}\|}\right\} 
    =: \{ \tilde{\bm{n}}_1, \tilde{\bm{n}}_2 \},
\label{eq:structural_vec_push}
\end{equation}
which we normalize, since we do not attribute any physical meaning with the vector’s length \cite{Reese2003,Boes2023}.
Consequently, two structural second-order tensors follow
\[
    \tilde{\bm{M}} \in 
    \left\{\tilde{\bm{n}}_1 \otimes \tilde{\bm{n}}_1,\ 
          \tilde{\bm{n}}_2 \otimes \tilde{\bm{n}}_2\right\} 
    =: \{\tilde{\bm{M}}_1,\tilde{\bm{M}}_2\}.
\]

With the elastic right Cauchy--Green tensor $\bm{C}_e = \bm{F}_e^T\bm{F}_e$ and the inelastic left Cauchy--Green tensor $\bm{B}_i = \bm{F}_i^T\bm{F}_i$, the following \textit{similarity} holds
\begin{equation}
    \bm{C}_e^k\,\bm{B}_i^r\,\tilde{\bm{M}}_1^s\,\tilde{\bm{M}}_2^t = 
    \bm{F}_i^{-T}\,
    \big(\bm{C}\bm{C}_i^{-1}\big)^k\,\bm{C}_i^{\,r}
    \left(\tfrac{\bm{C}_i\bm{M}}{\bm{C}_i : \bm{M}}\right)^{\!s}
    \left(\tfrac{\bm{M}\bm{C}_i^{-1}}{\bm{C}_i^{-1} : \bm{M}}\right)^{\!t}
    \bm{F}_i^T,
    \quad k,r\in\mathbb{Z},\quad s,t\in\mathbb{N}_0.
\label{eq:invariants_map_psi}
\end{equation}
Thus every invariant relative to the intermediate configuration can be expressed in referential quantities while remaining independent of the choice of intermediate configuration. 
Since $\tilde{\bm M}_1$ and $\tilde{\bm M}_2$ are idempotent, we may restrict $s,t\in\{0,1\}$ without loss.\footnote{Mixed-variant mappings are not discussed in this contribution; see Appendix~\ref{app:Helmholtz}.}

Using \eqref{eq:invariants_map_psi}, we define the isotropic set $\mathcal{S}_\psi$ and the anisotropic set $\mathcal{A}_\psi$. 
In line with \cite{Zheng1994}, the isotropic irreducible basis in the intermediate configuration is
\[
\bm{C}_e,\ \mathrm{cof}\bm{C}_e,\ J_e,\ 
\bm{B}_i,\ \mathrm{cof}\bm{B}_i,\ J_i,\ 
\bm{C}_e\bm{B}_i,\ \mathrm{cof}\bm{C}_e\,\mathrm{cof}\bm{B}_i,\ 
\mathrm{cof}\bm{C}_e\,\bm{B}_i,\ \bm{C}_e\,\mathrm{cof}\bm{B}_i,
\]
which maps to the reference configuration as
\[
\bm{C}\bm{C}_i^{-1},\quad 
\tfrac{1}{J_i^2}\bm{C}_i\,\mathrm{cof}\bm{C},\quad 
\tfrac{J}{J_i},\quad 
\bm{C}_i,\quad \mathrm{cof}\bm{C}_i,\quad J,\quad 
\bm{C},\quad \mathrm{cof}\bm{C},\quad 
\tfrac{1}{J_i^2}\bm{C}_i\,\mathrm{cof}\bm{C}\,\bm{C}_i,\quad 
J_i^2\,\bm{C}\bm{C}_i^{-2},
\]
where $J_i$ has been replaced by $J$ to directly account for the volumetric growth condition \eqref{eq:growth_J}.
The trace of each second-order argument is convex in either $\bm F$ or $\mathrm{cof}\bm F$, as they are of the form $f(\bm F,\bm G)=\mathrm{tr}(\bm F^T\bm F\,\bm G)$ (or analogously with $\mathrm{cof}\,\bm F$) with $\bm G\succeq 0$; see Appendix~\ref{app:Helmholtz}. 
In addition, the scalar invariants $J/J_i$ and $J$ are convex in $J$. 
Hence, the entire irreducible basis is polyconvex and defines $\mathcal{S}_\psi$.
Notably, the seventh and eighth mixed invariants of elastic and inelastic arguments simplify to total deformation, which gives them a clear physical interpretation.

For the anisotropic set $\mathcal{A}_\psi$, we restrict ourselves to the mapping $\tilde{\bm M}_1$ for simplicity, though both can be included (e.g.\ letting the neural network select the more appropriate one). 
In line with \cite{Zheng1994}, we introduce
\[
\bm{C}_e\tilde{\bm M}_1,\quad 
\mathrm{cof}\bm{C}_e\tilde{\bm M}_1,\quad 
\bm{B}_i\tilde{\bm M}_1,\quad 
\mathrm{cof}\bm{B}_i\tilde{\bm M}_1,\quad 
\bm{C}_e(\bm{B}_i \,\#\, \tilde{\bm M}_1),
\]
which map to
\[
\tfrac{\bm{C}\bm{M}}{\bm{C}_i : \bm{M}},\quad 
\tfrac{1}{J_i^2}\,\bm{C}_i\,\mathrm{cof}\bm{C}\,\tfrac{\bm{C}_i\bm{M}}{\bm{C}_i:\bm{M}},\quad 
\tfrac{\bm{C}_i^2\bm{M}}{\bm{C}_i:\bm{M}},\quad 
\tfrac{J_i^2\bm{M}}{\bm{C}_i:\bm{M}},\quad 
\bm{C}\bm{C}_i^{-1}\Big(\bm{C}_i \,\#\, \tfrac{\bm{M}\bm{C}_i}{\bm{C}_i:\bm{M}}\Big),
\]
with $\#$ the outer product of second-order tensors\footnote{For materials with two distinct in-plane preferred directions, the outer product of the corresponding structural tensors can be used to compute the structural tensor being normal to the plane they span; cf. \cite{Boes2023}.}; see Appendix~\ref{app:Helmholtz}. 
This unusual choice replaces the classical invariant $\bm{C}_e\bm{B}_i\tilde{\bm M}_1$, which is generally not polyconvex.

To normalize the free energy and stresses, we introduce the convex functions
\[
\beta_1(\bm A) := \mathrm{tr}\bm A - 3 - \ln\det\bm A, \quad
\beta_2(A) := A - 1 - \ln A, \quad
\beta_3(\bm A) := \ln\cosh(\max(\mathrm{tr}\bm A - 1,0)).
\]
For $\beta_1$, one has $\beta_1=\sum_{i=1}^3 (A_i - 1 - \ln A_i)$ if $\bm A\in\mathrm{Sym}_+^{3\times 3}$ with the eigenvalues $A_i$. 
All $\beta_j$ are nonnegative, convex, and vanish at the normalized state ($\bm A=\bm I$ for $\beta_1$, $A=1$ for $\beta_2$, and $\mathrm{tr}\bm A=1$ for $\beta_3$).
Further, $\beta_3$ is non-decreasing.
Figure~\ref{fig:normalization_func} illustrates the normalization functions $\beta_1$, $\beta_2$, and $\beta_3$.

We then define the isotropic invariant set
\begin{equation}
\begin{aligned}
    S_1  &= \beta_1(\bm C \bm C_i^{-1}), 
    &\quad S_2  &= \beta_1\!\left(\tfrac{1}{J_i^2}\bm C_i \mathrm{cof}\bm C\right), 
    &\quad S_3  &= \beta_2\!\left(\tfrac{J}{J_i}\right), 
    &\quad S_4  &= \beta_1(\bm C_i), \\[0.5em]
    S_5  &= \beta_1(\mathrm{cof}\,\bm C_i), 
    &\quad S_6  &= \beta_2(J), 
    &\quad S_7  &= \beta_1(\bm C), 
    &\quad S_8  &= \beta_1(\mathrm{cof}\,\bm C), \\[0.5em]
    S_9  &= \beta_3\!\left(\tfrac{1}{J_i^2}\bm C_i\,\mathrm{cof}\bm C\,\bm C_i\right), 
    &\quad S_{10} &= \beta_3(J_i^2 \bm C \bm C_i^{-2}).
\end{aligned}
\end{equation}
Each $S_i$ is polyconvex\footnote{For $f(J)=-\ln(a J^b)=-\ln a - b \ln J$ with $J>0$ and $a>0$, one has $f''(J)=b/J^2$. Since $J^2>0$, the function is convex iff $b\geq 0$ (strictly convex for $b>0$).
The invariant $f(\bm{C},\bm{G})=\mathrm{tr}(\bm{C}\bm{G})$ is polyconvex if $\bm{G}\succeq 0$, same holds when $\bm{C}$ is replaced by its cofactor; see Appendix~\ref{app:Helmholtz}.}. 

Analogously, the anisotropic set is
\begin{equation}
\begin{aligned}
    A_1 &= \beta_1\!\left(\tfrac{\bm C\bm M}{\bm C_i : \bm M}\right), 
    &\quad A_2 &= \beta_3\!\left(\tfrac{1}{J_i^2}\,\bm C_i\,\mathrm{cof}\bm C\,
                      \tfrac{\bm C_i\bm M}{\bm C_i:\bm M}\right), 
    &\quad A_3 &= \beta_3\!\left(\tfrac{\bm C_i^2 \bm M}{\bm C_i:\bm M}\right), \\[0.5em]
    A_4 &= \beta_3\!\left(\tfrac{J_i^2\bm M}{\bm C_i:\bm M}\right), 
    &\quad A_5 &= \beta_3\!\left(\bm C\bm C_i^{-1}
            \Big( \bm C_i \,\#\, \tfrac{\bm M\bm C_i}{\bm C_i:\bm M}\Big)\right).
\end{aligned}
\end{equation}
Again, each $A_i$ is polyconvex. 

Finally, the Helmholtz free energy is expressed as
\begin{equation}
\psi = \psi\left(\mathcal{T}_\psi\right), \quad \mathcal{T}_\psi(\bm C,\bm C_i,\bm M) := \left(\mathcal{S}_\psi(\bm C,\bm C_i),\,
                                 \mathcal{A}_\psi(\bm C,\bm C_i,\bm M))\right).
\label{eq:Helmholtz_Tset}
\end{equation}
We follow the energy normalization condition, but slightly relax the stress normalization condition. 
That is, we allow for individual normalization conditions by choosing the $\beta_i$ functions.
However, if mixed invariants involving both elastic and inelastic parts are not employed, our choices for $\beta_i$ are consistent with the classical assumption of an elastic, stress-free state.
%
\subsection{Dual potential}
\label{sec:DualPotential}
We introduce the set $\mathcal{T}_\varphi$~\eqref{eq:dual_potential_T} as a basis for describing the inelastic evolution of solids. 
Similarly to the Helmholtz free energy, the set is formulated in terms of invariants. 
In contrast to $\psi$, which must be polyconvex in $(\bm F,\mathrm{cof}\bm F,J)$ (rather than merely in its invariants), thermodynamic consistency of the dual potential does not require convexity of the chosen invariants. 
This reflects the different mathematical roles of the two potentials; see \cite{Gluege2017,Janeka2018} for theoretical discussions of non-convex potentials and \cite{Argyris1974} for experimental evidence of non-convex yield surfaces in prestressed concrete.

The isotropic invariants $\mathcal{S}_\varphi$ are given in Equation~\eqref{eq:stress_invars_isotropic}. 
For the anisotropic contribution, we permit both push-forwards of the structural vector in~\eqref{eq:structural_vec_push}. 
Following \cite{Zheng1994}, this leads to
\begin{equation}
\begin{aligned}
    A_{11} &= \mathrm{dev}\,\bm{\Sigma} : \tilde{\bm{M}}_1, 
    &\quad A_{12} &= \mathrm{dev}\,\bm{\Sigma} : \tilde{\bm{M}}_2, 
    &\quad A_{21} &= \sqrt{\tfrac{1}{2}\,\bigl(\mathrm{dev}\,\bm{\Sigma}\bigr)^2 : \tilde{\bm{M}}_1}, \\[0.5em]
    A_{22} &= \sqrt{\tfrac{1}{2}\,\bigl(\mathrm{dev}\,\bm{\Sigma}\bigr)^2 : \tilde{\bm{M}}_2}, 
    &\quad A_{3}  &= \bm{\Sigma} : \mathrm{sym}(\tilde{\bm{n}}_1 \otimes \tilde{\bm{n}}_2), 
    &\quad A_{4}  &= \sqrt{\tfrac{1}{2}\,\bm{\Sigma}^2 : \mathrm{sym}(\tilde{\bm{n}}_1 \otimes \tilde{\bm{n}}_2)} ,
\end{aligned}
\end{equation}
and we set $\mathcal{A}_\varphi := \left(A_{11},A_{12},A_{21},A_{22},A_{3},A_{4}\right)$. 
All these invariants are positively $1$-homogeneous in $\bm{\Sigma}$.

As discussed earlier, the Mandel stress $\bm{\Sigma}$ shares its eigenvalues with the Kirchhoff stress under suitable conditions, while the Cauchy stress admits a clearer physical interpretation (being related by the determinant $J$). 
Using the positive $1$-homogeneity of $\mathcal{T}_\varphi$ in $\bm{\Sigma}$, the evolution equation~\eqref{eq:Evolution} can be written as
\begin{equation}
    \bm D_i \in \partial_{\bm\Sigma}\,\varphi\!\left(\tfrac{\mathcal{T}_\varphi}{J}\right)
    = \tfrac{1}{J}\,\Big[\partial_{\mathbf z}\varphi(\mathbf z)\Big]_{\mathbf z=\mathcal{T}_\varphi/J} : \partial_{\bm\Sigma}\mathcal{T}_\varphi ,
\end{equation}
where the gradient of $\varphi$ is evaluated at $\mathcal{T}_\varphi/J$. 
Thermodynamic consistency is preserved, because for each component $T_k$ of $\mathcal{T}_\varphi$,
\[
(\partial_{\bm\Sigma}T_k):\bm\Sigma \;=\; T_k, \quad k=1,\hdots,9
\]
hence,
\[
\tfrac{1}{J}\,\langle\partial_{\bm\Sigma}\mathcal{T}_\varphi, \bm{\Sigma} \rangle
\;=\; \tfrac{\mathcal{T}_\varphi}{J}.
\]

\paragraph{Co-rotated intermediate configuration.}
Direct computation of $\bm{\Sigma}$, $\tilde{\bm n}_1$, and $\tilde{\bm n}_2$ would require the inelastic part $\bm F_i$ of the deformation gradient, which is non-unique. 
One may either establish a mapping from intermediate to referential invariants (analogous to Equation~\eqref{eq:invariants_map_psi}) or adopt a co-rotated intermediate configuration \cite{Holthusen2023} (similarly exploited in \cite{Liu2025}) that maps all non-unique quantities to unique counterparts via (cf. Figure~\ref{fig:map})
\[
\bar{(\bullet)} := \bm R_i^T \star (\bullet) = \bm R_{i*}^T \star (\bullet)_{*},
\]
where $\bm R_i$ and $\bm R_{i*}$ denote the rotational parts associated with $\bm F_i$ and $\bm{F}_{i*}$, respectively. 
This yields
\begin{equation}
    \bar{\bm{\Sigma}} = -\,2\,\bm U_i\,\frac{\partial\psi}{\partial\bm C_i}\,\bm U_i, 
    \quad 
    \bar{\bm{D}}_i = \mathrm{sym}(\dot{\bm U}_i\bm U_i^{-1})
    = \tfrac{1}{2}\,\bm U_i^{-1}\,\dot{\bm C}_i\,\bm U_i^{-1},
    \quad
    \bar{\bm{n}}_1 = \frac{\bm{U}_i\,\bm{n}}{\|\bm{U}_i\,\bm{n}\|},
    \quad
    \bar{\bm{n}}_2 = \frac{\mathrm{cof}\bm{U}_i\,\bm{n}}{\|\mathrm{cof}\bm{U}_i\,\bm{n}\|},
    \quad
\end{equation} 
An attractive feature of this representation is that the functional form of constitutive relations—such as the invariant set—remains unchanged under the transformation.

\section{Constitutive Neural Network for Anisotropic Inelasticity}
\label{sec:ConstiNN}
Building on the constitutive fundamentals in Section~\ref{sec:Fundamentals}, the invariant sets $\mathcal{T}_\psi$ and $\mathcal{T}_\varphi$ introduced in Sections~\ref{sec:Helmholtz} and \ref{sec:DualPotential}, and the network architectures of Section~\ref{sec:NetworkTypes}, we now combine neural networks with the constitutive framework of anisotropic inelasticity. 
Our aim is to discover the Helmholtz free energy and the dual potential in a thermodynamically consistent manner that respects the material principles. 
In fact, the approach to discovering the Helmholtz free energy is closely aligned with the Physics-Augmented Neural Networks (PANNs) framework \cite{Klein2022,Klein2022b,Linden2023}.
Furthermore, we seek to replace the solution of the implicit time discretization scheme by a recurrent architecture, in analogy to \cite{Asad2023,Rosenkranz2024}.
While the small strain regime is investigated in \cite{Rosenkranz2024} and the finite strain regime is studied in \cite{Asad2023}, both assume an additive decomposition of the total strain and internal variables.

\paragraph{Helmholtz free energy.}
The Helmholtz free energy is represented by an Input Convex Neural Network (ICNN) that is polyconvex with respect to $(\bm{F},\mathrm{cof}\bm{F},J)$. 
However, we pass $\mathcal{T}_\psi$ to the ICNN, which already provides a convex representation of $(\bm{F},\mathrm{cof}\bm{F},J)$.
This motivates the slight adaptation of the ICNN architecture in Equation~\eqref{eq:ICNN_adapted} compared to the original \eqref{eq:ICNN}.
To enforce the volumetric growth conditions~\eqref{eq:growth_J}--\eqref{eq:growth_Je}, we augment the network with a convex penalty contribution $\psi_{\mathrm{gr}}$ that prevents the discovery of spurious solutions without volumetric dependence. 
Employing the strictly convex function $\beta_2(A) = A - 1 - \ln A$
the energy takes the form
\begin{equation}
    \psi = \mathcal{N}_c(\bullet,\,\mathcal{T}_\psi) 
    + \psi_{\mathrm{gr}}(J,\tfrac{J}{J_i}),
\label{eq:psi_as_network}
\end{equation}
with
\begin{equation}
    \psi_{\mathrm{gr}}(J,\tfrac{J}{J_i}) 
    = \lambda_{\mathrm{gr}} \big( \beta_2(J) + \beta_2(\tfrac{J}{J_i}) \big),
\end{equation}
where $\lambda_{\mathrm{gr}}$ is a penalty parameter, fixed to $10^{-4}$ in all simulations. 
In Equation~\eqref{eq:psi_as_network}, $\bullet$ denotes the dependence of $\psi$ on the non-convex quantities. 

\paragraph{Dual potential.}
For the dual potential, we employ the composition of ICNNs and Input Monotonic Neural Networks proposed in \eqref{eq:composition_network}:
\begin{equation}
    \varphi = \mathcal{N}_\circ(\bullet,\,(\mathcal{T}_\varphi,-\mathcal{T}_\varphi)),
\label{eq:phi_as_network}
\end{equation}
where $\bullet$ again denotes the dependence on non-convex quantities. 
Unlike the free energy, $\varphi$ must only be convex with respect to the invariants themselves, not with respect to their arguments, i.e., the thermodynamic driving force $\bm{\Sigma}$. 
To recover the structure of the original ICNN~\eqref{eq:ICNN}, we pass both $\mathcal{T}_\varphi$ and $-\mathcal{T}_\varphi$ through the convex branch of the network. 
To guarantee non-negativity of the ICNN output within the composition, the final activation function is chosen as the ReLU. 
Due to the invariants’ definition, the zero-value constraint is automatically satisfied.

\paragraph{Time discretization.}
In the inelastic setting, the evolution problem must be discretized in time $t \in [t_n, t_{n+1}]$. 
For the finite strain regime, the exponential integrator scheme is well-suited \cite{Vladimirov2008}, as it consistently preserves the underlying kinematics of the evolution of $\bm{U}_i$ \cite{Holthusen2025growth}. 
The \textit{explicit} integration step reads
\begin{equation}
    \bm{C}_{i_{n+1}} = \bm{U}_{i_n}\, \exp(2\, \Delta t\, \bm{D}_{i_n})\, \bm{U}_{i_n}, 
    \qquad \bm{U}_{i_{n+1}} = \sqrt{\bm{C}_{i_{n+1}}},
    \label{eq:explicit_evo}
\end{equation}
with $\Delta t := t_{n+1}-t_n$. 
The corresponding \textit{implicit} scheme requires solving
\begin{equation}
    \mathbf{r} := \bm{C}_{i_{n}} - \bm{U}_{i_{n+1}}\, \exp(-2\, \Delta t\, \bm{D}_{i_{n+1}})\, \bm{U}_{i_{n+1}} \;\overset{!}{=}\; \bm{0},
    \label{eq:implicit_evo}
\end{equation}
where $\bm{D}_{i_{n+1}}$ depends nonlinearly on $\bm{U}_{i_{n+1}}$.

\begin{figure}
    \centering
    \includegraphics{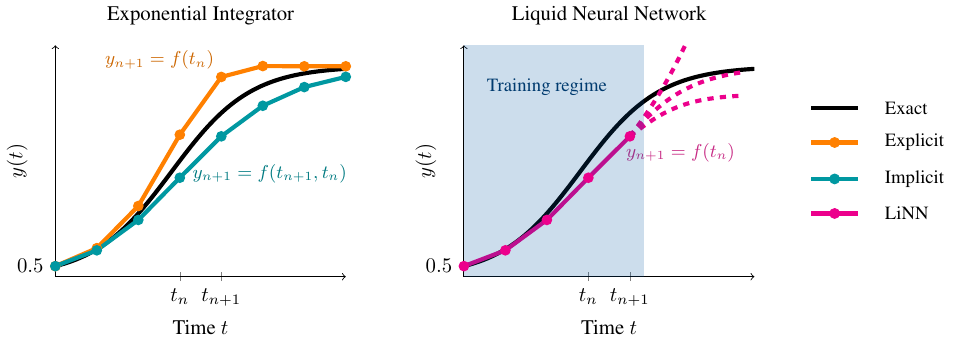}
\caption{
Schematic of time discretizations of the ordinary differential equation $\dot{y} = A(y)\,y$ with $A(y) = r(1 - y/k)$, where $r = 0.7$ and $k = 10$. 
The exponential integrator with time step $\Delta t = 1.5$ shows the explicit scheme $y_{n+1} = \exp(\Delta t\,A(y_n))\,y_n$ and the implicit scheme $y_{n+1} = \exp(\Delta t\,A(y_{n+1}))\,y_n$, the latter must be solved numerically. 
The Liquid Neural Network (LiNN) approximates $y_{n+1}$ by minimizing the loss $\mathcal{L}=(y_{n+1} - \exp(\Delta t\,A(y_{n+1}))\,y_n)^2$. 
Within the training domain, the LiNN attains accuracy comparable to the implicit scheme while remaining explicit; outside this domain, predictive reliability is not ensured.
}
\label{fig:time_plot}
\end{figure}

Following \cite{Asad2023,Rosenkranz2024}, we predict the internal variable by an auxiliary neural network during training. 
This facilitates stable training of the physically relevant networks for $\psi$ and $\varphi$. 
We adopt an architecture based on LiNNs~\eqref{eq:LiNN_EM}; Figure~\ref{fig:time_plot} illustrates the different time discretization schemes based on the explicit and implicit exponential integrators as well as the LiNN. 
The hidden states $\mathbf{h}\in\mathbb{R}^6$ are initialized as $\mathbf{h}_0=\mathbf{0}$. 
We then learn $\bm{\alpha}=\mathcal{N}_\alpha$ and $\mathbf{f}=\mathcal{N}_f$ via
\begin{equation}
(\mathbf{h}_n,(\bm{E}_{n+1},\bm{D}_i^{\mathrm{trial}})) \mapsto \mathcal{N}_\alpha(\mathbf{h}_n,(\bm{E}_{n+1},\bm{D}_i^{\mathrm{trial}})), 
\quad 
(\mathbf{h}_n,(\bm{E}_{n+1},\bm{D}_i^{\mathrm{trial}})) \mapsto \mathcal{N}_f(\mathbf{h}_n,(\bm{E}_{n+1},\bm{D}_i^{\mathrm{trial}})),
\label{eq:Na_Nf}
\end{equation}
where $\bm{E}_{n+1}$ denotes the Green--Lagrange strain tensor and $\bm{D}_i^{\mathrm{trial}}$ is the trial value of $\bm{D}_i$, i.e., $\bm{D}_i$ evaluated with the current $\bm{C}_{n+1}$ but freezing $\bm{U}_{i_{n+1}}=\bm{U}_{i_n}$. 
This choice follows the well-established trial step procedure in plasticity, ensuring that the input is close to the target output, which--to our experience--in practice stabilizes training. 
A systematic study of this effect is beyond the current scope.

Evaluating $\mathcal{N}_\alpha$ and $\mathcal{N}_f$ in Equation~\eqref{eq:Na_Nf} allows us to update the hidden state according to Equation~\eqref{eq:LiNN_EM}. 
Since these states do not generally coincide with the independent components of $\bm{U}_i \in \mathrm{Sym}_+^{3\times 3}$, positive definiteness must be enforced. 
For this purpose, we employ the Cholesky decomposition of $\bm{U}_i$ and map $\mathbf{h}$ to the lower-triangular factor of $\bm{U}_i$; see Appendix~\ref{app:numerics}.

\paragraph{Loss function.}
During training, the weights and biases $\bm{\theta} := \{\mathbf{W}_\psi, \mathbf{b}_\psi, \mathbf{W}_\varphi, \mathbf{b}_\varphi, \mathbf{W}_h, \mathbf{b}_h\}$
of the three networks for $\psi$, $\varphi$, and the hidden states $\mathbf{h}$ are determined by solving the minimization problem
\begin{equation}
    \bm{\theta}^\ast = \arg\min_{\bm{\theta}} \; \mathcal{L}(\bm{\theta}),
\end{equation}
where the total loss is additively split into\footnote{A regularization term can be added if desired.}
\begin{equation}
    \mathcal{L}(\bm{\theta}) = \mathcal{L}_{\mathrm{stress}}(\bm{\theta}) 
    + \lambda_{\mathrm{evo}}\,\mathcal{L}_{\mathrm{evo}}(\bm{\theta}).
    \label{eq:loss}
\end{equation}
Here, $\mathcal{L}_{\mathrm{stress}}$ measures the discrepancy between predicted and experimentally observed stresses, while $\mathcal{L}_{\mathrm{evo}}$ quantifies the residual of the implicit evolution equation~\eqref{eq:implicit_evo}. 
The penalty parameter $\lambda_{\mathrm{evo}}$, accounting for the relative magnitudes of the two losses, is heuristically set to $1000$. 
Both contributions are evaluated via mean squared error. 
For the stress contribution we obtain
\begin{equation}
    \mathcal{L}_{\mathrm{stress}}(\bm{\theta}) 
    = \frac{1}{B \, T \, O} \sum_{b=1}^B \sum_{t=1}^T \sum_{o=1}^O 
\left( \underline{S}_{b,t,o}(\bm{\theta}) - \underline{\hat{S}}_{b,t,o} \right)^2,
\label{eq:loss_stress}
\end{equation}
with the number of batches $B$, time steps $T$, independent stress components $O=6$, the predicted normalized second Piola--Kirchhoff stress components $\underline{S}(\bm{\theta})$, and the normalized experimental stresses $\underline{\hat{S}}$. 
Normalization is performed by dividing the raw experimental stress $\hat{\bm{S}}$ by the absolute maximum $\underline{S}$ across all batches and time steps, i.e., $\hat{\underline{\bm{S}}}=\hat{\bm{S}}/\underline{S}$.
Thus, after training we have to multiply $\underline{\bm{S}}$ by $\underline{S}$ to obtain $\bm{S}$. 
Analogously, the evolution loss is defined as
\begin{equation}
    \mathcal{L}_{\mathrm{evo}}(\bm{\theta}) 
    = \frac{1}{B \, T \, O} \sum_{b=1}^B \sum_{t=1}^T \sum_{o=1}^O 
\left( r_{b,t,o}(\bm{\theta}) \right)^2,
\end{equation}
where $r(\bm{\theta})$ denotes the independent components of the residual of the implicit evolution equation~\eqref{eq:implicit_evo}.

\section{Numerical results}
\label{sec:results}
In this section, we present a comprehensive evaluation of the proposed framework for discovering inelastic material behavior at finite strains. 
The analysis proceeds in three steps. 
First, we describe the generation of artificial training data using a classical constitutive model (Section~\ref{sec:results_generation}). 
Second, we examine the predictive capabilities of the discovered networks at the material point level, both for isotropic and anisotropic materials, and compare them to recurrent neural network baselines that lack physics priors (Section~\ref{sec:results_material_point}). 
Finally, we assess whether the favorable material point performance translates to unseen structural boundary value problems, thereby testing the robustness and generalization capability of the framework (Section~\ref{sec:results_structural}).

For clarity, the same neural architectures are employed across all studies. 
The network representing the Helmholtz free energy \eqref{eq:psi_as_network} is defined as
\[
\mathcal{N}_c:\mathbb{R}^{15}\rightarrow\mathbb{R},\quad 
\mathcal{T}_\psi \ 
\textcolor{rwth1}{\xrightarrow{\exp} 16 
\xrightarrow{\mathrm{Softplus}} 16 
\xrightarrow{\mathrm{Softplus}} 16 
\xrightarrow{\mathrm{Softplus}} 16 
\xrightarrow{\mathrm{Softplus}}}\ \psi,
\]
while the network for the dual potential \eqref{eq:phi_as_network} is designed as (cf.~Figure~\ref{fig:composition_network} for the color code)
\[
\mathcal{N}_\circ:\mathbb{R}^{18}\rightarrow\mathbb{R},\quad 
(\mathcal{T}_\varphi,-\mathcal{T}_\varphi)\ 
\textcolor{rwth1}{\xrightarrow{\exp} 16 
\xrightarrow{\mathrm{ReLU}} 16 
\xrightarrow{\mathrm{ReLU}} 16 
\xrightarrow{\mathrm{ReLU}}}\ 
\textcolor{rwth9}{4}\ 
\textcolor{rwth8}{\xrightarrow{\tanh} 16 
\xrightarrow{\tanh} 16 
\xrightarrow{\mathrm{Softplus}} 16 
\xrightarrow{\mathrm{Linear}}}\ \varphi.
\]

During training, the internal variables are predicted by an auxiliary Liquid Neural Network (LiNN), which serves to stabilize the evolution dynamics
\[
\mathcal{N}_\alpha:\mathbb{R}^{18}\rightarrow\mathbb{R}^6,\quad 
(\mathbf{h}_n,(\bm{E}_{n+1},\bm{D}_i^{\mathrm{trial}}))\ 
\xrightarrow{\mathrm{GELU}} 12 
\xrightarrow{\mathrm{GELU}} 12 
\xrightarrow{\mathrm{GELU}} 8 
\xrightarrow{\mathrm{GELU}} 8 
\xrightarrow{\mathrm{ReLU}} \bm{\alpha},
\]
and
\[
\mathcal{N}_f:\mathbb{R}^{18}\rightarrow\mathbb{R}^6,\quad 
(\mathbf{h}_n,(\bm{E}_{n+1},\bm{D}_i^{\mathrm{trial}}))\ 
\xrightarrow{\mathrm{GELU}} 12 
\xrightarrow{\mathrm{GELU}} 12 
\xrightarrow{\mathrm{GELU}} 8 
\xrightarrow{\mathrm{GELU}} 8 
\xrightarrow{\mathrm{Linear}} \mathbf{f},
\]
where $\mathrm{GELU}$ refers to the Gaussian Error Linear Unit activation function \cite{Hendrycks2023}.

All networks are implemented in \texttt{Flax} on top of \texttt{JAX} \cite{jax2018github}. 
Optimization uses the ADAM algorithm from \texttt{Optax} with a learning rate of $10^{-3}$. 
To avoid instabilities due to excessively large updates, global-norm gradient clipping with a threshold of $10^{-3}$ is applied, similar to \cite{Holthusen2025}. 
Finite element meshes are generated in \texttt{Gmsh} \cite{geuzaine2009gmsh}, and post-processing is performed in \texttt{ParaView} \cite{ayachit2015paraview}.

\subsection{Generation of training data}
\label{sec:results_generation}
\begin{figure}
    \centering
    \includegraphics[]{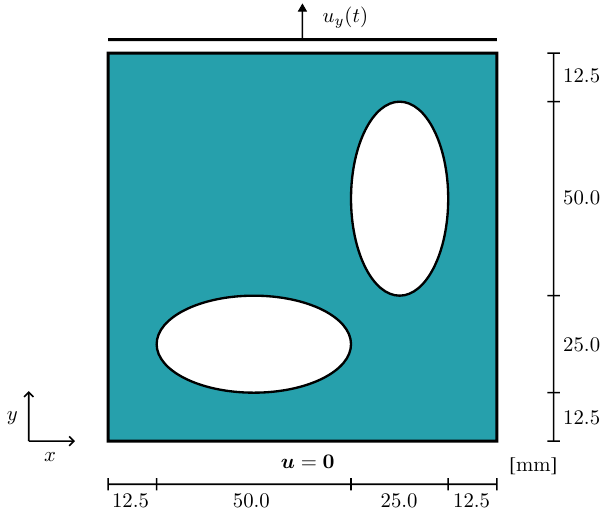}
    \caption{
        \textbf{Plate with two elliptic holes.}
        Geometry and boundary conditions of the BVP used to generate the training data set, adapted from \cite{Flaschel2022}.
        The bottom edge is clamped, while on the top edge a vertical displacement $u_y(t)$ is prescribed. 
        The displacements in $x$ and $z$ directions are free on the top, but the $z$-displacement is fixed at the front and rear surfaces.
        The plate has a thickness of $6\,\si{\milli\metre}$ (three hexahedral elements across the thickness); in-plane, $710$ elements are used.}
    \label{fig:PWTEH}
\end{figure}
\begin{figure}
    \centering
    \includegraphics{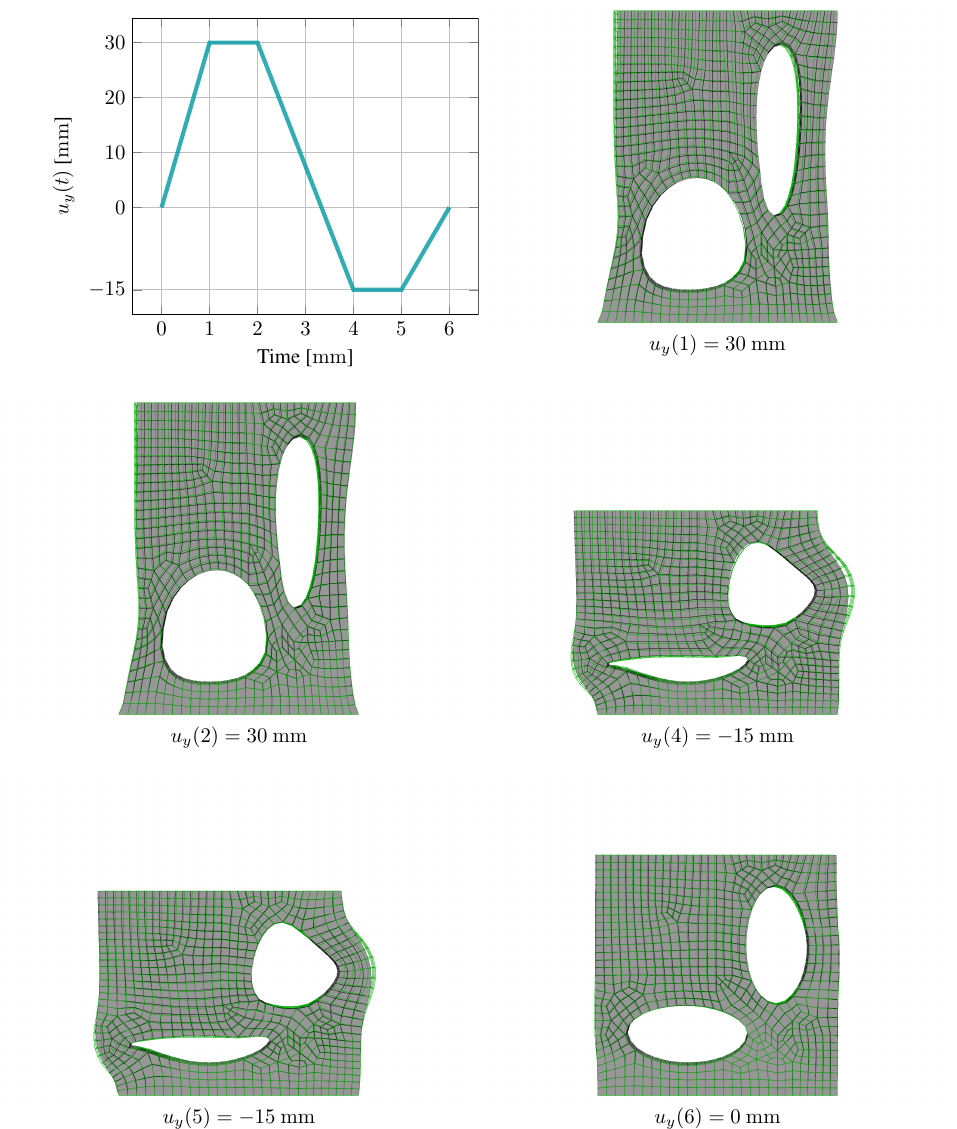}
    \caption{Prescribed loading program and representative deformed configurations for isotropic (green mesh) and anisotropic (gray body/black mesh) materials. 
    The comparison highlights the influence of anisotropy under otherwise identical loading. 
    At the end of the program, although the external displacement returns to its initial value, residual strains remain due to inelasticity and the unloaded configurations are not identical; see also Figure~\ref{fig:PWTEH_Graduy}.}
    \label{fig:PWTEH_stages}
\end{figure}
\begin{figure}
    \centering
    \includegraphics{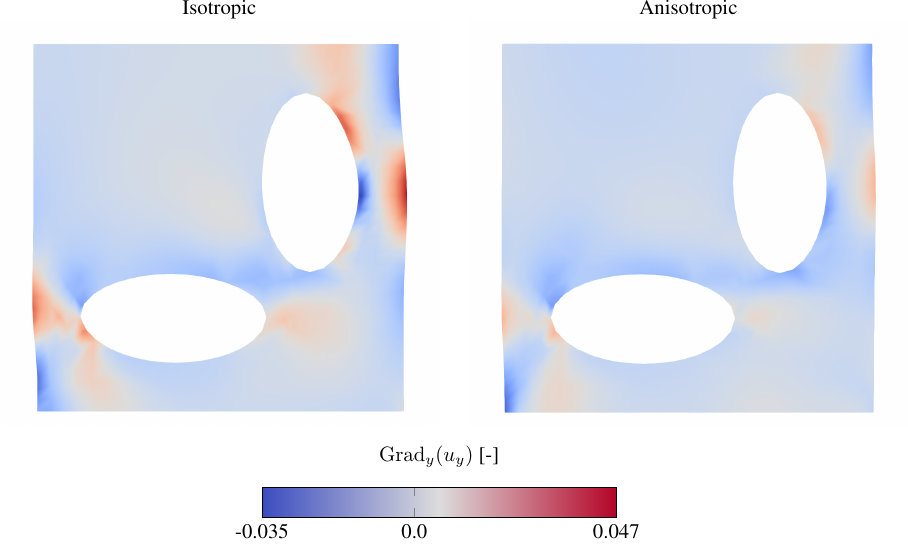}
    \caption{Gradient of the vertical displacement $u_y$ along the $y$-direction at the end of the loading program with $u_y(6) = 0~\si{\milli\metre}$. 
For a purely elastic material, the gradient vanishes throughout the specimen. 
In contrast, due to inelastic effects, a residual deformation remains for both the isotropic and anisotropic materials.}
    \label{fig:PWTEH_Graduy}
\end{figure}
\paragraph{Reference material model.}
To generate sufficiently rich and controlled data sets, we employ a classical constitutive model in which Helmholtz free energy and dual potential are additively split into isotropic and anisotropic parts
\begin{equation}
    \psi = \psi_{\mathrm{iso}} + \psi_{\mathrm{ani}}, 
    \quad 
    \varphi = \varphi_{\mathrm{iso}} + \varphi_{\mathrm{ani}}.
    \label{eq:reference_model}
\end{equation}
For the isotropic study, anisotropic terms are omitted. 
The functional forms follow \cite{Schroeder2008} for the free energy and \cite{Reese1998} for the dual potential; detailed equations are given in Appendix~\ref{app:reference_model}. 
Rheologically, the model corresponds to a three-element rheological system: an equilibrium spring in parallel with a non-equilibrium spring–dashpot branch. 
Noteworthy, anisotropy is introduced via a metric tensor rather than a structural tensor, and the dual potential uses a mixed-variant mapping without normalized vector lengths. 
Thus, the networks must approximate the constitutive response rather than reproducing it analytically.

Material parameters are listed in Table~\ref{tab:material_parameters}. 
For time integration of the evolution equations, we use the explicit scheme \eqref{eq:explicit_evo}. 
Additionally, in the anisotropic case, the preferred direction is
\[
\bm{n} = (1/\sqrt{2},\, 1/\sqrt{2},\, 0)^T.
\]

\begin{table}[h!]
  \caption{Material parameters of the reference model used to generate the artificial training data set. 
  The model can be interpreted as an equilibrium branch ($\mathrm{eq}$) in parallel with a non-equilibrium branch ($\mathrm{neq}$).}
  \centering
  \begin{tabular}{c c c | c c c | c c | c c | c | c}
    \multicolumn{6}{c |}{Isotropic} & \multicolumn{5}{c|}{Anisotropic} & Relaxation Time\\ \hline\hline
    \multicolumn{3}{c|}{$\mathrm{eq}$} &  \multicolumn{3}{c |}{$\mathrm{neq}$} & \multicolumn{2}{c|}{$\mathrm{eq}$} &  \multicolumn{2}{c|}{$\mathrm{neq}$} & & \\
    $a$ & $b$ & $c$ & $a$ & $b$ & $c$ & $\alpha$ & $\eta$ & $\alpha$ & $\eta$ & $\beta$ & $\tau$\\ \hline
    \si{\mega\pascal} & \si{\mega\pascal} & \si{\mega\pascal} & \si{\mega\pascal} & \si{\mega\pascal} & \si{\mega\pascal} & $-$ & \si{\mega\pascal} & $-$ & \si{\mega\pascal} & $-$ & \si{\second} \\
    80 & 100 & 100 & 40 & 50 & 50 & 2 & 10 & 2 & 10 & 2 & 12
  \end{tabular}
  \label{tab:material_parameters}
\end{table}

\paragraph{Boundary value problem and loading.}
The simulation setup is illustrated in Figure~\ref{fig:PWTEH}. 
The plate is discretized with 2130 eight-node hexahedral elements and loaded over 120 time steps of size $\Delta t = 0.05\,\si{\second}$, resulting in a total duration of $6\,\si{\second}$. 
The prescribed displacement $u_y(t)$ consists of three linear loading/unloading phases with two hold phases at constant displacement in between. 
Representative deformed configurations are shown in Figure~\ref{fig:PWTEH_stages}. 
The isotropic and anisotropic specimens deform differently under identical loading, and during the hold phase the anisotropic case is more prone to relax due to anisotropic dissipation. 
At the end of the program, both specimens exhibit residual deformations, characteristic of inelasticity; see Figure~\ref{fig:PWTEH_Graduy}.

\paragraph{Training data extraction and strategy.}
Stress–stretch data are extracted from one quadrature point per element. 
Here, we use only a subset of 1111 out of 2130 elements, yielding $133{,}320$ data pairs (1111 elements with 120 time steps). 
In contrast to elastic problems, clustering the invariant space is not feasible \cite{Kalina2021}, since the internal variables are not known a priori. 
Hence, the raw stress–stretch data are employed directly.
An approach to cluster also in the case of inelasticity is desirable for real world application but beyond the scope of this contribution.

As reported in the literature \cite{Holthusen2025}, direct training on the full dataset proved unstable. 
We therefore adopt a two-stage procedure: first, pre-train for 500 epochs on a single element and the first 40 time steps; second, continue training with the best parameters from pre-training on the full dataset. 
This strategy significantly improves stability. 

\begin{figure}
    \centering
    \includegraphics{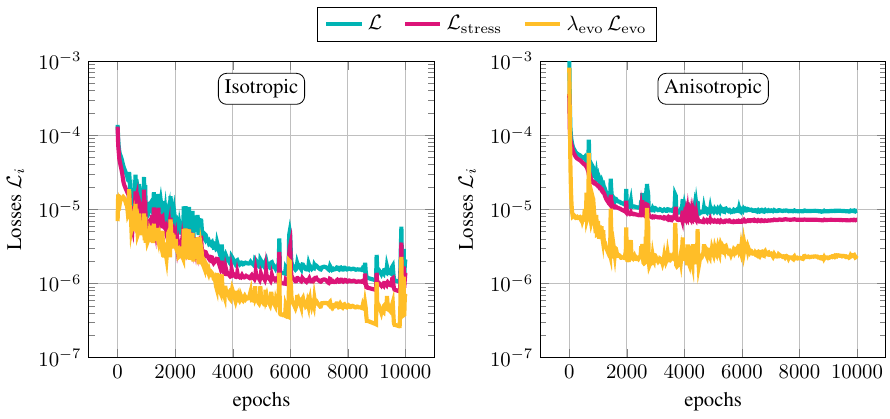}
    \caption{Training losses over $10{,}000$ epochs for isotropic and anisotropic data sets. 
    The penalty parameter is $\lambda_{\mathrm{evo}}=1000$. 
    No additional regularization is applied beyond gradient clipping.}
    \label{fig:PWTEH_losses}
\end{figure}

After pre-training, both isotropic and anisotropic networks are trained for 10,000 epochs. 
The resulting losses are shown in Figure~\ref{fig:PWTEH_losses}. 
The isotropic model converges to a loss of order $10^{-6}$, while the anisotropic one stabilizes around $10^{-5}$. 
The auxiliary loss related to the evolution equation remains below the stress loss but, due to the penalty $\lambda_{\mathrm{evo}}=1000$, neither contribution dominates. 
Normalization uses maximum stresses $\underline{S}=1406.429\,\si{\mega\pascal}$ (isotropic) and $\underline{S}=1014.538\,\si{\mega\pascal}$ (anisotropic); cf. Equation~\eqref{eq:loss_stress}.

\FloatBarrier

\subsection{Material point study}
\label{sec:results_material_point}
We next assess the performance of the discovered networks at the material point level. 
This study provides a controlled setting in which the predictive accuracy of the constitutive framework can be analyzed in detail, both for isotropic and anisotropic materials. 
At the same time, it allows for a systematic comparison with purely data-driven recurrent neural networks that do not incorporate constitutive knowledge, thereby highlighting the effect of embedding physical structure into the network architecture.

To this end, we consider two recurrent baselines. 
The first is an Elman-type recurrent neural network \cite{Elman1990}, denoted $\mathcal{N}_{\mathrm{RNN}}$, which was already used in \cite{Holthusen2025} for inelastic Constitutive Neural Networks. 
The second is a Liquid Neural Network, denoted $\mathcal{N}_{\mathrm{LiNN}}$. 
Both serve as references representing classical recurrent approaches without physics priors. 
Our proposed physics-embedded models are denoted $\mathcal{N}_{\mathrm{Consti}}^{\mathrm{LiNN}}$, where the inelastic variables are predicted by a LiNN, and $\mathcal{N}_{\mathrm{Consti}}^{\mathrm{Exp}}$, where the evolution is integrated explicitly via the exponential update rule \eqref{eq:explicit_evo}. 
It is important to note that $\mathcal{N}_{\mathrm{LiNN}}$, which directly predicts stresses, must not be confused with the auxiliary LiNN in $\mathcal{N}_{\mathrm{Consti}}^{\mathrm{LiNN}}$. 
Architectural details of the recurrent baselines are provided in Appendix~\ref{app:recurrent}. 
In all recurrent models, the number of hidden states propagated through time is set to six.

\begin{figure}
    \centering
    \includegraphics{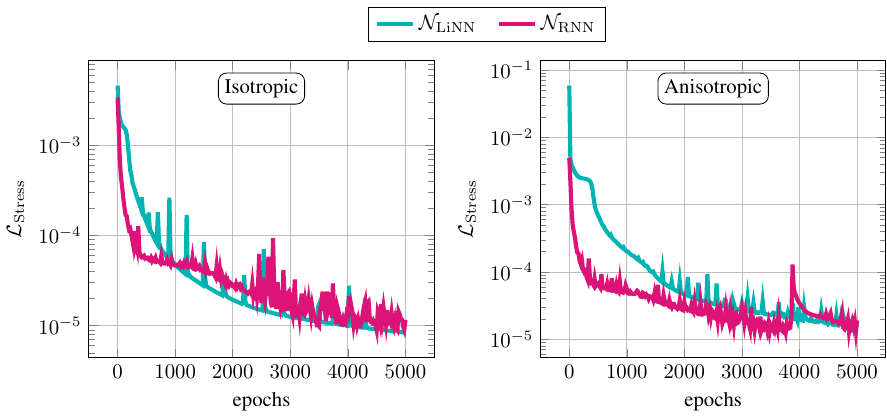}
    \caption{Training losses over the epochs for isotropic and anisotropic data sets for the plain recurrent neural networks.
    The total number of epochs is $5{,}000$.
    The losses correspond to Equation~\eqref{eq:loss_stress}, which here coincides with the total loss.
    Neither gradient clipping nor additional regularization is applied.}
\label{fig:loss_LiNN_RNN}
\end{figure}

For a fair comparison, the RNN and LiNN baselines are trained until the same loss value is reached after $5{,}000$ epochs under both isotropic and anisotropic conditions. 
The evolution of the training losses is presented in Figure~\ref{fig:loss_LiNN_RNN}. 
In each case, the stress discrepancy decreases to magnitudes on the order of $10^{-5}$, which is comparable to the performance of the physics-embedded networks shown in Figure~\ref{fig:PWTEH_losses}. 
As in the previous section, stresses are normalized during preprocessing to ensure balanced training.

\subsubsection{Isotropic response}
\label{sec:results_material_point_isotropic}
We begin with the isotropic case, i.e.\ $\psi=\psi_{\mathrm{iso}}$ and $\varphi=\varphi_{\mathrm{iso}}$ according to Equation~\eqref{eq:reference_model}. 
The analysis proceeds in two steps: first we examine training performance on an element (ID 111) contained in the dataset, and subsequently we test the predictive capability on an element (ID 53) not used during training.

\begin{figure}
    \centering
    \includegraphics{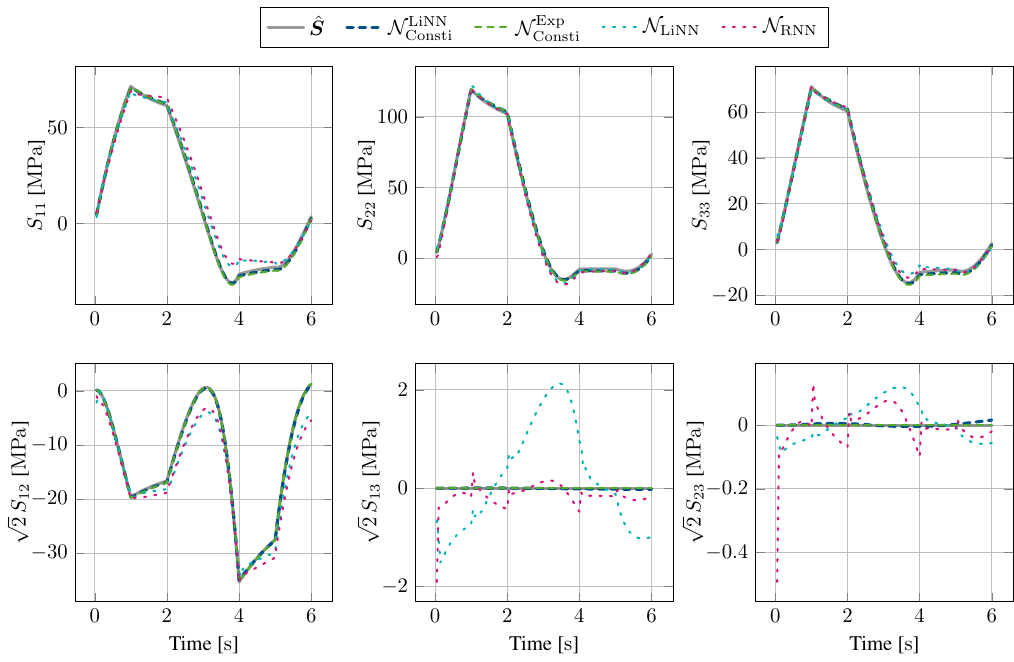}
    \caption{Training results for $\mathcal{N}_{\mathrm{Consti}}^{\mathrm{LiNN}}$, $\mathcal{N}_{\mathrm{Consti}}^{\mathrm{Exp}}$, $\mathcal{N}_{\mathrm{RNN}}$, and $\mathcal{N}_{\mathrm{LiNN}}$.
    The reference model is isotropic, i.e., $\psi=\psi_{\mathrm{iso}}$ and $\varphi=\varphi_{\mathrm{iso}}$ with material parameters given in Table~\ref{tab:material_parameters}.
    Element ID: 111. Curves start with the first loading step at $t=0.05$.
    Of all four networks, only the $\mathcal{N}_{\mathrm{Consti}}^{\mathrm{Exp}}$ network correctly captures the zero shear stress in the off-plane directions.}
    \label{fig:material_point_isotropic_train}
\end{figure}

\paragraph{Training.}
Figure~\ref{fig:material_point_isotropic_train} displays the stress components predicted for one representative training element. 
Both physics-embedded variants, $\mathcal{N}_{\mathrm{Consti}}^{\mathrm{LiNN}}$ and $\mathcal{N}_{\mathrm{Consti}}^{\mathrm{Exp}}$, reproduce the reference stresses with very high accuracy. 
The only noticeable deviation occurs in the out-of-plane shear component $S_{23}$, where $\mathcal{N}_{\mathrm{Consti}}^{\mathrm{LiNN}}$ produces an artificial stress increase. 
This artifact, however, disappears when the explicit exponential integrator is used. 
Apart from this effect, the results of both variants are almost indistinguishable, indicating that the auxiliary LiNN mainly stabilizes the update of inelastic variables while having only a marginal influence on the stress prediction itself. 
This observation is consistent with the findings of \cite{Rosenkranz2024}.

In contrast, the purely recurrent baselines show clear deficiencies. 
While $\mathcal{N}_{\mathrm{RNN}}$ and $\mathcal{N}_{\mathrm{LiNN}}$ capture the dominant normal stress components qualitatively, their predictions deviate significantly for shear stresses. 
Notably, oscillations occur in the in-plane shear and in the minor shear components, indicating a lack of robustness when stresses evolve along more complex trajectories. 
Although increasing the number of hidden states or employing more complex recurrent architectures may improve performance to some degree, the constitutive networks consistently achieve superior results while guaranteeing physically admissible stress states by construction.

\begin{figure}
    \centering
    \includegraphics{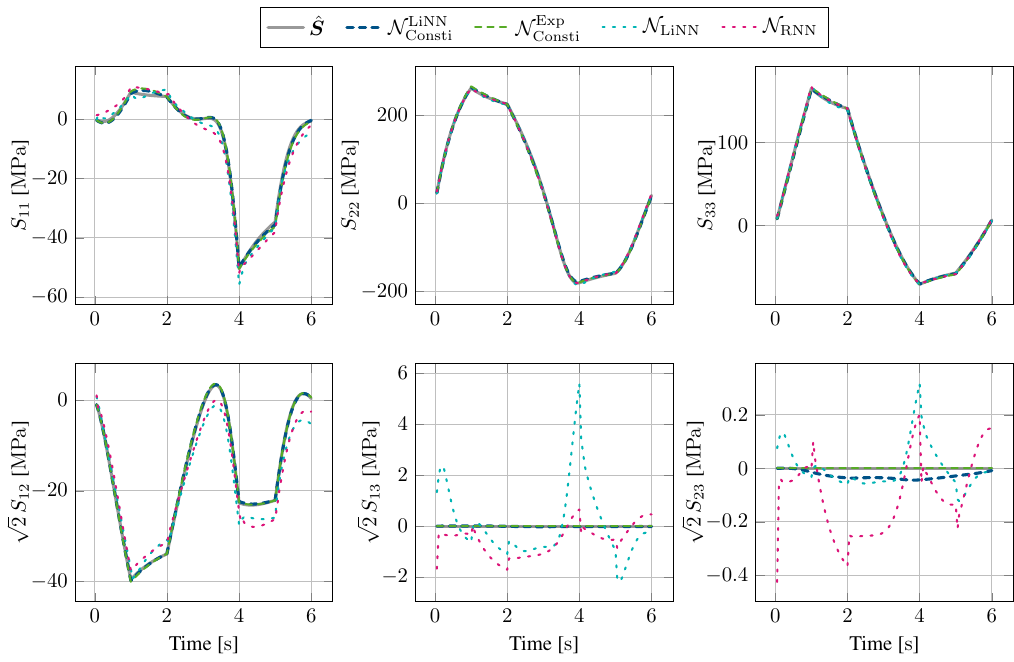}
    \caption{Testing results for $\mathcal{N}_{\mathrm{Consti}}^{\mathrm{LiNN}}$, $\mathcal{N}_{\mathrm{Consti}}^{\mathrm{Exp}}$, $\mathcal{N}_{\mathrm{RNN}}$, and $\mathcal{N}_{\mathrm{LiNN}}$.
    The reference model is isotropic, i.e., $\psi=\psi_{\mathrm{iso}}$ and $\varphi=\varphi_{\mathrm{iso}}$ with material parameters given in Table~\ref{tab:material_parameters}.
    Element ID: 53. Curves start with the first loading step at $t=0.05$.
    Of all four networks, only the $\mathcal{N}_{\mathrm{Consti}}^{\mathrm{Exp}}$ network correctly captures the zero shear stress in the off-plane directions.}
    \label{fig:material_point_isotropic_test}
\end{figure}

\paragraph{Testing.}
The generalization capability of the networks is evaluated on an unseen element not included in the training data (element ID 53). 
The results are presented in Figure~\ref{fig:material_point_isotropic_test}. 
Once again, the constitutive networks reproduce the reference stresses with high fidelity. 
Only small deviations appear, most notably in $S_{11}$, while the remaining stress components follow the reference solution almost perfectly. 
The artificial rise in the out-of-plane shear observed for $\mathcal{N}_{\mathrm{Consti}}^{\mathrm{LiNN}}$ during training is also present here, but is absent in $\mathcal{N}_{\mathrm{Consti}}^{\mathrm{Exp}}$. 
The recurrent baselines capture the overall stress evolution but continue to show oscillations in the shear components and less accurate predictions in the in-plane shear. 
Nevertheless, no severely nonphysical stress–strain responses, such as stress decreasing with increasing strain, were observed in these tests—failure modes that have been reported for plain RNNs in the literature \cite{Holthusen2025,Wang2023}. 
In this regard, an advantage of more advanced networks is clearly that embedding constitutive structure yields a much more reliable predictive model.

\subsubsection{Anisotropic response}
\label{sec:results_material_point_anisotropic}
We now turn to the anisotropic case, where the free energy and dual potential include anisotropic contributions, $\psi=\psi_{\mathrm{iso}}+\psi_{\mathrm{ani}}$ and $\varphi=\varphi_{\mathrm{iso}}+\varphi_{\mathrm{ani}}$; see Equation~\eqref{eq:reference_model}. 
As before, we analyze both training and testing performance using the same element IDs as in the isotropic study for direct comparability.

\begin{figure}
    \centering
    \includegraphics{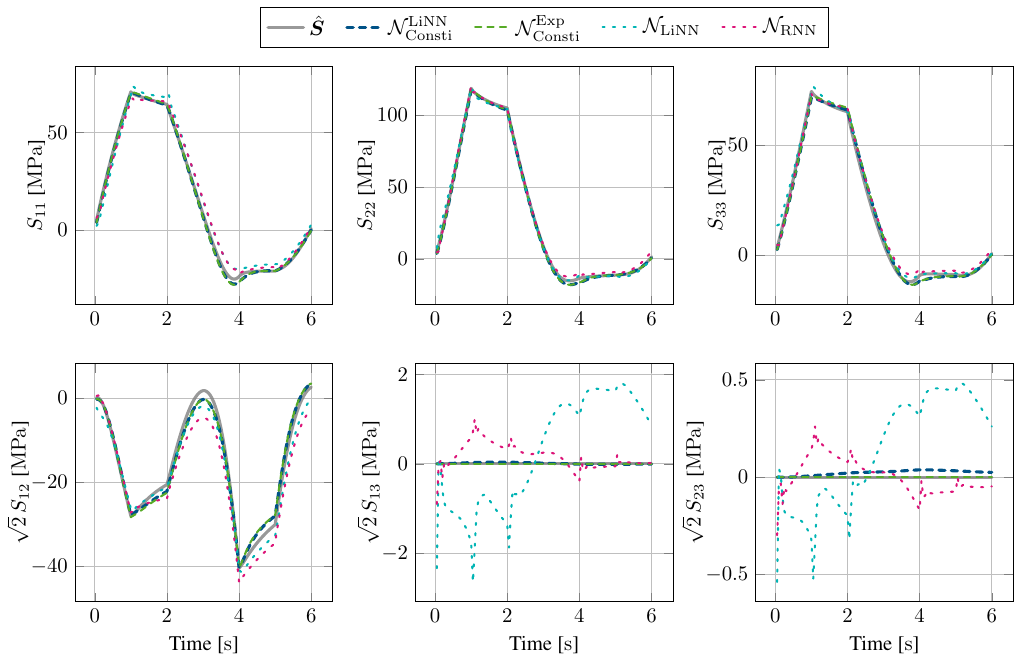}
    \caption{Training results for $\mathcal{N}_{\mathrm{Consti}}^{\mathrm{LiNN}}$, $\mathcal{N}_{\mathrm{Consti}}^{\mathrm{Exp}}$, $\mathcal{N}_{\mathrm{RNN}}$, and $\mathcal{N}_{\mathrm{LiNN}}$.
    The reference model is anisotropic, i.e., $\psi=\psi_{\mathrm{iso}}+\psi_{\mathrm{ani}}$ and $\varphi=\varphi_{\mathrm{iso}}+\varphi_{\mathrm{ani}}$ with parameters given in Table~\ref{tab:material_parameters}.
    Element ID: 111. Curves start with the first loading step at $t=0.05$.
    Of all four networks, only the $\mathcal{N}_{\mathrm{Consti}}^{\mathrm{Exp}}$ network correctly captures the zero shear stress in the off-plane directions.}
    \label{fig:material_point_anisotropic_train}
\end{figure}

\paragraph{Training.}
The training results are shown in Figure~\ref{fig:material_point_anisotropic_train}. 
Here, the constitutive networks again outperform the recurrent baselines across most stress components. 
However, discrepancies are visible in the in-plane shear $S_{12}$, which is not captured with the same accuracy as the normal stresses. 
This reduced accuracy can be attributed to the difference in how anisotropy is modeled: in the reference material, anisotropy is introduced via a metric tensor, whereas the neural networks rely on invariants in terms of a second-order structural tensor in their input representation. 
In addition, we recognize once more an artificial stress increase when employing $\mathcal{N}_{\mathrm{Consti}}^{\mathrm{LiNN}}$, which, however, vanishes with the explicit integration scheme. 
Despite this mismatch, the physics-embedded approaches retain stability and robustness, while the recurrent baselines suffer from pronounced oscillations, especially in the shear components.

\begin{figure}
    \centering
    \includegraphics{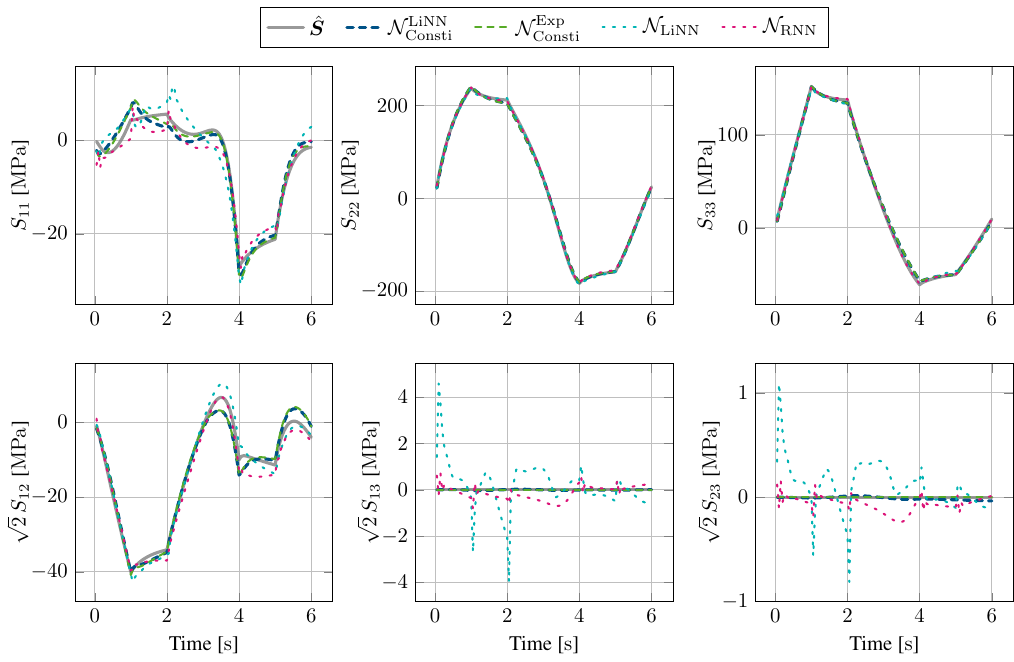}
    \caption{Testing results for $\mathcal{N}_{\mathrm{Consti}}^{\mathrm{LiNN}}$, $\mathcal{N}_{\mathrm{Consti}}^{\mathrm{Exp}}$, $\mathcal{N}_{\mathrm{RNN}}$, and $\mathcal{N}_{\mathrm{LiNN}}$.
    The reference model is anisotropic, i.e., $\psi=\psi_{\mathrm{iso}}+\psi_{\mathrm{ani}}$ and $\varphi=\varphi_{\mathrm{iso}}+\varphi_{\mathrm{ani}}$ with parameters given in Table~\ref{tab:material_parameters}.
    Element ID: 53. Curves start with the first loading step at $t=0.05$.
    Of all four networks, only the $\mathcal{N}_{\mathrm{Consti}}^{\mathrm{Exp}}$ network correctly captures the zero shear stress in the off-plane directions.}
    \label{fig:material_point_anisotropic_test}
\end{figure}

\paragraph{Testing.}
The testing results for element ID 53 are displayed in Figure~\ref{fig:material_point_anisotropic_test}. 
The dominant normal stresses, in particular $S_{22}$ and $S_{33}$, are predicted with high accuracy. 
For $S_{11}$, however, systematic deviations are observed across all models. 
The in-plane shear remains challenging, with none of the models reproducing it fully accurately. 
Yet, a decisive difference emerges: only the recurrent baselines exhibit oscillations in the remaining shear stresses, whereas the constitutive networks provide smooth and physically consistent responses. 
Interestingly, $\mathcal{N}_{\mathrm{Consti}}^{\mathrm{LiNN}}$ and $\mathcal{N}_{\mathrm{Consti}}^{\mathrm{Exp}}$ deliver almost identical predictions, underlining that the auxiliary LiNN has little influence on accuracy but plays an important role in stabilizing training. 
Indeed, training $\mathcal{N}_{\mathrm{Consti}}^{\mathrm{Exp}}$ alone frequently diverged to $\mathrm{NaN}$ losses, whereas the auxiliary LiNN enabled stable convergence.

\medskip
In summary, the material point study demonstrates that the proposed physics-embedded networks provide a significant improvement over purely recurrent baselines. 
They achieve superior accuracy in both isotropic and anisotropic settings, eliminate oscillations in the shear response, and ensure admissible stress states by construction. 
The auxiliary LiNN stabilizes training, particularly in the anisotropic finite-strain regime, but has only marginal influence on the final stress predictions. 
These findings establish a solid basis for assessing the networks in more complex structural applications, which will be the focus of the next section.

\subsection{Structural level study}
\label{sec:results_structural}
Having validated the proposed networks at the material point scale, we now turn to structural boundary value problems (BVPs). 
This step is essential, because structural simulations combine the responses of many integration points, involve complex stress redistributions, and require a consistent tangent operator to ensure convergence of the finite element method. 
Thus, structural studies provide a demanding test of whether the advantages observed at the local scale carry over to engineering-scale applications.

\begin{figure}
    \centering
    \includegraphics{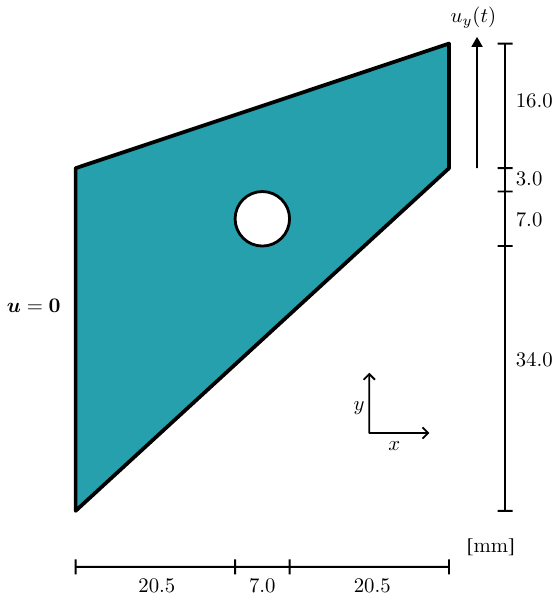}
    \caption{\textbf{Cook's membrane with circular hole.}
    Schematic illustration of the BVP used to evaluate predictive capability at the structural level.
    The left edge is fully clamped, while on the right edge a vertical displacement $u_y(t)$ is prescribed; other components are left free.
    A circular hole of diameter $7\,\si{\milli\metre}$ introduces stress concentrations.
    The plate has a thickness of $6\,\si{\milli\metre}$, discretized with three hexahedral elements across the thickness and 504 elements in-plane.}
    \label{fig:Cook}
\end{figure}

We select a variant of the classical Cook’s membrane test \cite{Cook1981}, here extended to a fully three-dimensional geometry with a central hole (Figure~\ref{fig:Cook}). 
The chosen configuration induces a bending-dominated stress state with strong stress redistributions around the hole, making it a challenging yet well-established benchmark for constitutive models. 
At the structural level, we focus on the network $\mathcal{N}_{\mathrm{Consti}}^{\mathrm{Exp}}$, where the inelastic evolution is integrated explicitly. 
This choice simplifies the computation of the consistent material tangent, which can be obtained efficiently via forward-mode automatic differentiation (\texttt{jax.jacfwd}), and avoids the need to invoke the implicit function theorem when solving for the inelastic variables implicitly.

\subsubsection{Isotropic response}
\label{sec:results_structural_isotropic}
\begin{figure}
    \centering
    \includegraphics{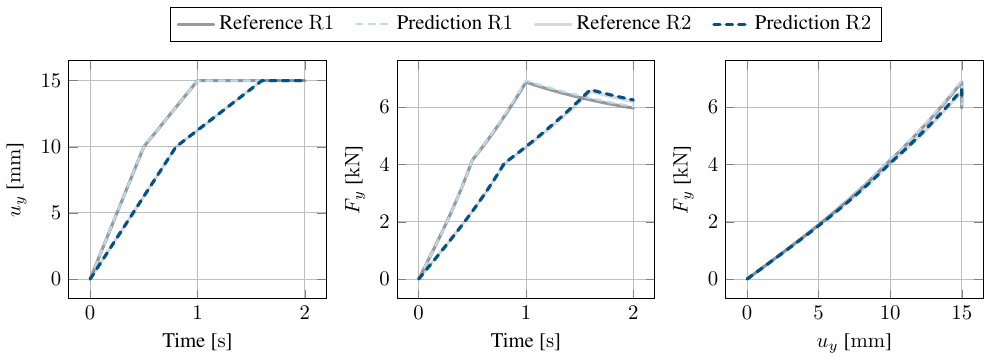}
    \caption{The loading program and resulting measured forces at the right surface (cf. Figure~\ref{fig:Cook}) for the isotropic material.
    Left: Prescribed displacement at the right surface in vertical direction at two different loading rates $\mathrm{R}1$ and $\mathrm{R}2$.
    Middle: Measured reaction force over time at both loading rates.
    Right: Force-displacement curves for both loading rates. Due to the inelastic effects, the stiffness between the two loading rates varies.}
    \label{fig:results_cook_isotropic}
\end{figure}

\paragraph{Boundary value problem.}
To investigate rate dependence and relaxation effects, we consider two linear loading phases of different magnitude, followed by a hold phase at constant displacement. 
For program $\mathrm{R}1$, the total loading time of $2\,\si{\second}$ is divided into 40 increments of $\Delta t = 0.05\,\si{\second}$, while for program $\mathrm{R}2$ we use 50 increments of $\Delta t = 0.04\,\si{\second}$. 
This setup allows us to test whether the network correctly reproduces both the immediate rate effects and the delayed relaxation behavior.
Figure~\ref{fig:results_cook_isotropic} show the two load programs applied.

\paragraph{Results.}
The global reaction forces measured at the right edge are plotted in Figure~\ref{fig:results_cook_isotropic}. 
The constitutive network matches the reference solution very closely. 
In particular, it reproduces the increased apparent stiffness under the faster loading rate $\mathrm{R}1$, as well as the gradual decrease in reaction force during the hold phase, which arises from inelastic relaxation. 
The corresponding force–displacement curves confirm this behavior: higher loading rates yield higher reaction forces for the same imposed displacement, and this trend is captured accurately by the network.

\begin{figure}
    \centering
    \includegraphics{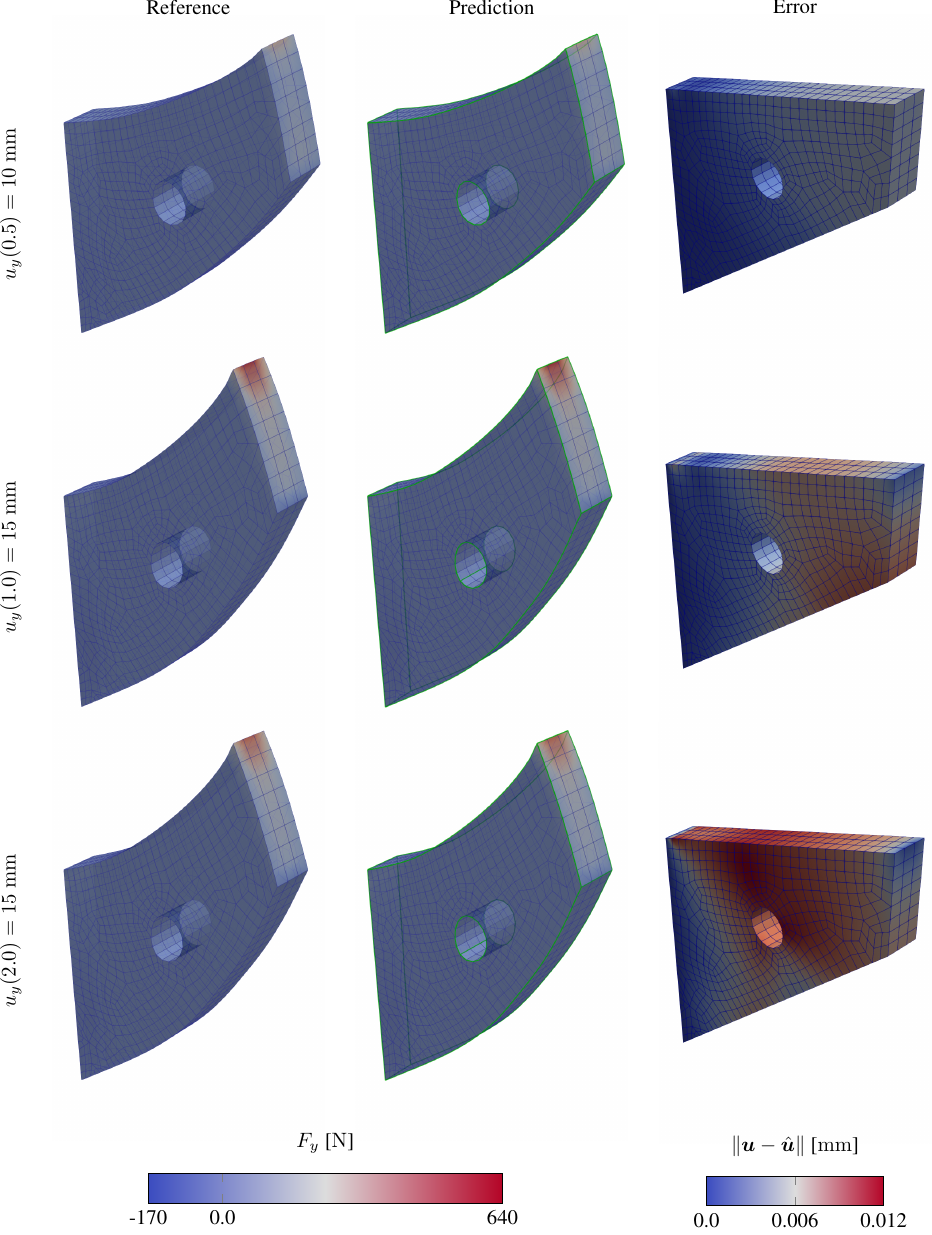}
    \caption{Contour plots of the Cook’s membrane test with isotropic material for the first loading rate.
    Rows represent selected load steps (cf. Figure~\ref{fig:results_cook_isotropic}). 
    Columns 1–2 show nodal force contours of the reference and neural network prediction, with green lines marking the reference deformation. 
    The last column displays the Euclidean norm of the displacement error $\hat{\bm{u}}-\bm{u}$, where $\hat{\bm{u}}$ corresponds to the reference and $\bm{u}$ to the prediction.}
    \label{fig:results_cook_isotropic_contour}
\end{figure}

A closer examination of the spatial response is given in Figure~\ref{fig:results_cook_isotropic_contour}, which presents contour plots of nodal forces and displacement errors for program $\mathrm{R}1$. 
The nodal force distributions predicted by the network are in good agreement with the reference solution, and the outer feature edges of the deformed geometry coincide almost identical. 
The Euclidean norm of the displacement error $\|\hat{\bm{u}}-\bm{u}\|$ remains below $0.012\,\si{\milli\metre}$ throughout the simulation, which is negligible relative to the prescribed displacement. 
The small residual errors on the right boundary arise from discrepancies in the unconstrained $x$- and $z$-directions. 
Overall, both global and local measures demonstrate that the discovered network provides an accurate and physically consistent prediction for isotropic materials.
These observations are consistently found for the loading rate $\mathrm{R}2$, which is not shown here for brevity.

\subsubsection{Anisotropic response}
\label{sec:results_structural_anisotropic}
\paragraph{Boundary value problem.}
We now repeat the Cook’s membrane study for an anisotropic material. 
To deliberately test generalization beyond the training conditions, the preferred direction is rotated to 
\[
\bm{n} = \left(\tfrac{1}{\sqrt{3}},\,\tfrac{1}{\sqrt{3}},\,\tfrac{1}{\sqrt{3}}\right)^T,
\]
thereby coupling all three spatial axes. 
The same loading programs $\mathrm{R}1$ and $\mathrm{R}2$ are applied, consisting of two linear loading phases followed by a hold phase.

\begin{figure}
    \centering
    \includegraphics{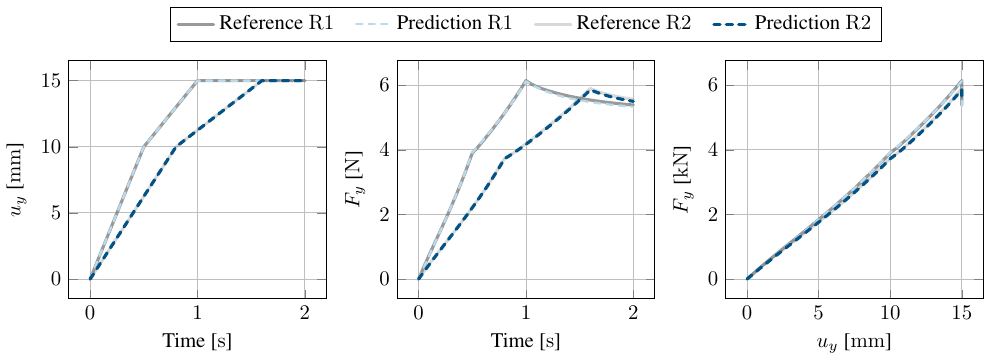}
    \caption{The loading program and resulting measured forces at the right surface (cf. Figure~\ref{fig:Cook}) for the anisotropic material.
    Left: Prescribed displacement at the right surface in vertical direction at two different loading rates $\mathrm{R}1$ and $\mathrm{R}2$.
    Middle: Measured reaction force over time at both loading rates.
    Right: Force-displacement curves for both loading rates. Due to the inelastic effects, the stiffness between the two loading rates varies.}
    \label{fig:results_cook_anisotropic}
\end{figure}

\paragraph{Results.}
The reaction forces are shown in Figure~\ref{fig:results_cook_anisotropic}. 
As in the isotropic case, higher loading rates lead to higher apparent stiffness and larger reaction forces, and relaxation is visible during the hold phase. 
However, compared to the isotropic material, both peak forces and stiffnesses are consistently lower. 
This reduction reflects the physical effect of misalignment: the preferred direction must reorient toward the loading axis before it can effectively resist deformation. 
Consequently, the specimen undergoes significant twisting and out-of-plane deflection, which is not present in the isotropic response.

\begin{figure}
    \centering
    \includegraphics{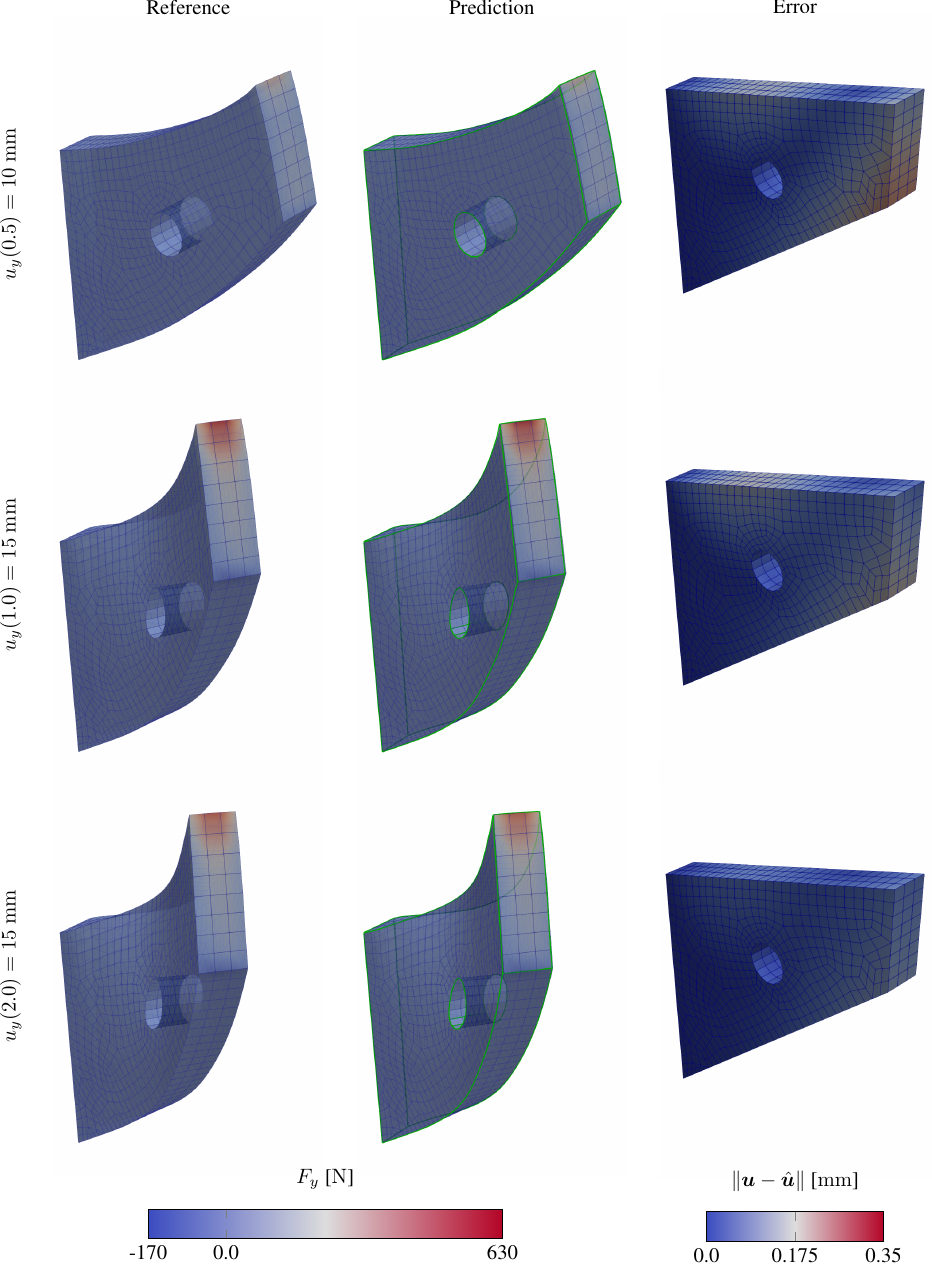}
    \caption{Contour plots of the Cook’s membrane test with anisotropic material for the first loading rate.
    The specimen twists around the $y$-axis due to anisotropy, in contrast to the isotropic case (Figure~\ref{fig:results_cook_isotropic_contour}). 
    Rows represent selected load steps (cf. Figure~\ref{fig:results_cook_anisotropic}). Columns 1–2 show nodal force contours of the reference and neural network prediction, with green lines marking the reference deformation. 
    The last column displays the Euclidean norm of the displacement error $\hat{\bm{u}}-\bm{u}$, where $\hat{\bm{u}}$ corresponds to the reference and $\bm{u}$ to the prediction.}
    \label{fig:results_cook_anisotropic_contour}
\end{figure}

These deformation patterns are clearly captured in the contour plots of Figure~\ref{fig:results_cook_anisotropic_contour}. 
The structure twists around the $y$-axis during the hold phase, and the network prediction follows the reference solution both qualitatively and quantitatively. 
Although the displacement errors are larger than in the isotropic case, reaching up to $0.35\,\si{\milli\metre}$ at intermediate times, the overall deformation mode is reproduced faithfully. 
Notably, the anisotropic case also exhibits stronger relaxation in the displacement field: even at constant load, the structure continues to twist as the preferred direction gradually align with the principal loading direction. 
This time-dependent reorientation is captured convincingly by the network.

\medskip
In summary, the structural study demonstrates that the proposed framework generalizes well from material point training to unseen structural BVPs. 
For isotropic materials, both global reaction forces and local deformation patterns are reproduced with very high accuracy. 
For anisotropic materials, even under rotated preferred directions and complex bending-dominated loading, the network captures the essential features of the response, including rate dependence, relaxation, and twisting deformation. 
These results underline the robustness of embedding constitutive structure into neural networks and confirm their potential for predictive structural simulations.
\FloatBarrier
\section{Discussion}
\label{sec:discussion}
We have presented a framework for designing a dual potential that governs the amount of dissipation while consistently embedding anisotropy into both the Helmholtz free energy and the dual potential itself. 
The central theoretical extensions of our approach are twofold: 
(i) we propose a dual potential that is not necessarily convex; yet satisfies the dissipation inequality a priori, and 
(ii) we define the integrity sets of the energy and the dual potential under consideration of different consistent mappings of the structural tensor to the intermediate configuration.
In this course, we discuss the entire set of invariants including those that combine elastic and inelastic tensors.

Beyond the theoretical formulation of anisotropic inelasticity at finite strains, we translated the constitutive relations into neural network architectures. 
Specifically, we employed a modified version of Input Convex Neural Networks (ICNNs) and introduced Input Monotonic Neural Networks (IMNNs) to extend the dual potential beyond convex formulations. 
For both the energy and the dual potential, we constructed suitable sets of invariants, which serve as physics-embedded inputs to the corresponding neural networks.

In line with \cite{Asad2023,Rosenkranz2024}, we eliminate the need for classical time-integration schemes in updating the inelastic variables by using neural network prediction.
To this end, we incorporated recurrent Liquid Neural Networks (LiNNs) into the overall architecture.
We designed a trial-step procedure as input to the LiNN and enforced by construction that the inelastic stretches remain positive definite. 

The framework was first evaluated at the material-point level, where we compared its performance and accuracy against recurrent neural architectures without physics priors. 
Subsequently, we validated the discovered models at the structural scale by predicting deformation and reaction forces for unseen boundary value problems. 
Across all investigated cases, the framework produced accurate predictions even outside the training regime and exhibited stable performance. 
Nevertheless, several aspects require further discussion and refinement. We address some of these aspects below.

\paragraph{Choice of invariants.}
We proposed two distinct invariant sets to represent anisotropic material behavior: one for the Helmholtz free energy (deformation invariants) and another for the dual potential (stress invariants). 
The choice of invariants, however, is not unique. 
Alternative formulations include Landau invariants \cite{Guan2026} or generalized invariants \cite{Anassi2024}, the latter being shown to capture material response with accuracy comparable to larger invariant sets while reducing redundancy. 
This property is particularly relevant in the neural network context \cite{Martonova2025}, as smaller invariant sets can reduce network complexity and mitigate uncertainty in weight optimization. 
In our case, the invariants for the energy were designed to be consistent with polyconvexity. 
Yet, recent work by \cite{Wiedemann2023} introduced a singular-value-based representation capable of modeling stress plateaus in purely elastic materials, which was already implemented within a neural architecture \cite{Geuken2025}. 
However, the integration of anisotropy into such approaches remains unresolved. 
While the literature on deformation invariants is extensive, significantly fewer contributions address the invariants underlying dual potentials. 
Most existing designs adhere to convexity in stress space, which is sufficient but not necessary. 
Thus, there is considerable room to explore more expressive invariant sets. 
In this respect, neural networks might offer a promising tool for discovering such invariants, similar to how pattern recognition in transformer architectures has been leveraged in mathematics \cite{Charton2024}.

\paragraph{Data availability.}
Training the proposed networks required synthetic data, as high-fidelity data sets for anisotropic inelasticity at finite strains remain scarce. 
While the generality of the framework is an advantage, it also necessitates larger training data sets. 
Increasing the number of invariants directly raises the information entropy required to identify the underlying material behavior in a (near-)unique manner. 
Whether it is feasible to generate such extensive data sets or whether adaptive multiscale data-generation schemes \cite{Prume2025} will be necessary remains an open question. 
Additionally, experimental data typically consist of force–displacement measurements rather than stress–stretch data, which we employed for training. 
In this context, data-driven identification methods \cite{Stainier2019}, which were used in the context of physics-augmented neural network training strategies \cite{Linden2025}, may provide an alternative route.
Furthermore, initialization is crucial for training neural networks, which becomes more challenging with greater non-linearity.
In this regard, the material fingerprinting method \cite{Flaschel2025} may help identify appropriate initial values and mitigate the need for pre-training.
However, their applicability to anisotropic inelasticity has not yet been clarified.

\paragraph{Inelastic phenomena.}
Our study was restricted to viscoelasticity at finite strains, primarily because it represents the simplest inelastic phenomenon without requiring additional constraints such as yield conditions or growth laws. 
Nevertheless, recent efforts have also focused on plasticity \cite{Boes2023,Flaschel2022,Jadoon2025,Eghtesand2024,Dettmer2024} and growth \cite{Holthusen2025growth}, although the rigorous incorporation of anisotropy in these works remains limited, and convexity-based potentials are typically employed. 
Our methodology is designed to extend seamlessly to other inelastic mechanisms governed by multiplicative decomposition. 
Notably, the proposed non-convex dual potential can be directly adapted to other types of inelasticity such as damage models, which are also formulated via potentials. 
The challenge in incorporating increasingly complex inelastic phenomena lies in ensuring their unique discovery, which directly depends on data availability and information entropy. 
An interesting research direction is to explore whether auxiliary neural networks for internal variables can stabilize training in such cases, particularly for plasticity.

\paragraph{Interpretability and regularization.}
In this work, we prioritized expressivity over interpretability. 
While our networks were designed to satisfy physical and constitutive constraints by construction, their interpretability is limited compared to, for instance, Constitutive Artificial Neural Networks \cite{Linka2023} and their inelastic extensions \cite{Holthusen2024}. 
We did not employ sparsity-promoting regularization such as Lasso, which could enhance interpretability, nor did we systematically investigate dropout or similar techniques that may prevent over-specialization of network pathways.
To this end, clustering methods similar to those in \cite{Flaschel2023a} might help increase the interpretability of dense networks by clustering different pathways into a common pathway without loss of accuracy.
A central question remains whether the analytical form of the entire network is essential, given that invariants already carry interpretable physical meaning as deformation or stress measures. 
We also did not examine whether the network architecture can reduce to a purely elastic case, which must be included as a limiting scenario. 
For isotropic inelasticity, this reduction was studied in \cite{Holthusen2025}, where Lasso regularization effectively discovered a vanishing dual potential. 
Interestingly, comparable behavior was observed without regularization, albeit at the expense of interpretability. 
Finally, emerging approaches such as neural architecture search \cite{Poyser2024} may play a crucial role in solid mechanics, both for determining whether inelasticity is relevant and for uncovering anisotropy when it is not explicitly known a priori.

\section{Conclusion}
\label{sec:conclusion}

We have proposed a methodology to consistently embed physics into neural networks for modeling anisotropic inelasticity at finite strains based on the design of a dual potential. 
The approach generalizes beyond convex formulations, ensures the dissipation inequality a priori, and integrates anisotropy consistently through invariant-based representations of the Helmholtz free energy and the dual potential. 
By embedding constitutive principles into adapted neural architectures, including Input Monotonic Neural Networks and Liquid Neural Networks, we achieved stable and accurate predictions both at the material-point and structural level, even outside the training regime.  

While the methodology offers a flexible foundation transferable to various inelastic phenomena, its success critically depends on the availability of informative data.
Future research may address the systematic interpretation of the discovered networks, different data-generation strategies, and extensions to more complex inelastic mechanisms such as plasticity. 
\FloatBarrier

\appendix

\section{Additions}
\label{app:A}
\subsection{Input Monotonic Neural Network}
\label{app:IMNN}
In the following, we discuss the monotonicity properties of the Input Monotonic Neural Network formulation.  
Let $\mathbf{s} \in \mathbb{R}^n$ denote the input parameters and $\mathbf{t} \in \mathbb{R}^m$ the network output.  
The mapping from one layer to the next is defined as
\[
\mathbf{p}_{l+1} \;=\; g_l\!\left( \mathbf{W}_l^p \mathbf{p}_l + \mathbf{W}_l^s \mathbf{s} + \mathbf{b}_l \right), 
\qquad l = 0,\dots,L-1,
\]
with $\mathbf{p}_0 = \mathbf{s}$. Here, $\mathbf{W}_l^p$ and $\mathbf{W}_l^s$ are weight matrices with non-negative entries, $\mathbf{b}_l$ are bias vectors, and the activation functions $g_l$ are chosen to be monotonically increasing functions (such as ReLU, softplus, or sigmoid).

This structure directly implies that the overall network mapping $\mathbf{s} \mapsto \mathbf{t}$ is monotonically non-decreasing with respect to each input component. The reasoning is straightforward:  
if one increases any component of the input vector $\mathbf{s}$, the affine terms $\mathbf{W}_l^p \mathbf{p}_l + \mathbf{W}_l^s \mathbf{s} + \mathbf{b}_l$ in each layer cannot decrease, because all entries of $\mathbf{W}_l^p$ and $\mathbf{W}_l^s$ are non-negative. The subsequent application of a monotonically increasing activation function preserves this order. Consequently, the intermediate states $\mathbf{p}_l$ remain ordered in the same way for all layers. The final output $\mathbf{t} = \mathbf{p}_L$ therefore inherits this property.

From a differential perspective, whenever the activation functions are differentiable, the partial derivatives of the outputs with respect to the inputs satisfy
\[
\frac{\partial t_i}{\partial s_j} \;\geq\; 0 \qquad \text{for all } i,j.
\]
This means that no output component decreases when any input parameter increases.  
In other words, the network as defined above provides a mapping that is monotone: the response $\mathbf{t}$ cannot decrease under an increase of any component of the parameter vector $\mathbf{s}$.

\subsection{Numerical details}
\label{app:numerics}
\paragraph{Computation of roots.}
In the computation of the square and cubic roots of the stress invariants, 
we are particularly concerned with their derivatives. 
However, these are not defined at zero. 
To circumvent this issue, following \cite{Holthusen2025}, 
we slightly modify the definitions as
\[
\sqrt{x} := \frac{x}{(x+\epsilon)^{1/2}}, 
\qquad 
\sqrt[3]{x} := \frac{x}{(|x|+\epsilon)^{2/3}}.
\]
Throughout this work, we set $\epsilon = 0.01$. 
With this choice, one obtains
\[
\left.\frac{\mathrm{d}}{\mathrm{d}x}\sqrt{x}\,\right|_{x=0} = 10.0,
\qquad 
\left.\frac{\mathrm{d}}{\mathrm{d}x}\sqrt[3]{x}\,\right|_{x=0} \approx 21.54435.
\]

\paragraph{Cholesky decomposition.}
Within the time discretization of the inelastic variable, we employ a recurrent Liquid Neural Network (LiNN) architecture to predict the inelastic stretches 
$\bm{U}_i \in \mathrm{Sym}_+^{3\times 3}$ at $t_{n+1}$. 
Thus, we must ensure that the hidden states $\mathbf{h}$ propagated by the LiNN are consistently mapped to a symmetric positive definite tensor. 
Since every $\bm{U}_i$ admits a Cholesky factorization,
\[
\bm{U}_i = \bm{L}\,\bm{L}^T,
\]
with $\bm{L}$ a real lower triangular matrix with strictly positive diagonal entries $L_{ii}>0$, we parameterize $\bm{L}$ through six independent states $\mathbf{h} = (h_1,\ldots,h_6)$
\[
\bm{L} = \begin{pmatrix}
    f(h_1) & 0 & 0 \\
    h_4 & f(h_2) & 0 \\
    h_6 & h_5 & f(h_3)
\end{pmatrix},
\]
where
\[
f(x) = \frac{1}{\ln(2)} \ln\!\big(1+\exp(\ln(2)\,x)\big)
\]
is a scaled softplus activation ensuring $f(x)>0$ for all $x$ and $f(0)=1$. 
With the initialization $\mathbf{h}=\mathbf{0}$, we obtain $\bm{L}=\bm{I}$ and hence $\bm{U}_i=\bm{I}$.

\subsection{Thermodynamic consistency}
\label{app:consistency}
\paragraph{Isotropic materials.}
We define the stress invariants
\[
S_1 := \operatorname{tr}\bm{\Sigma},\qquad
S_2' := \tfrac12\,\mathrm{dev}\bm{\Sigma}:\mathrm{dev}\bm{\Sigma},\qquad
S_3' := \tfrac13\,\operatorname{tr}\!\big((\mathrm{dev}\,\bm{\Sigma})^3\big),
\]
together with $S_2 := \sqrt{S_2'}$ and $S_3 := \sqrt[3]{S_3'}$.  
Set
\[
\mathcal{S}_\varphi(\bm{\Sigma}) := \big(S_1(\bm{\Sigma}),\,S_2(\bm{\Sigma}),\,S_3(\bm{\Sigma})\big),\qquad
\varphi(\bm{\Sigma}) := \omega(\mathcal{S}_\varphi(\bm{\Sigma})),
\]
where $\omega:\mathbb{R}^3\to\mathbb{R}$ is convex, non-negative, and satisfies $\omega(\mathbf{0})=0$.
The derivatives are
\[
\frac{\partial S_1}{\partial \bm{\Sigma}}=\bm{I},\qquad
\frac{\partial S_2}{\partial \bm{\Sigma}}=\frac{1}{2\sqrt{S_2'}}\,\mathrm{dev}\,\bm{\Sigma},\qquad
\frac{\partial S_3}{\partial \bm{\Sigma}}=\frac{1}{3 S_3'^{2/3}}\,\mathrm{dev}\!\big((\mathrm{dev}\,\bm{\Sigma})^2\big).
\]
Since $S_1,S_2,S_3$ are homogeneous of degree $1$, it follows that
\[
\frac{\partial S_k}{\partial \bm{\Sigma}}:\bm{\Sigma} = S_k \quad (k=1,2,3).
\]
By the chain rule,
\[
\partial_{\bm{\Sigma}}\varphi
= \sum_{k}\frac{\partial \omega}{\partial S_k}\,\frac{\partial S_k}{\partial \bm{\Sigma}},
\quad\Rightarrow\quad
\bm{\Sigma}:\partial_{\bm{\Sigma}}\varphi
= \big\langle \mathcal{S}_\varphi,\,\partial_{\mathcal{S}_\varphi}\omega \big\rangle.
\]
With $\bm{D}_i \in \partial_{\bm{\Sigma}}\varphi$, the reduced dissipation inequality reads
\[ 
\bm{\Sigma}:\bm{D}_i = \big\langle \mathcal{S}_\varphi,\,\partial_{\mathcal{S}_\varphi}\omega \big\rangle \;\ge\; \omega(\mathcal{S}_\varphi)-\omega(\mathbf{0}) = \omega(\mathcal{S}_\varphi) \;\ge\; 0, 
\]
ensuring non-negative dissipation.
For a more detailed derivation, see e.g. \cite{Holthusen2025}.
If $\varphi=(\zeta\circ\omega)(\mathcal{S}_\varphi)$ with $\zeta:\mathbb{R}\to\mathbb{R}$ monotone increasing, then
\[
\bm{\Sigma}:\partial_{\bm{\Sigma}}\varphi
= \frac{\partial\zeta}{\partial\omega}\,\big\langle \mathcal{S}_\varphi,\,\partial_{\mathcal{S}_\varphi}\omega \big\rangle \;\ge 0,
\]
so thermodynamic consistency is still preserved.

\paragraph{Anisotropic materials.}
The push-forward $\tilde{\bm{M}}=\bm{F}_i \star \bm{M}$ represents either a covariant or contravariant mapping of the structural vector; see Equation~\eqref{eq:structural_vec_push}.
Following \cite{Zheng1994}, we may form the irreducible basis in terms of the mixed invariants of the stress and the structural vectors
\[
\bm{\Sigma} : \tilde{\bm{M}}_j = \mathrm{dev}\bm{\Sigma} : \tilde{\bm{M}}_j + \frac{S_1}{3},\quad
\bm{\Sigma}^2 : \tilde{\bm{M}}_j = (\mathrm{dev}\bm{\Sigma})^2 : \tilde{\bm{M}}_j + 2\frac{S_1}{3}\,\mathrm{dev}\bm{\Sigma} : \tilde{\bm{M}}_j + \frac{S_1^2}{9},
\]
\[
\tilde{\bm{n}}_1 \cdot \bm{\Sigma}\,\tilde{\bm{n}}_2 = \bm{\Sigma} : \mathrm{sym}(\tilde{\bm{n}}_1 \otimes \tilde{\bm{n}}_2),\quad
\tilde{\bm{n}}_1 \cdot \bm{\Sigma}^2\,\tilde{\bm{n}}_2 = \bm{\Sigma}^2 : \mathrm{sym}(\tilde{\bm{n}}_1 \otimes \tilde{\bm{n}}_2),
\]
where $\tilde{\bm{M}}_j :=\tilde{\bm{n}}_j\otimes\tilde{\bm{n}}_j$ with $k=1,2$.
We introduce
\[
\mathcal{A}_\varphi(\bm{\Sigma},\tilde{\bm{n}}_1,\tilde{\bm{n}}_2) :=
(A_{1j},A_{2j},A_{3},A_{4})_{j=1,2},
\]
with
\[
A_{1j} := \mathrm{dev}\bm{\Sigma} : \tilde{\bm{M}}_j,\quad
A_{2j} := \sqrt{\tfrac12 (\mathrm{dev}\bm{\Sigma})^2 : \tilde{\bm{M}}_j},\quad
A_3 := \bm{\Sigma} : \mathrm{sym}(\tilde{\bm{n}}_1 \otimes \tilde{\bm{n}}_2),\quad
A_4 := \sqrt{\tfrac12\,\bm{\Sigma}^2 : \mathrm{sym}(\tilde{\bm{n}}_1 \otimes \tilde{\bm{n}}_2)}.
\]
Each of these invariants is homogeneous of degree $1$.  
We set
\[
\varphi(\bm{\Sigma},\tilde{\bm{n}}_1,\tilde{\bm{n}}_2) := (\zeta \circ \omega)(\mathcal{T}_\varphi(\bm{\Sigma},\tilde{\bm{n}}_1,\tilde{\bm{n}}_2)), \quad \mathcal{T}_\varphi(\bm{\Sigma},\tilde{\bm{n}}_1,\tilde{\bm{n}}_2) := (\mathcal{S}_\varphi(\bm{\Sigma}),\mathcal{A}_\varphi(\bm{\Sigma},\tilde{\bm{n}}_1,\tilde{\bm{n}}_2)),
\]
with $\omega:\mathbb{R}^9\to\mathbb{R}$ convex, non-negative, and $\omega(\mathbf{0})=0$.  
The chain rule gives
\[
\bm{\Sigma}:\partial_{\bm{\Sigma}}\varphi
= \frac{\partial\zeta}{\partial\omega}\,\big\langle \mathcal{T}_\varphi,\,\partial\omega_{\mathcal{T}_\varphi} \big\rangle \;\ge 0,
\]
ensuring thermodynamic consistency as in the isotropic case.

\subsection{Helmholtz free energy}
\label{app:Helmholtz}
\paragraph{Mixed-variant mappings.}
In Equation~\eqref{eq:invariants_map_psi}, for $k=r=0$ and $s=t=1$,
\[
\tilde{\bm{M}}_1\,\tilde{\bm{M}}_2
= \frac{1}{\gamma}\,\bm F_i\,\bm M\,\bm F_i^{-1},
\]
i.e., proportional to the mixed-variant map $\bm{F}_i\star\bm{M}=\bm F_i\bm M\bm F_i^{-1}$, where
\[
\gamma := (\bm C_i:\bm M)\,(\bm C_i^{-1}:\bm M)
      = (\bm n\!\cdot\!\bm C_i\bm n)\,(\bm n\!\cdot\!\bm C_i^{-1}\bm n) \;\ge\; 1
\]
by the Cauchy--Schwarz inequality, with equality if $\bm n$ is an eigenvector of $\bm C_i$.
If one wishes to mimic $s=t=-1$ by definition (not inversion), set
\[
\Big(\frac{\bm{C}_i\bm{M}}{\bm{C}_i : \bm{M}}\Big)^{-1} := (\bm{C}_i : \bm{M})\,\bm{M}\bm{C}_i^{-1}, 
\qquad
\Big(\frac{\bm{M}\bm{C}_i^{-1}}{\bm{C}_i^{-1} : \bm{M}}\Big)^{-1} := (\bm{C}_i^{-1} : \bm{M})\,\bm{C}_i\bm{M},
\]
which yields, for $k=r=0$ and $s=t=-1$,
\[
\tilde{\bm{M}}_1^{-1}\,\tilde{\bm{M}}_2^{-1}
= \gamma\,\bm F_i^{-T}\,\bm M\,\bm F_i^{T},
\]
i.e., proportional to the alternative mixed-variant map.

\paragraph{Polyconvexity.}
To construct a polyconvex Helmholtz free energy, one must employ invariants that are convex in 
$(\bm F, \mathrm{cof}\bm F, J)$. 
As an example, consider
\[
f(\bm F,\bm G) := \mathrm{tr}(\bm F^T \bm F \,\bm G).
\]
Its second Gâteaux derivative in direction $\bm H\in\mathbb{M}^{3\times 3}$ is
\[
\mathfrak{D}^2_{\bm F}[f(\bm F,\bm G)](\bm H,\bm H)
= \frac{\mathrm{d}^2}{\mathrm{d}\epsilon^2}\,\mathrm{tr}\,\big((\bm F+\epsilon\bm H)^T(\bm F+\epsilon\bm H)\,\bm G\big)\Big|_{\epsilon=0}
= 2\,\bm G : \bm H^T\bm H .
\]
Thus, $f$ is convex in $\bm F$ whenever $\bm G\succeq 0$ (positive semidefinite) \cite{Schroeder2008}. 
Analogously, replacing $\bm F$ by $\mathrm{cof}\,\bm F$ shows the same condition for convexity in the cofactor.

\paragraph{Polyconvexity of mixed-variant mappings.}
Let $\bm{F}_i \star \bm{M}$ denote a mixed-variant mapping consistent with the principle of indifference to the choice of intermediate configuration, e.g., $\bm{F}_i \star \bm{M} = \bm{F}_i\bm{M}\bm{F}_i^{-1}$, $\bm{F}_i \star \bm{M} = \bm{F}_i^{-T}\bm{M}\bm{F}_i^{T}$ or any analogous form.
Then the function $f(\bm{F},\bm{F}_i \star \bm{M})=\mathrm{tr}(\bm{F}^T\bm{F}\,\bm{F}_i \star \bm{M}) = \mathrm{tr}(\bm{F}^T\bm{F}\,\mathrm{sym}(\bm{F}_i \star \bm{M}))$ cannot be guaranteed to be polyconvex \textit{a priori} as $\mathrm{sym}(\bm{F}_i \star \bm{M})$ is not guaranteed to be positive (semi-)definite.

\paragraph{Outer tensor product of second order tensors.}
Let $\#$ denote the outer tensor product of two second-order tensors,
\[
(\bm{a} \otimes \bm{b}) \# (\bm{c} \otimes \bm{d})
:= (\bm{a} \times \bm{c}) \otimes (\bm{b} \times \bm{d}),
\]
where $\times$ denotes the vector cross product.
Let $\bm{B} \in \mathrm{Sym}_+^{3\times 3}$ and $\bm{G} = \bm{g} \otimes \bm{g}$. 
Then their outer product can be written as
\[
\bm{B} \# \bm{G}
= \sum_{i=1}^3 B_i \, (\bm{p}_i \otimes \bm{p}_i) \# \bm{G}
= \sum_{i=1}^3 B_i \, (\bm{p}_i \times \bm{g}) \otimes (\bm{p}_i \times \bm{g}),
\]
which is positive semi-definite, with $B_i$ and $\bm{p}_i$ denoting the eigenvalues and eigenvectors of $\bm{B}$, respectively.
More generally, the outer tensor product of any two tensors $\bm{A}, \bm{B} \in \mathbb{M}^{3\times 3}$ is given by
\[
\bm{A} \# \bm{B} 
= \big(\operatorname{tr}\bm{A}\,\operatorname{tr}\bm{B} - \operatorname{tr}(\bm{A}\bm{B})\big)\,\bm{I} 
+ \bm{B}^T\bm{A}^T + \bm{A}^T\bm{B}^T 
- (\operatorname{tr}\bm{A})\,\bm{B}^T - (\operatorname{tr}\bm{B})\,\bm{A}^T,
\]
which yields a symmetric tensor whenever both arguments are symmetric.
In addition, for any $\bm{P}\in \mathrm{GL}(3,\mathbb{R})$ one obtains the covariance property
\[
(\bm{P}\bm{A}\bm{P}^{-1}) \# (\bm{P}\bm{B}\bm{P}^{-1})
= \bm{P}^{-T} (\bm{A} \# \bm{B}) \bm{P}^T.
\]
In view of these properties and of the Helmholtz free energy in Section~\ref{sec:Helmholtz}, 
the invariant
\[
\operatorname{tr}\!\left( \bm{C}_e \,(\bm{B}_i \# \tilde{\bm{M}}) \right)
= \operatorname{tr}\!\left( \bm{C}\,\bm{F}_i^{-1}(\bm{B}_i \# \tilde{\bm{M}})\bm{F}_i^{-T} \right)
\]
is polyconvex, since $\bm{F}_i^{-1}(\bm{B}_i \# \tilde{\bm{M}})\bm{F}_i^{-T}$ is positive semi-definite.
Finally, using Equation~\eqref{eq:invariants_map_psi}, we obtain
\[
\operatorname{tr}\!\left( \bm{C}\,\bm{F}_i^{-1}(\bm{B}_i \# \tilde{\bm{M}})\bm{F}_i^{-T} \right)
= \operatorname{tr}\!\left( \bm{C}_i^{-1}\bm{C}\,(\bm{C}_i \# \bm{C}_i\bm{M}) \right)
= \operatorname{tr}\!\left( \bm{C}\bm{C}_i^{-1} \,(\bm{C}_i \# \bm{M}\bm{C}_i) \right).
\]
\subsection{Reference material model}
\label{app:reference_model}
As reference material model, we adopt a rheological formulation with three elements. 
The model consists of an equilibrium branch, represented by a spring element, arranged in parallel to a non-equilibrium branch, consisting of a spring–dashpot series connection. 
Accordingly, the Helmholtz free energy $\psi$ and the dual potential $\varphi$ are additively decomposed into isotropic and anisotropic contributions:
\[
    \psi = 
    \underbrace{
        \psi_{\mathrm{iso}}^{\mathrm{eq}}(\bm{C};a^{\mathrm{eq}},b^{\mathrm{eq}},c^{\mathrm{eq}})
        + \psi_{\mathrm{iso}}^{\mathrm{neq}}(\bm{C}_e;a^{\mathrm{neq}},b^{\mathrm{neq}},c^{\mathrm{neq}})
    }_{=:\psi_{\mathrm{iso}}}
    +
    \underbrace{
        \psi_{\mathrm{ani}}^{\mathrm{eq}}(\bm{C},\bm{G};\alpha^{\mathrm{eq}},\eta^{\mathrm{eq}})
        + \psi_{\mathrm{ani}}^{\mathrm{neq}}(\bm{C}_e,\tilde{\bm{G}};\alpha^{\mathrm{neq}},\eta^{\mathrm{neq}})
    }_{=:\psi_{\mathrm{ani}}},
\]
\[
    \varphi = \varphi_{\mathrm{iso}}(\bm{\Sigma};a^{\mathrm{neq}},b^{\mathrm{neq}},c^{\mathrm{neq}}, \tau) 
    + \varphi_{\mathrm{ani}}(\bm{\Sigma},\tilde{\bm{M}}_3;\eta^{\mathrm{neq}}, \tau).
\]
Here, $a$, $b$, $c$, $\alpha$, and $\eta$ denote material parameters, while 
$\bm{G}$ represents a metric tensor that accounts for the preferred direction 
\cite{Schroeder2008}, and must not be confused with a second-order structural tensor. 
Assuming the preferred direction aligns with $\bm{e}_1$, the reference metric reads
\[
    \bm{G}_0 = 
    \begin{pmatrix}
        \beta^2 & 0 & 0 \\
        0 & \tfrac{1}{\beta} & 0 \\
        0 & 0 & \tfrac{1}{\beta}
    \end{pmatrix},
\]
where $\beta$ is an anisotropy parameter.  
To obtain $\bm{G}$ aligned with an arbitrary preferred direction $\bm{n} \in \mathbb{S}^2$, 
we employ Rodrigues’ rotation formula. A rotation matrix $\bm{R} \in \mathrm{SO}(3)$ is constructed such that $\bm{e}_1$ is mapped onto $\bm{n}$, with axis and angle defined by
\[
\bm{k} = \bm{e}_1 \times \bm{n}, \qquad \hat{\bm{k}} = \frac{\bm{k}}{\|\bm{k}\|}, 
\qquad \theta = \arccos \!\left( \bm{e}_1 \cdot \bm{n} \right).
\]
The corresponding rotation matrix is
\[
\bm{R} = \bm{I} + \sin\theta \,[\hat{\bm{k}}]_\times 
    + (1 - \cos\theta)\,[\hat{\bm{k}}]_\times^2,
\]
with $[\hat{\bm{k}}]_\times$ denoting the skew-symmetric cross-product operator. 
The metric tensor is then given by
\[
    \bm{G} = \bm{R}\,\bm{G}_0\,\bm{R}^T.
\]

For the push-forward to the intermediate configuration, we adopt
\[
    \tilde{\bm{G}} = \bm{F}_i \,\bm{G}\,\bm{F}_i^T,
    \qquad
    \tilde{\bm{M}}_3 = \bm{F}_i\,(\bm{n}\otimes\bm{n})\,\bm{F}_i^{-1}.
\]

For the isotropic Helmholtz free energy, we employ the form proposed in \cite{Schroeder2008}:
\[
    \psi_{\mathrm{iso}}^{\mathrm{eq/neq}}(\bm{A}) 
    = a\,\mathrm{tr}\bm{A} + b\,\mathrm{tr}\,\mathrm{cof}\bm{A} + c\,\det\bm{A} 
    - \tfrac{d_1}{2}\,\ln\det\bm{A},
    \qquad d_1 = 2a + 4b + 2c.
\]
Further, for the anisotropic Helmholtz free energy, we adopt
\[
    \psi_{\mathrm{ani}}^{\mathrm{eq/neq}}(\bm{A},\bm{B}) 
    = \frac{\eta}{\alpha\,(\mathrm{tr}\bm{B})^\alpha}
      \left( (\mathrm{tr}\bm{A}\bm{B})^\alpha 
           + (\mathrm{tr}\,\mathrm{cof}\bm{A}\,\bm{B})^\alpha \right)
      - \tfrac{d_2}{2}\,\ln\det\bm{A},
    \qquad d_2 = 2\eta.
\]

In analogy to \cite{Reese1998}, the isotropic part of the dual potential is chosen as
\[
    \varphi_{\mathrm{iso}}(\bm{\Sigma}) 
    = \frac{1}{\tau}\left(\frac{1}{18}\,\frac{(\mathrm{tr}\bm{\Sigma})^2}{c} 
    + \frac{2}{a+b}\,\left(\frac{1}{2}\,\mathrm{tr}\,\mathrm{dev}\bm{\Sigma}\,\mathrm{dev}\bm{\Sigma}\right)\right),
\]
while the anisotropic contribution is given by
\[
    \varphi_{\mathrm{ani}}(\bm{\Sigma},\tilde{\bm{M}}_3) 
    = \frac{1}{\tau}\left(\frac{1}{2\eta}\,\left( (\mathrm{tr}\,\bm{\Sigma}\tilde{\bm{M}}_3)^2 
    + \mathrm{tr}\,(\bm{\Sigma}^2\tilde{\bm{M}}_3) \right)\right).
\]
\subsection{Plain recurrent neural networks}
\label{app:recurrent}

\paragraph{Recurrent Neural Network (RNN).}
The recurrent neural network employed in this work is formulated in discrete time with input $\mathbf{x}_t \in \mathbb{R}^{n_x}$, hidden state $\mathbf{h}_t \in \mathbb{R}^{n_h}$, and output $\mathbf{y}_t \in \mathbb{R}^{n_y}$.
At each time step $t$, the hidden state is updated by concatenating the input and the previous hidden state,
\[
    \mathbf{z}_t^{(0)} = 
    \begin{bmatrix}
        \mathbf{x}_t \\
        \mathbf{h}_{t-1}
    \end{bmatrix} \in \mathbb{R}^{n_x+n_h}.
\]
This vector is processed by a sequence of dense layers with nonlinear activations,
\[
    \mathbf{z}_t^{(\ell)} = \phi_\ell \!\left( \mathbf{W}_\ell \mathbf{z}_t^{(\ell-1)} + \mathbf{b}_\ell \right), 
    \quad \ell = 1,\dots,L,
\]
with $\mathbf{W}_\ell \in \mathbb{R}^{n_\ell \times n_{\ell-1}}$, $\mathbf{b}_\ell \in \mathbb{R}^{n_\ell}$, and nonlinear activation $\phi_\ell(\cdot)$.
The final transformation maps onto the hidden state space of size $n_h$,
\[
    \mathbf{h}_t = \mathbf{W}_h \mathbf{z}_t^{(L)} + \mathbf{b}_h, 
    \quad \mathbf{W}_h \in \mathbb{R}^{n_h \times n_L}.
\]
The output is computed as a linear mapping of the hidden state,
\[
    \mathbf{y}_t = \mathbf{W}_{hy}\mathbf{h}_t + \mathbf{b}_y, 
    \quad \mathbf{W}_{hy} \in \mathbb{R}^{n_y \times n_h}.
\]
This flexible cell structure allows one to define arbitrary numbers of hidden layers and activation functions inside the recurrent update.

In compact form, the chosen RNN topology reads
\[
\mathcal{N}_{\mathrm{RNN}}:\mathbb{R}^{n_x+n_h}\rightarrow\mathbb{R}^{n_h},\quad
(\mathbf{x}_t,\mathbf{h}_{t-1})\
\xrightarrow{\mathrm{GELU}} 64
\xrightarrow{\mathrm{GELU}} 64
\xrightarrow{\mathrm{GELU}} 64
\xrightarrow{\mathrm{GELU}} 64
\xrightarrow{\mathrm{GELU}} 64
\xrightarrow{\mathrm{Linear}} \mathbf{h}_t,
\]
with $n_x=7$ (strain-like features $\bm{C}$ and time increment $\Delta t$) and $n_h=6$.
The output mapping is
\[
\mathbf{y}_t = \mathbf{W}_{hy}\mathbf{h}_t + \mathbf{b}_y \in \mathbb{R}^{6}.
\]

\paragraph{Liquid Neural Network (LiNN).}
The Liquid Neural Network builds upon the same input–hidden–output structure with 
$\mathbf{x}_t \in \mathbb{R}^{n_x}$, $\mathbf{h}_t \in \mathbb{R}^{n_h}$, and $\mathbf{y}_t \in \mathbb{R}^{n_y}$.
The key difference lies in the hidden state update, which introduces adaptive dynamics through two subnetworks:
a \emph{source network} $\mathbf{f}(\mathbf{x}_t, \mathbf{h}_{t-1}) \in \mathbb{R}^{n_h}$ and a \emph{modulation network} $\bm{\alpha}(\mathbf{x}_t, \mathbf{h}_{t-1}) \in \mathbb{R}^{n_h}$.
Both subnetworks take the concatenated input–state vector
\[
    \mathbf{z}_t^{(0)} = 
    \begin{bmatrix}
        \mathbf{x}_t \\
        \mathbf{h}_{t-1}
    \end{bmatrix}
    \in \mathbb{R}^{n_x+n_h},
\]
as input and are constructed entirely from \emph{centered layers}, i.e.
\[
    \mathrm{CL}(z) = \phi(\mathbf{W}z+\mathbf{b}) - \phi(\mathbf{b}),
\]
which stabilize training by removing trivial bias-induced offsets.
The hidden state update follows an explicit Euler integration scheme,
\[
    \mathbf{h}_t = (1-\bm{\alpha})\odot \mathbf{h}_{t-1} + \Delta t\,\mathbf{f},
\]
where $\odot$ denotes the Hadamard product and $\Delta t$ is a prescribed time step.
The output is obtained by
\[
\mathbf{y}_t = \mathbf{W}_{hy}\mathbf{h}_t + \mathbf{b}_y \in \mathbb{R}^{6}.
\]

In compact form, the chosen LiNN topologies are
\[
\mathcal{N}_f:\mathbb{R}^{n_x+n_h}\rightarrow\mathbb{R}^{n_h},\quad
(\mathbf{x}_t,\mathbf{h}_{t-1})\
\xrightarrow{\mathrm{GELU}} 64
\xrightarrow{\mathrm{GELU}} 64
\xrightarrow{\mathrm{GELU}} 64
\xrightarrow{\mathrm{GELU}} 64
\xrightarrow{\mathrm{Linear}} \mathbf{f},
\]
\[
\mathcal{N}_\alpha:\mathbb{R}^{n_x+n_h}\rightarrow\mathbb{R}^{n_h},\quad
(\mathbf{x}_t,\mathbf{h}_{t-1})\
\xrightarrow{\mathrm{GELU}} 64
\xrightarrow{\mathrm{GELU}} 64
\xrightarrow{\mathrm{GELU}} 64
\xrightarrow{\mathrm{GELU}} 64
\xrightarrow{\mathrm{ReLU}} \bm{\alpha}.
\]
Here, $n_x=7$ (strain-like features $\bm{C}$ and $\Delta t$) and $n_h=6$.
Although $\Delta t$ already appears in the update scheme, we include it also as network input for consistency with the RNN case.
\section{Declarations}
%
\subsection{Acknowledgements}
First of all, the authors would like to thank Ahmad Awad for his help with implementing the finite element framework in \texttt{JAX}.
Hagen Holthusen gratefully acknowledge financial support of the projects 495926269 and 417002380 by the Deutsche Forschungsgemeinschaft.
Further, this work was supported by the NSF CMMI Award 2320933 Automated Model Discovery for Soft Matter and by the ERC Advanced Grant 101141626 DISCOVER to Ellen Kuhl.
%
%
\subsection{Conflict of interest}
The authors of this work certify that they have no affiliations with or involvement in any organization or entity with any financial interest (such as honoraria; participation in speakers’ bureaus; membership, employment, consultancies, stock ownership, or other equity interest; and expert testimony or patent-licensing arrangements), or non-financial interest (such as personal or professional relationships, affiliations, knowledge or beliefs) in the subject matter or materials discussed in this manuscript.
%
\subsection{Availability of data, source code and material}
Our data, source code and examples of the \texttt{JAX}/\texttt{Flax} implementations of the neural networks and the finite element framework are accessible to the public at \url{https://doi.org/10.5281/zenodo.17199965}.
%
\subsection{Contributions by the authors}
\textbf{Hagen Holthusen:} Conceptualization, Methodology, Software, Validation, Investigation, Formal Analysis, Data Curation, Writing - Original Draft, Writing - Review \& Editing, Funding acquisition \\
\textbf{Ellen Kuhl:} Writing - Review \& Editing, Supervision, Funding acquisition
%
\subsection{Statement of AI-assisted tools usage}
The authors acknowledge the use of OpenAI’s ChatGPT, an AI language model, for assistance in generating and refining text. 
The authors reviewed, edited, and take full responsibility for the content and conclusions of this work

\bibliographystyle{unsrt}  
\bibliography{references}  

\begin{thebibliography}{100}

\bibitem{lu2021physics}
Lu~Lu, Raphael Pestourie, Wenjie Yao, Zhicheng Wang, Francesc Verdugo, and Steven~G Johnson.
\newblock Physics-informed neural networks with hard constraints for inverse design.
\newblock {\em SIAM Journal on Scientific Computing}, 43(6):B1105--B1132, 2021.

\bibitem{JADOON2025106161}
Asghar~A. Jadoon, Karl~A. Kalina, Manuel~K. Rausch, Reese Jones, and Jan~Niklas Fuhg.
\newblock Inverse design of anisotropic microstructures using physics-augmented neural networks.
\newblock {\em Journal of the Mechanics and Physics of Solids}, 203:106161, 2025.

\bibitem{GHOULI2025106210}
Saeid Ghouli, Moritz Flaschel, Siddhant Kumar, and Laura {De Lorenzis}.
\newblock A topology optimisation framework to design test specimens for one-shot identification or discovery of material models.
\newblock {\em Journal of the Mechanics and Physics of Solids}, 203:106210, 2025.

\bibitem{STOLLBERG2025117808}
Jonathan Stollberg, Tarun Gangwar, Oliver Weeger, and Dominik Schillinger.
\newblock Multiscale topology optimization of functionally graded lattice structures based on physics-augmented neural network material models.
\newblock {\em Computer Methods in Applied Mechanics and Engineering}, 438:117808, 2025.

\bibitem{DANOUN2024117192}
Aymen Danoun, Etienne Prulière, and Yves Chemisky.
\newblock Fe-lstm: A hybrid approach to accelerate multiscale simulations of architectured materials using recurrent neural networks and finite element analysis.
\newblock {\em Computer Methods in Applied Mechanics and Engineering}, 429:117192, 2024.

\bibitem{Kalina2023}
Karl~A. Kalina, Lennart Linden, J\"{o}rg Brummund, and Markus K\"{a}stner.
\newblock Fe\textsuperscript{ANN}: an efficient data-driven multiscale approach based on physics-constrained neural networks and automated data mining.
\newblock {\em Computational Mechanics}, 71(5):827–851, February 2023.

\bibitem{Abueidda2024}
Diab~W. Abueidda and Mostafa~E. Mobasher.
\newblock I-fenn for thermoelasticity based on physics-informed temporal convolutional network (pi-tcn).
\newblock {\em Computational Mechanics}, 74(6):1229–1259, April 2024.

\bibitem{fuhg2024polyconvexneuralnetworkmodels}
Jan~N. Fuhg, Asghar Jadoon, Oliver Weeger, D.~Thomas Seidl, and Reese~E. Jones.
\newblock Polyconvex neural network models of thermoelasticity, 2024.

\bibitem{kehls2025autoencoderbasednonintrusivemodelorder}
Jannick Kehls, Ellen Kuhl, Tim Brepols, Kevin Linka, and Hagen Holthusen.
\newblock Autoencoder-based non-intrusive model order reduction in continuum mechanics, 2025.

\bibitem{hesthaven2018non}
Jan~S Hesthaven and Stefano Ubbiali.
\newblock Non-intrusive reduced order modeling of nonlinear problems using neural networks.
\newblock {\em Journal of Computational Physics}, 363:55--78, 2018.

\bibitem{Flaschel2021}
Moritz Flaschel, Siddhant Kumar, and Laura {De Lorenzis}.
\newblock Unsupervised discovery of interpretable hyperelastic constitutive laws.
\newblock {\em Computer Methods in Applied Mechanics and Engineering}, 381:113852, 2021.

\bibitem{Linka2021}
Kevin Linka, Markus Hillgärtner, Kian~P. Abdolazizi, Roland~C. Aydin, Mikhail Itskov, and Christian~J. Cyron.
\newblock Constitutive artificial neural networks: A fast and general approach to predictive data-driven constitutive modeling by deep learning.
\newblock {\em Journal of Computational Physics}, 429:110010, 2021.

\bibitem{Klein2022}
Dominik~K. Klein, Mauricio Fernández, Robert~J. Martin, Patrizio Neff, and Oliver Weeger.
\newblock Polyconvex anisotropic hyperelasticity with neural networks.
\newblock {\em Journal of the Mechanics and Physics of Solids}, 159:104703, 2022.

\bibitem{CHIACHIO2022104333}
Manuel Chiachío, María Megía, Juan Chiachío, Juan Fernandez, and María~L. Jalón.
\newblock Structural digital twin framework: Formulation and technology integration.
\newblock {\em Automation in Construction}, 140:104333, 2022.

\bibitem{duede2024oilwaterdiffusion}
Eamon Duede, William Dolan, André Bauer, Ian Foster, and Karim Lakhani.
\newblock Oil \& water? diffusion of ai within and across scientific fields, 2024.

\bibitem{Hajkowicz_2023}
Stefan Hajkowicz, Conrad Sanderson, Sarvnaz Karimi, Alexandra Bratanova, and Claire Naughtin.
\newblock Artificial intelligence adoption in the physical sciences, natural sciences, life sciences, social sciences and the arts and humanities: A bibliometric analysis of research publications from 1960-2021.
\newblock {\em Technology in Society}, 74, August 2023.

\bibitem{Ghaboussi1991}
J.~Ghaboussi, J.~H. Garrett, and X.~Wu.
\newblock Knowledge‐based modeling of material behavior with neural networks.
\newblock {\em Journal of Engineering Mechanics}, 117(1):132--153, 1991.

\bibitem{RAISSI2019686}
M.~Raissi, P.~Perdikaris, and G.E. Karniadakis.
\newblock Physics-informed neural networks: A deep learning framework for solving forward and inverse problems involving nonlinear partial differential equations.
\newblock {\em Journal of Computational Physics}, 378:686--707, 2019.

\bibitem{MASI2021104277}
Filippo Masi, Ioannis Stefanou, Paolo Vannucci, and Victor Maffi-Berthier.
\newblock Thermodynamics-based artificial neural networks for constitutive modeling.
\newblock {\em Journal of the Mechanics and Physics of Solids}, 147:104277, 2021.

\bibitem{MASI2022115190}
Filippo Masi and Ioannis Stefanou.
\newblock Multiscale modeling of inelastic materials with thermodynamics-based artificial neural networks (tann).
\newblock {\em Computer Methods in Applied Mechanics and Engineering}, 398:115190, 2022.

\bibitem{Linka2023}
Kevin Linka and Ellen Kuhl.
\newblock A new family of constitutive artificial neural networks towards automated model discovery.
\newblock {\em Computer Methods in Applied Mechanics and Engineering}, 403:115731, 2023.

\bibitem{Linden2023}
Lennart Linden, Dominik~K. Klein, Karl~A. Kalina, J\"{o}rg Brummund, Oliver Weeger, and Markus K\"{a}stner.
\newblock Neural networks meet hyperelasticity: A guide to enforcing physics.
\newblock {\em Journal of the Mechanics and Physics of Solids}, 179:105363, October 2023.

\bibitem{Flaschel2022}
Moritz Flaschel, Siddhant Kumar, and Laura De~Lorenzis.
\newblock Discovering plasticity models without stress data.
\newblock {\em npj Computational Materials}, 8(1), April 2022.

\bibitem{Xu2025}
Haotian Xu, Moritz Flaschel, and Laura De~Lorenzis.
\newblock Discovering non-associated pressure-sensitive plasticity models with euclid.
\newblock {\em Advanced Modeling and Simulation in Engineering Sciences}, 12(1), January 2025.

\bibitem{MARINO2023104643}
Enzo Marino, Moritz Flaschel, Siddhant Kumar, and Laura {De Lorenzis}.
\newblock Automated identification of linear viscoelastic constitutive laws with euclid.
\newblock {\em Mechanics of Materials}, 181:104643, 2023.

\bibitem{FLASCHEL2023115867}
Moritz Flaschel, Siddhant Kumar, and Laura {De Lorenzis}.
\newblock Automated discovery of generalized standard material models with euclid.
\newblock {\em Computer Methods in Applied Mechanics and Engineering}, 405:115867, 2023.

\bibitem{Halphen1975}
Bernard Halphen and Quoc~Son Nguyen.
\newblock Sur les mat{\'e}riaux standard g{\'e}n{\'e}ralis{\'e}s.
\newblock {\em Journal de m{\'e}canique}, 14(1):39--63, 1975.

\bibitem{HACKL1997667}
Klaus Hackl.
\newblock Generalized standard media and variational principles in classical and finite strain elastoplasticity.
\newblock {\em Journal of the Mechanics and Physics of Solids}, 45(5):667--688, 1997.

\bibitem{FLASCHEL2025106103}
Moritz Flaschel, Paul Steinmann, Laura {De Lorenzis}, and Ellen Kuhl.
\newblock Convex neural networks learn generalized standard material models.
\newblock {\em Journal of the Mechanics and Physics of Solids}, 200:106103, 2025.

\bibitem{HUANG2022104856}
Shenglin Huang, Zequn He, Bryan Chem, and Celia Reina.
\newblock Variational onsager neural networks (vonns): A thermodynamics-based variational learning strategy for non-equilibrium pdes.
\newblock {\em Journal of the Mechanics and Physics of Solids}, 163:104856, 2022.

\bibitem{Holthusen2024}
Hagen Holthusen, Lukas Lamm, Tim Brepols, Stefanie Reese, and Ellen Kuhl.
\newblock Theory and implementation of inelastic constitutive artificial neural networks.
\newblock {\em Computer Methods in Applied Mechanics and Engineering}, 428:117063, 2024.

\bibitem{BENADY2024116967}
Antoine Benady, Emmanuel Baranger, and Ludovic Chamoin.
\newblock Unsupervised learning of history-dependent constitutive material laws with thermodynamically-consistent neural networks in the modified constitutive relation error framework.
\newblock {\em Computer Methods in Applied Mechanics and Engineering}, 425:116967, 2024.

\bibitem{Rosenkranz2024}
Max Rosenkranz, Karl~A. Kalina, J{\"o}rg Brummund, WaiChing Sun, and Markus K{\"a}stner.
\newblock Viscoelasticty with physics-augmented neural networks: model formulation and training methods without prescribed internal variables.
\newblock {\em Computational Mechanics}, 74(6):1279--1301, Dec 2024.

\bibitem{Dettmer2024}
Wulf~G. Dettmer, Eugenio~J. Muttio, Reem Alhayki, and Djordje Perić.
\newblock A framework for neural network based constitutive modelling of inelastic materials.
\newblock {\em Computer Methods in Applied Mechanics and Engineering}, 420:116672, 2024.

\bibitem{ASAD2023116463}
Faisal As’ad and Charbel Farhat.
\newblock A mechanics-informed deep learning framework for data-driven nonlinear viscoelasticity.
\newblock {\em Computer Methods in Applied Mechanics and Engineering}, 417:116463, 2023.

\bibitem{Holthusen2025}
Hagen Holthusen, Kevin Linka, Ellen Kuhl, and Tim Brepols.
\newblock A generalized dual potential for inelastic constitutive artificial neural networks: A jax implementation at finite strains.
\newblock {\em Journal of the Mechanics and Physics of Solids}, page 106337, 2025.

\bibitem{Jadoon2025}
Asghar~Arshad Jadoon, Knut~Andreas Meyer, and Jan~Niklas Fuhg.
\newblock Automated model discovery of finite strain elastoplasticity from uniaxial experiments.
\newblock {\em Computer Methods in Applied Mechanics and Engineering}, 435:117653, 2025.

\bibitem{Boes2025}
Birte Boes, Jaan-Willem Simon, and Hagen Holthusen.
\newblock Accounting for plasticity: An extension of inelastic constitutive artificial neural networks, 2025.

\bibitem{DAMMA2025117937}
Franz Dammaß, Karl~A. Kalina, and Markus Kästner.
\newblock Neural networks meet phase-field: A hybrid fracture model.
\newblock {\em Computer Methods in Applied Mechanics and Engineering}, 440:117937, 2025.

\bibitem{Holthusen2025growth}
Hagen Holthusen, Tim Brepols, Kevin Linka, and Ellen Kuhl.
\newblock Automated model discovery for tensional homeostasis: Constitutive machine learning in growth and remodeling.
\newblock {\em Computers in Biology and Medicine}, 186:109691, March 2025.

\bibitem{jadoon2025thermodynamicallyconsistenthybridpermutationinvariant}
Asghar~A. Jadoon, Ravi~G. Patel, Brian~N. Granzow, Reese~E. Jones, D.~Thomas Seidl, and Jan~N. Fuhg.
\newblock Thermodynamically consistent hybrid and permutation-invariant neural yield functions for anisotropic plasticity, 2025.

\bibitem{Amos2017}
Brandon Amos, Lei Xu, and J.~Zico Kolter.
\newblock Input convex neural networks, 2017.

\bibitem{KUMAR2025118159}
Sanjeev Kumar, D.~Thomas Seidl, Brian~N. Granzow, Jin Yang, and Jan~Niklas Fuhg.
\newblock A comparative study of calibration techniques for finite strain elastoplasticity: Numerically-exact sensitivities for femu and vfm.
\newblock {\em Computer Methods in Applied Mechanics and Engineering}, 444:118159, 2025.

\bibitem{Jones2022}
Reese~E. Jones, Ari~L. Frankel, and K.~L. Johnson.
\newblock A neural ordinary differential equation framework for modeling inelastic stress response via internal state variables.
\newblock {\em Journal of Machine Learning for Modeling and Computing}, 3(3):1–35, 2022.

\bibitem{jones2025attentionbasedneuralordinarydifferential}
Reese~E. Jones and Jan~N. Fuhg.
\newblock An attention-based neural ordinary differential equation framework for modeling inelastic processes, 2025.

\bibitem{BORKOWSKI2022106678}
L.~Borkowski, C.~Sorini, and A.~Chattopadhyay.
\newblock Recurrent neural network-based multiaxial plasticity model with regularization for physics-informed constraints.
\newblock {\em Computers \& Structures}, 258:106678, 2022.

\bibitem{LI2025105325}
Jin-Zhao Li, Zhi-Ping Guan, Jiong-Rui Chen, and Hui-Chao Jin.
\newblock A long short-term memory-based constitutive modeling framework for capturing strain path dependence in plastic deformation.
\newblock {\em Mechanics of Materials}, 205:105325, 2025.

\bibitem{GUO2025118358}
Binyao Guo, Zihan Lin, and QiZhi He.
\newblock History-aware neural operator: Robust data-driven constitutive modeling of path-dependent materials.
\newblock {\em Computer Methods in Applied Mechanics and Engineering}, 447:118358, 2025.

\bibitem{Maurizi2022}
Marco Maurizi, Chao Gao, and Filippo Berto.
\newblock Predicting stress, strain and deformation fields in materials and structures with graph neural networks.
\newblock {\em Scientific Reports}, 12(1), December 2022.

\bibitem{patel2025generalautomatedmethodbuilding}
Ravi~G. Patel, Reese~E. Jones, D.~Thomas Seidl, Brian~N. Granzow, and Jan~N. Fuhg.
\newblock A general, automated method for building structural tensors of arbitrary order for anisotropic function representations, 2025.

\bibitem{TAC2022115248}
Vahidullah Tac, Francisco {Sahli Costabal}, and Adrian~B. Tepole.
\newblock Data-driven tissue mechanics with polyconvex neural ordinary differential equations.
\newblock {\em Computer Methods in Applied Mechanics and Engineering}, 398:115248, 2022.

\bibitem{FUHG2022114915}
Jan~N. Fuhg and Nikolaos Bouklas.
\newblock On physics-informed data-driven isotropic and anisotropic constitutive models through probabilistic machine learning and space-filling sampling.
\newblock {\em Computer Methods in Applied Mechanics and Engineering}, 394:114915, 2022.

\bibitem{KALINA2025117725}
Karl~A. Kalina, Jörg Brummund, WaiChing Sun, and Markus Kästner.
\newblock Neural networks meet anisotropic hyperelasticity: A framework based on generalized structure tensors and isotropic tensor functions.
\newblock {\em Computer Methods in Applied Mechanics and Engineering}, 437:117725, 2025.

\bibitem{MCCULLOCH2024461}
Jeremy~A. McCulloch and Ellen Kuhl.
\newblock Automated model discovery for textile structures: The unique mechanical signature of warp knitted fabrics.
\newblock {\em Acta Biomaterialia}, 189:461--477, 2024.

\bibitem{Tac2022}
Vahidullah Tac, Vivek~D. Sree, Manuel~K. Rausch, and Adrian~B. Tepole.
\newblock Data-driven modeling of the mechanical behavior of anisotropic soft biological tissue.
\newblock {\em Engineering with Computers}, 38(5):4167–4182, September 2022.

\bibitem{Osman2025}
Osman G\"ultekin, Ahmad Moeineddin, Barış Cansız, Krunoslav Sveric, Axel Linke, and Michael Kaliske.
\newblock A physics-informed neural network model for the anisotropic hyperelasticity of the human passive myocardium.
\newblock {\em International Journal for Numerical Methods in Engineering}, 126(14):e70067, 2025.

\bibitem{Martonov2025}
Denisa Martonová, Sigrid Leyendecker, Gerhard~A. Holzapfel, and Ellen Kuhl.
\newblock Discovering dispersion: How robust is automated model discovery for human myocardial tissue?
\newblock {\em Biomechanics and Modeling in Mechanobiology}, August 2025.

\bibitem{Vervenne2025}
Thibault Vervenne, Mathias Peirlinck, Nele Famaey, and Ellen Kuhl.
\newblock Constitutive neural networks for main pulmonary arteries: discovering the undiscovered.
\newblock {\em Biomechanics and Modeling in Mechanobiology}, 24(2):615–634, February 2025.

\bibitem{fuhg2024reviewdatadrivenconstitutivelaws}
Jan~Niklas Fuhg, Govinda~Anantha Padmanabha, Nikolaos Bouklas, Bahador Bahmani, WaiChing Sun, Nikolaos~N. Vlassis, Moritz Flaschel, Pietro Carrara, and Laura~De Lorenzis.
\newblock A review on data-driven constitutive laws for solids, 2024.

\bibitem{watson2025machinelearningphysicsknowledge}
Joe Watson, Chen Song, Oliver Weeger, Theo Gruner, An~T. Le, Kay Pompetzki, Ahmed Hendawy, Oleg Arenz, Will Trojak, Miles Cranmer, Carlo D'Eramo, Fabian Bülow, Tanmay Goyal, Jan Peters, and Martin~W. Hoffman.
\newblock Machine learning with physics knowledge for prediction: A survey, 2025.

\bibitem{Germain1983}
P.~Germain, Q.~S. Nguyen, and P.~Suquet.
\newblock Continuum thermodynamics.
\newblock {\em Journal of Applied Mechanics}, 50(4b):1010–1020, December 1983.

\bibitem{Rockafellar1970}
Ralph~Tyrell Rockafellar.
\newblock {\em Convex Analysis}.
\newblock Princeton University Press, Princeton, 1970.

\bibitem{Sill1997MonotonicN}
Joseph Sill.
\newblock Monotonic networks.
\newblock In {\em Neural Information Processing Systems}, 1997.

\bibitem{Daniels2010}
Hennie Daniels and Marina Velikova.
\newblock Monotone and partially monotone neural networks.
\newblock {\em IEEE Transactions on Neural Networks}, 21(6):906–917, June 2010.

\bibitem{you2017deeplatticenetworkspartial}
Seungil You, David Ding, Kevin Canini, Jan Pfeifer, and Maya Gupta.
\newblock Deep lattice networks and partial monotonic functions, 2017.

\bibitem{Asad2023}
Faisal As’ad and Charbel Farhat.
\newblock A mechanics-informed neural network framework for data-driven nonlinear viscoelasticity.
\newblock In {\em AIAA SCITECH 2023 Forum}. American Institute of Aeronautics and Astronautics, January 2023.

\bibitem{Hasani2021}
Ramin Hasani, Mathias Lechner, Alexander Amini, Daniela Rus, and Radu Grosu.
\newblock Liquid time-constant networks.
\newblock {\em Proceedings of the AAAI Conference on Artificial Intelligence}, 35(9):7657--7666, May 2021.

\bibitem{Cox2002}
S.M. Cox and P.C. Matthews.
\newblock Exponential time differencing for stiff systems.
\newblock {\em Journal of Computational Physics}, 176(2):430–455, March 2002.

\bibitem{Itskov2004}
Mikhail Itskov.
\newblock On the application of the additive decomposition of generalized strain measures in large strain plasticity.
\newblock {\em Mechanics Research Communications}, 31(5):507--517, 2004.

\bibitem{Eckart1948}
Carl Eckart.
\newblock The thermodynamics of irreversible processes. iv. the theory of elasticity and anelasticity.
\newblock {\em Physical Review}, 73(4):373–382, February 1948.

\bibitem{Kroener1959}
Ekkehart Kr\"{o}ner.
\newblock Allgemeine kontinuumstheorie der versetzungen und eigenspannungen.
\newblock {\em Archive for Rational Mechanics and Analysis}, 4(1):273–334, January 1959.

\bibitem{Sidoroff1974}
Fran\c~cois Sidoroff.
\newblock Un mod\`ele visco\'elastique non lin\'eaire avec configuration interm\'ediaire.
\newblock {\em J. M\'ecanique}, 13:679--713, 1974.

\bibitem{Rodriguez1994}
Edward~K. Rodriguez, Anne Hoger, and Andrew~D. McCulloch.
\newblock Stress-dependent finite growth in soft elastic tissues.
\newblock {\em Journal of Biomechanics}, 27(4):455–467, April 1994.

\bibitem{Christ2009}
Daniel Christ and Stefanie Reese.
\newblock A finite element model for shape memory alloys considering thermomechanical couplings at large strains.
\newblock {\em International Journal of Solids and Structures}, 46(20):3694–3709, October 2009.

\bibitem{Casey2016}
James Casey.
\newblock A convenient form of the multiplicative decomposition of the deformation gradient.
\newblock {\em Mathematics and Mechanics of Solids}, 22(3):528–537, August 2016.

\bibitem{Boehler1979}
Jean-Paul Boehler.
\newblock A simple derivation of representations for non-polynomial constitutive equations in some cases of anisotropy.
\newblock {\em ZAMM-Journal of Applied Mathematics and Mechanics/Zeitschrift f{\"u}r Angewandte Mathematik und Mechanik}, 59(4):157--167, 1979.

\bibitem{Spencer1984}
A.~J.~M. Spencer.
\newblock {\em Constitutive Theory for Strongly Anisotropic Solids}, page 1–32.
\newblock Springer Vienna, 1984.

\bibitem{Schroeder2003}
J\"{o}rg Schr\"{o}der and Patrizio Neff.
\newblock Invariant formulation of hyperelastic transverse isotropy based on polyconvex free energy functions.
\newblock {\em International Journal of Solids and Structures}, 40(2):401–445, January 2003.

\bibitem{Ball1976}
John~M. Ball.
\newblock Convexity conditions and existence theorems in nonlinear elasticity.
\newblock {\em Archive for Rational Mechanics and Analysis}, 63(4):337–403, December 1976.

\bibitem{Ball1977}
John~M Ball.
\newblock Constitutive inequalities and existence theorems in nonlinear elastostatics.
\newblock In {\em Nonlinear analysis and mechanics: Heriot-Watt symposium}, volume~1, pages 187--241. Pitman London, 1977.

\bibitem{Marsden1994}
Jerrold~E Marsden and Thomas~JR Hughes.
\newblock {\em Mathematical foundations of elasticity}.
\newblock Courier Corporation, 1994.

\bibitem{Svendsen2001}
Bob Svendsen.
\newblock On the modelling of anisotropic elastic and inelastic material behaviour at large deformation.
\newblock {\em International Journal of Solids and Structures}, 38(52):9579–9599, December 2001.

\bibitem{Coleman1961}
Bernard~D. Coleman and Walter Noll.
\newblock Foundations of linear viscoelasticity.
\newblock {\em Reviews of Modern Physics}, 33(2):239–249, April 1961.

\bibitem{Coleman1963}
Bernard~D. Coleman and Walter Noll.
\newblock The thermodynamics of elastic materials with heat conduction and viscosity.
\newblock {\em Archive for Rational Mechanics and Analysis}, 13(1):167–178, December 1963.

\bibitem{Coleman1967}
Bernard~D. Coleman and Morton~E. Gurtin.
\newblock Thermodynamics with internal state variables.
\newblock {\em The Journal of Chemical Physics}, 47(2):597–613, July 1967.

\bibitem{Dettmer2004}
Wulf Dettmer and Stefanie Reese.
\newblock On the theoretical and numerical modelling of armstrong–frederick kinematic hardening in the finite strain regime.
\newblock {\em Computer Methods in Applied Mechanics and Engineering}, 193(1–2):87–116, January 2004.

\bibitem{Reese2021}
S.~Reese, T.~Brepols, M.~Fassin, L.~Poggenpohl, and S.~Wulfinghoff.
\newblock Using structural tensors for inelastic material modeling in the finite strain regime – a novel approach to anisotropic damage.
\newblock {\em Journal of the Mechanics and Physics of Solids}, 146:104174, January 2021.

\bibitem{Rice1971}
J~R Rice.
\newblock Inelastic constitutive relations for solids: An internal-variable theory and its application to metal plasticity.
\newblock {\em J. Mech. Phys. Solids}, 19(6):433--455, November 1971.

\bibitem{Germain1998}
Paul Germain.
\newblock Functional concepts in continuum mechanics.
\newblock {\em Meccanica}, 33(5):433–444, October 1998.

\bibitem{Dafalias1987}
Y.~F. Dafalias.
\newblock Issues on the constitutive formulation at large elastoplastic deformations, part 1: Kinematics.
\newblock {\em Acta Mechanica}, 69(1–4):119–138, December 1987.

\bibitem{Reese2003}
Stefanie Reese.
\newblock Meso-macro modelling of fibre-reinforced rubber-like composites exhibiting large elastoplastic deformation.
\newblock {\em International Journal of Solids and Structures}, 40(4):951–980, February 2003.

\bibitem{Boes2023}
Birte Boes, Jaan-Willem Simon, Stefanie Reese, and Hagen Holthusen.
\newblock A novel continuum mechanical framework for decoupled material behavior in thickness and in-plane directions.
\newblock {\em Computer Methods in Applied Mechanics and Engineering}, 415:116192, October 2023.

\bibitem{Zheng1994}
Q.-S. Zheng.
\newblock Theory of representations for tensor functions—a unified invariant approach to constitutive equations.
\newblock {\em Applied Mechanics Reviews}, 47(11):545–587, November 1994.

\bibitem{Gluege2017}
Rainer Gl\"{u}ge and Sara Bucci.
\newblock Does convexity of yield surfaces in plasticity have a physical significance?
\newblock {\em Mathematics and Mechanics of Solids}, 23(9):1364–1373, August 2017.

\bibitem{Janeka2018}
Adam Janečka and Michal Pavelka.
\newblock Non-convex dissipation potentials in multiscale non-equilibrium thermodynamics.
\newblock {\em Continuum Mechanics and Thermodynamics}, 30(4):917–941, April 2018.

\bibitem{Argyris1974}
J.H. Argyris, G.~Faust, J.~Szimmat, E.P. Warnke, and K.J. Willam.
\newblock Recent developments in the finite element analysis of prestressed concrete reactor vessels.
\newblock {\em Nuclear Engineering and Design}, 28(1):42–75, July 1974.

\bibitem{Holthusen2023}
Hagen Holthusen, Christiane Rothkranz, Lukas Lamm, Tim Brepols, and Stefanie Reese.
\newblock Inelastic material formulations based on a co-rotated intermediate configuration—application to bioengineered tissues.
\newblock {\em Journal of the Mechanics and Physics of Solids}, 172:105174, March 2023.

\bibitem{Liu2025}
Zeng Liu, Rogelio Ortigosa, Antonio~J. Gil, and Javier Bonet.
\newblock Large strain constitutive modelling of soft compressible and incompressible solids: Generalised isotropic and anisotropic viscoelasticity.
\newblock {\em Journal of the Mechanics and Physics of Solids}, 203:106194, 2025.

\bibitem{Klein2022b}
Dominik~K. Klein, Rogelio Ortigosa, Jesús Martínez-Frutos, and Oliver Weeger.
\newblock Finite electro-elasticity with physics-augmented neural networks.
\newblock {\em Computer Methods in Applied Mechanics and Engineering}, 400:115501, 2022.

\bibitem{Vladimirov2008}
Ivaylo~N. Vladimirov, Michael~P. Pietryga, and Stefanie Reese.
\newblock On the modelling of non-linear kinematic hardening at finite strains with application to springback—comparison of time integration algorithms.
\newblock {\em International Journal for Numerical Methods in Engineering}, 75(1):1--28, 2008.

\bibitem{Hendrycks2023}
Dan Hendrycks and Kevin Gimpel.
\newblock Gaussian error linear units (gelus), 2023.

\bibitem{jax2018github}
James Bradbury, Roy Frostig, Peter Hawkins, Matthew~James Johnson, Chris Leary, Dougal Maclaurin, George Necula, Adam Paszke, Jake VanderPlas, Skye Wanderman-Milne, and Qiao Zhang.
\newblock Jax: composable transformations of {P}ython+{N}um{P}y programs.
\newblock {\em GitHub repository}, 2018.

\bibitem{geuzaine2009gmsh}
Christophe Geuzaine and Jean-Fran{\c{c}}ois Remacle.
\newblock Gmsh: A 3-d finite element mesh generator with built-in pre- and post-processing facilities.
\newblock {\em International Journal for Numerical Methods in Engineering}, 79(11):1309--1331, 2009.

\bibitem{ayachit2015paraview}
Utkarsh Ayachit.
\newblock The paraview guide: A parallel visualization application.
\newblock 2015.

\bibitem{Schroeder2008}
J.~Schröder, P.~Neff, and V.~Ebbing.
\newblock Anisotropic polyconvex energies on the basis of crystallographic motivated structural tensors.
\newblock {\em Journal of the Mechanics and Physics of Solids}, 56(12):3486--3506, 2008.

\bibitem{Reese1998}
Stefanie Reese and Sanjay Govindjee.
\newblock A theory of finite viscoelasticity and numerical aspects.
\newblock {\em International Journal of Solids and Structures}, 35(26):3455--3482, 1998.

\bibitem{Kalina2021}
Karl~A. Kalina, Lennart Linden, J\"{o}rg Brummund, Philipp Metsch, and Markus K\"{a}stner.
\newblock Automated constitutive modeling of isotropic hyperelasticity based on artificial neural networks.
\newblock {\em Computational Mechanics}, 69(1):213–232, October 2021.

\bibitem{Elman1990}
Jeffrey~L. Elman.
\newblock Finding structure in time.
\newblock {\em Cognitive Science}, 14(2):179--211, 1990.

\bibitem{Wang2023}
Lucy~M. Wang, Kevin Linka, and Ellen Kuhl.
\newblock Automated model discovery for muscle using constitutive recurrent neural networks.
\newblock {\em Journal of the Mechanical Behavior of Biomedical Materials}, 145:106021, 2023.

\bibitem{Cook1981}
Robert~D. Cook, David~S. Malkus, and Michael~E. Plesha.
\newblock {\em Concepts and Applications of Finite Element Analysis}.
\newblock John Wiley \& Sons, 2nd edition, 1981.

\bibitem{Guan2026}
Jiashen Guan, Xin Li, Hongyan Yuan, and Ju~Liu.
\newblock Hyperelastic modeling based on generalized landau invariants and multi-stage calibration.
\newblock {\em Journal of the Mechanics and Physics of Solids}, 206:106338, 2026.

\bibitem{Anassi2024}
Afshin Anssari-Benam, Alain Goriely, and Giuseppe Saccomandi.
\newblock Generalised invariants and pseudo-universal relationships for hyperelastic materials: A new approach to constitutive modelling.
\newblock {\em Journal of the Mechanics and Physics of Solids}, 193:105883, 2024.

\bibitem{Martonova2025}
Denisa Martonová, Alain Goriely, and Ellen Kuhl.
\newblock Generalized invariants meet constitutive neural networks: A novel framework for hyperelastic materials.
\newblock {\em Journal of the Mechanics and Physics of Solids}, page 106352, 2025.

\bibitem{Wiedemann2023}
David Wiedemann and Malte~A. Peter.
\newblock Characterization of polyconvex isotropic functions, 2023.

\bibitem{Geuken2025}
Gian-Luca Geuken, Patrick Kurzeja, David Wiedemann, and Jörn Mosler.
\newblock A novel neural network for isotropic polyconvex hyperelasticity satisfying the universal approximation theorem.
\newblock {\em Journal of the Mechanics and Physics of Solids}, 203:106209, 2025.

\bibitem{Charton2024}
François Charton, Jordan~S. Ellenberg, Adam~Zsolt Wagner, and Geordie Williamson.
\newblock Patternboost: Constructions in mathematics with a little help from ai, 2024.

\bibitem{Prume2025}
E.~Prume, C.~Gierden, M.~Ortiz, and S.~Reese.
\newblock Direct data-driven algorithms for multiscale mechanics.
\newblock {\em Computer Methods in Applied Mechanics and Engineering}, 433:117525, 2025.

\bibitem{Stainier2019}
Laurent Stainier, Adrien Leygue, and Michael Ortiz.
\newblock Model-free data-driven methods in mechanics: material data identification and solvers.
\newblock {\em Computational Mechanics}, 64(2):381–393, June 2019.

\bibitem{Linden2025}
Lennart Linden, Karl~A. Kalina, Jörg Brummund, Brain Riemer, and Markus Kästner.
\newblock A dual-stage constitutive modeling framework based on finite strain data-driven identification and physics-augmented neural networks, 2025.

\bibitem{Flaschel2025}
Moritz Flaschel, Denisa Martonová, Carina Veil, and Ellen Kuhl.
\newblock Material fingerprinting: A shortcut to material model discovery without solving optimization problems, 2025.

\bibitem{Eghtesand2024}
Adnan Eghtesad, Jingye Tan, Jan~Niklas Fuhg, and Nikolaos Bouklas.
\newblock Nn-evp: A physics informed neural network-based elasto-viscoplastic framework for predictions of grain size-aware flow response.
\newblock {\em International Journal of Plasticity}, 181:104072, 2024.

\bibitem{Flaschel2023a}
Moritz Flaschel, Huitian Yu, Nina Reiter, Jan Hinrichsen, Silvia Budday, Paul Steinmann, Siddhant Kumar, and Laura {De Lorenzis}.
\newblock Automated discovery of interpretable hyperelastic material models for human brain tissue with euclid.
\newblock {\em Journal of the Mechanics and Physics of Solids}, 180:105404, 2023.

\bibitem{Poyser2024}
Matt Poyser and Toby~P. Breckon.
\newblock Neural architecture search: A contemporary literature review for computer vision applications.
\newblock {\em Pattern Recognition}, 147:110052, 2024.

\end{thebibliography}


\end{document}